\documentclass[]{fairmeta}
\usepackage{graphicx}
\usepackage{amsmath}
\usepackage[version=4]{mhchem}
\usepackage{booktabs}
\usepackage{caption}

\usepackage[utf8]{inputenc} % allow utf-8 input
\usepackage[T1]{fontenc}    % use 8-bit T1 fonts
\usepackage{hyperref}       % hyperlinks
\usepackage{url}            % simple URL typesetting
\usepackage{booktabs}       % professional-quality tables
\usepackage{amsfonts}       % blackboard math symbols
\usepackage{nicefrac}       % compact symbols for 1/2, etc.
\usepackage{microtype}      % microtypography
\usepackage{xifthen}
\usepackage{multirow}
\usepackage{mciteplus}
\usepackage{epstopdf}
\usepackage{shellesc}

\usepackage{siunitx}
\usepackage{mhchem}
\usepackage{threeparttable}

\newcommand\mcc[1]{\multicolumn{1}{c}{#1}}  % centering a single table cell
\newcommand{\angs}{\text{\AA}}
\newcommand{\omol}{OMol25}

\newcommand{\kindatiny}{\fontsize{6pt}{7.2pt}\selectfont}

\newlength\savewidth\newcommand\shline{\noalign{\global\savewidth\arrayrulewidth
  \global\arrayrulewidth 0.5pt}\hline\noalign{\global\arrayrulewidth\savewidth}}
\newcommand{\tablestyle}[2]{%
    \fontfamily{ptm}\selectfont%
    \let\itold\it%
    \def\it{\itold \fontfamily{ptm}\selectfont}%
    \setlength{\tabcolsep}{#1}\renewcommand{\arraystretch}{#2}\centering\kindatiny%
    \let\citeold\cite%
    \renewcommand{\cite}[1]{\normalfont\fontfamily{ptm}\selectfont\tiny\citeold{##1}}%
}

\newcolumntype{x}[1]{>{\centering\arraybackslash}p{#1pt}}
\newcolumntype{y}[1]{>{\raggedright\arraybackslash}p{#1pt}}
\newcolumntype{z}[1]{>{\raggedleft\arraybackslash}p{#1pt}}
\newcolumntype{w}{>{\centering\arraybackslash}p{18pt}}
\newcolumntype{a}{>{\centering\arraybackslash}p{16pt}}

\definecolor{c0-title-bkg}{HTML}{ffffff}
\definecolor{c0-title-text}{HTML}{000000}
\definecolor{c0-item-bkg}{HTML}{ffffff}
\definecolor{c0-item-text}{HTML}{818589}

\definecolor{c1-title-bkg}{HTML}{d1e2dd}
\definecolor{c1-title-text}{HTML}{005953}
\definecolor{c1-item-bkg}{HTML}{e6efec}
\definecolor{c1-item-text}{HTML}{2d7b6d}

\definecolor{c2-title-bkg}{HTML}{cfe1e1}
\definecolor{c2-title-text}{HTML}{005760}
\definecolor{c2-item-bkg}{HTML}{e4eeed}
\definecolor{c2-item-text}{HTML}{24797b}

\definecolor{c3-title-bkg}{HTML}{cddfe5}
\definecolor{c3-title-text}{HTML}{330704}
\definecolor{c3-item-bkg}{HTML}{facac5}
\definecolor{c3-item-text}{HTML}{330704}

\definecolor{c4-title-bkg}{HTML}{cedce8}
\definecolor{c4-title-text}{HTML}{191b1c}
\definecolor{c4-item-bkg}{HTML}{e2edf4}
\definecolor{c4-item-text}{HTML}{191b1c}

\definecolor{c5-title-bkg}{HTML}{d0d9eb}
\definecolor{c5-title-text}{HTML}{0e1834}
\definecolor{c5-item-bkg}{HTML}{d0e6f9}
\definecolor{c5-item-text}{HTML}{0e1834}

\definecolor{c6-title-bkg}{HTML}{d3d5ed}
\definecolor{c6-title-text}{HTML}{220b1b}
\definecolor{c6-item-bkg}{HTML}{f5d8f1}
\definecolor{c6-item-text}{HTML}{220b1b}

\definecolor{c7-title-bkg}{HTML}{dad1ed}
\definecolor{c7-title-text}{HTML}{1c0b14}
\definecolor{c7-item-bkg}{HTML}{d58cb5}
\definecolor{c7-item-text}{HTML}{1c0b14}

\definecolor{c8-title-bkg}{HTML}{ded1ec}
\definecolor{c8-title-text}{HTML}{633273}
\definecolor{c8-item-bkg}{HTML}{D8BFD8}
\definecolor{c8-item-text}{HTML}{7c5997}

\definecolor{c9-title-bkg}{HTML}{e5d1eb}
\definecolor{c9-title-text}{HTML}{6c2f6b}
\definecolor{c9-item-bkg}{HTML}{f0e0f6}
\definecolor{c9-item-text}{HTML}{885591}

\definecolor{c10-title-bkg}{HTML}{ebd1e7}
\definecolor{c10-title-text}{HTML}{722e5f}
\definecolor{c10-item-bkg}{HTML}{f5e2f3}
\definecolor{c10-item-text}{HTML}{915487}

\definecolor{avg-title-bkg}{HTML}{f3f3f3}
\definecolor{avg-title-text}{HTML}{000000}
\definecolor{avg-item-bkg}{HTML}{f3f3f3}
\definecolor{avg-item-text}{HTML}{000000}

\newcommand{\addpadding}{%
  \rule{0pt}{\dimexpr\normalbaselineskip-0.5pt\relax}%
}

\newcommand{\ct}[2][c0]{\addpadding{\cellcolor{#1-item-bkg}\textcolor{#1-title-text}{#2}}}

\ExplSyntaxOn

\cs_generate_variant:Nn \coffin_typeset:Nnnnn {Nffff}

\NewDocumentCommand\rotbox{ O{l,H} D<>{0pt,0pt} m m}{
    \hcoffin_set:Nn \l_tmpa_coffin {#4}
    \coffin_rotate:Nn \l_tmpa_coffin {#3}
    \coffin_typeset:Nffff \l_tmpa_coffin 
        {\clist_item:nn{#1}{1}}
        {\clist_item:nn{#1}{2}}
        {\clist_item:nn{#2}{1}}
        {\clist_item:nn{#2}{2}}
}
\ExplSyntaxOff

\newlength{\ccustomlen}
\newcommand{\ccustom}[3][c0]{%
    \cellcolor{#1-item-bkg}{%
        \rotbox[l,t]{90}{%
            \parbox[t]{\ccustomlen}{%
                \ifthenelse{\isempty{#3}}{%
                    \mbox{%
                        \kindatiny\textcolor{#1-title-text}{#2}%
                    }%
                }{%
                    \kindatiny\textcolor{#1-title-text}{#2} \\%
                    \tiny{\textcolor{#1-item-text}{\it #3}}%
                }%
            }%
        }%
    }%
}

\newcommand{\cb}[3][c0]{%
    \setlength{\ccustomlen}{1.5cm}%
    \ccustom[#1]{#2}{#3}%
}

\usepackage{xspace}

\makeatother
\title{The Open Molecules 2025 (\omol) Dataset, Evaluations, and Models}

\author[1, *]{Daniel S. Levine}
\author[1, *]{Muhammed Shuaibi}
\author[2]{Michael G. Taylor}
\author[3,4]{Evan Walter Clark Spotte-Smith}
\author[5]{Muhammad R. Hasyim}
\author[1]{Kyle Michel}
\author[6]{Ilyes Batatia}
\author[6,7,8]{Gábor Csányi}
\author[1]{Misko Dzamba}
\author[9]{Peter Eastman}
\author[10]{Nathan C. Frey}
\author[1]{Xiang Fu}
\author[1]{Vahe Gharakhanyan}
\author[11,12,13]{Aditi S. Krishnapriyan}
\author[14,15]{Nitesh Kumar}
\author[10]{Joshua A. Rackers}
\author[11]{Sanjeev Raja}
\author[1]{Ammar Rizvi}
\author[16]{Andrew S. Rosen}
\author[1]{Zachary Ulissi}
\author[15,17]{Santiago Vargas} 
\author[1, \dagger]{C. Lawrence Zitnick}
\author[15,18, \dagger]{Samuel M. Blau}
\author[1, \dagger]{Brandon M. Wood}

\affiliation[1]{FAIR at Meta, San Francisco, CA, USA}
\affiliation[2]{Theoretical Division, Los Alamos National Laboratory, Los Alamos, NM, USA}
\affiliation[3]{School of Chemistry, University College Dublin, Dublin, Co. Dublin, Ireland}
\affiliation[4]{Department of Chemical Engineering, Carnegie Mellon University, Pittsburgh, PA, USA}
\affiliation[5]{Simons Center for Computational Physical Chemistry, New York University, New York, NY, USA}
\affiliation[6]{Engineering Laboratory, University of Cambridge, Cambridge, UK}
\affiliation[7]{Max Planck Institute of Polymer Research, Mainz, Germany}
\affiliation[8]{\AA ngstrom AI, Inc, Delaware, USA}
\affiliation[9]{Department of Chemistry, Stanford University, Stanford, CA, USA}
\affiliation[10]{Prescient Design, Genentech, New York, NY, USA}
\affiliation[11]{Department of Electrical Engineering and Computer Sciences, University of California, Berkeley, CA, USA}
\affiliation[12]{Department of Chemical and Biomolecular Engineering, University of California, Berkeley, CA, USA}
\affiliation[13]{Applied Mathematics and Computational Research, Lawrence Berkeley National Laboratory, Berkeley, CA, USA}
\affiliation[14]{Materials Sciences Division, Lawrence Berkeley National Laboratory, Berkeley, CA, USA}
\affiliation[15]{Energy Technologies Area, Lawrence Berkeley National Laboratory, Berkeley, CA, USA}
\affiliation[16]{Department of Chemical and Biological Engineering, Princeton University, Princeton, NJ, USA}
\affiliation[17]{Chemical Sciences Division, Lawrence Berkeley National Laboratory, Berkeley, CA, USA}
\affiliation[18]{Bakar Institute of Digital Materials for the Planet, University of California, Berkeley, CA, USA}

\contribution[*]{Co-first Author}
\contribution[\dagger]{Co-corresponding Author}

\abstract{Machine learning (ML) models hold the promise of transforming atomic simulations by delivering quantum chemical accuracy at a fraction of the computational cost. Realization of this potential would enable high-throughout, high-accuracy molecular screening campaigns to explore vast regions of chemical space and facilitate \textit{ab initio}-level simulations at sizes and time scales that were previously inaccessible. However, a fundamental challenge to creating ML models that perform well across molecular chemistry is the lack of comprehensive data for training. Despite substantial efforts in data generation, no large-scale molecular dataset exists that combines broad chemical diversity with a high level of accuracy. To address this gap, Meta FAIR introduces Open Molecules 2025 (\omol), a large-scale dataset composed of more than 140 million density functional theory (DFT) calculations at the $\omega$B97M-V/def2-TZVPD level of theory, representing billions of CPU core-hours of compute. \omol~uniquely blends elemental, chemical, and structural diversity including: 83 elements, a wide-range of intra- and intermolecular interactions, explicit solvation, variable charge/spin, conformers, and reactive structures. There are $\sim$83M unique molecular systems in \omol~covering small molecules, biomolecules, metal complexes, and electrolytes, including structures obtained from existing datasets. \omol~also greatly expands on the size of systems typically included in DFT datasets, with systems of up to 350 atoms. In addition to the public release of the data, we provide baseline models and a comprehensive set of model evaluations to encourage community engagement in developing the next-generation ML models for molecular chemistry.}

\metadata[Dataset]{\url{https://huggingface.co/facebook/OMol25}}
\metadata[Models]{\url{https://huggingface.co/facebook/OMol25}}
\metadata[Code]{\url{https://github.com/facebookresearch/fairchem}}
\correspondence{C.L.Z. (\email{zitnick@meta.com}), S.M.B. (\email{smblau@lbl.gov}), B.M.W. (\email{bmwood@meta.com})}

\let\oldaddcontentsline\addcontentsline
\renewcommand{\addcontentsline}[3]{}

\mciteErrorOnUnknownfalse

\begin{document}

\maketitle

\section{Introduction}
Molecular chemistry is a cornerstone of modern society, driving innovation in areas such as medicine~\cite{drug_design, Lombardino2004}, energy production~\cite{10.1021/cr068360d, 10.1021/ar980112j} and storage~\cite{Placke2017, MILLER2021249}, advanced computing~\cite{10.1021/acsomega.1c02912, 5389067}, agriculture~\cite{YAHAYA2023385, Leigh2004}, and more. Further progress in these areas hinges on the design and discovery of molecular systems with improved or novel properties~\cite{10.1126/science.aat2663}. Computational chemistry has become an essential tool to aid in this pursuit~\cite{10.1063/1.2911179}. It offers a means to efficiently explore the molecular design space and provide mechanistic insights, both of which are difficult to achieve with experimentation alone~\cite{10.1021/acscatal.7b03477, SpotteSmith2022, 10.1021/acs.chemrev.8b00399, GomezBombarelli2016, QU201556}. 

Over the past few decades, Density Functional Theory (DFT) has become the prevailing quantum chemistry modeling tool~\cite{Becke_dft_1993, Lee_dft_1988, mardirossian_thirty_2017} due to its ability to model complex atomic interactions very generally with a reasonable compromise between accuracy and computational efficiency. However, its computational complexity remains a major limitation, which scales approximately cubically with the number of electrons. This inhibits routine, large-scale screening campaigns and severely limits its use for long time-scale simulations or structures containing more than a few hundred atoms.

Recently, Machine Learning Interatomic Potentials (MLIPs) that act as DFT surrogates have emerged as a means to approach the accuracy of DFT at a small fraction of the required computation~\cite{C6SC05720A, C7SC04934J}. This progress has been driven in part by modeling innovation, moving from small descriptor-based neural networks or kernel methods to much larger graph neural networks \cite{Behler_2007_nn, schutt2017_schnet, gasteiger2021_gemnet, batzner20223_nequip, batatia2022_mace, liao_equiformerv2_2024, qu2025_escaip, fu_2025_esen, SO3LR}, which currently represent the state of the art. As models increase in size and accuracy, further advances in learning general representations depend on having access to large-scale, high-quality data.

Small-scale ($< \mathcal{O}(1M)$) datasets, such as MD-17~\cite{Chmiela_2017_md17}, MD-22~\cite{md22}, and QM9~\cite{Ram_2014_qm9}, helped launch the field but are limited to a few atom types (\textit{e.g.}, C, H, O, N, and F) and have narrow chemical diversity. More recently, there have been a number of $\mathcal{O}(1-10M)$ dataset efforts that expand chemical and structural diversity and, to a lesser extent, elemental diversity \cite{devereux_extending_2020, schreiner_transition1x_2022, zhang_ani-1xbb_2025, qm7-x,  christensen_orbnet_2021, Eastman_spice_2023,eastman_nutmeg_2024, axelrod_geom_2022, Ganscha2025}. However, the vast majority of calculations involve only charge neutral, isolated organic molecules with a relatively small number of atoms ($N < 50$). An ideal molecular dataset would contain a mix of system sizes that have high elemental, chemical, and structural diversity with awareness of charge and spin~\cite{yuan2025foundationmodelsatomisticsimulation}. The reason no such large-scale $\mathcal{O}(100M)$ molecular dataset exists is the prohibitive computational expense of DFT.

Beyond models and training data, building useful evaluations is another important component of driving improvements in MLIPs. In the past, evaluations have focused on total energy, forces, and structure metrics such as the mean absolute error (MAE) computed on random in-distribution splits of the data. While these metrics can be informative, they fail to capture whether a model is actually useful for practical chemistry applications. Additional metrics are needed to determine whether models respect certain physical properties such as energy conservation~\cite{bigi_2025_darkside, fu_2025_esen, amin2025towards}, as well as generalize to out-of-distribution data. Thorough and well-motivated assessments of MLIPs on practically-relevant, domain-informed tasks would help to clarify for the community where current deficiencies lie and where sufficient accuracy has already been achieved. There have been recent efforts in this area for molecular systems \cite{brew_wiggle150_2025}, but more complete efforts are needed.

In this paper, we present the Open Molecules 2025 (\omol) Dataset, a large-scale resource for training molecular chemistry machine learning models. \omol{} comprises over 140 million DFT single-point calculations containing up to 350 atoms at a high level of DFT theory ($\omega$B97M-V/def2-TZVPD)~\cite{mardirossian_omegab97m, rappoport_2010_def2tzvp, rappoport_2015_def2tzvpd}. \omol{} draws from diverse chemistry disciplines including biochemistry, electrochemistry, and organic and inorganic chemistry with all of the first 83 elements represented. The dataset provides a wide sampling of chemical complexity, including systems with varying charge and spin states, explicit solvation, reactivity, and various intermolecular interactions. Structural diversity is incorporated through a variety of sampling techniques, such as classical and MLIP-based molecular dynamics (MD) and conformer sampling. Additionally, we recompute a number of existing datasets to ensure a consistent level of DFT theory across previous efforts. Beyond the dataset, we introduce a series of evaluation tasks that represent common objectives in computational chemistry, such as conformer ranking, ionization energies, spin gaps, and distance scaling. We evaluate baseline models trained on \omol{} on these tasks to provide insight on where future model development should focus. We provide all of the \omol~data with a CC BY 4.0 license, and model weights with a commercially permissive license (with some geographic and acceptable use restrictions). We have also released a public leaderboard to inspire innovation on ML models for molecular chemistry.

\begin{figure}
\includegraphics[width=1.0\linewidth]{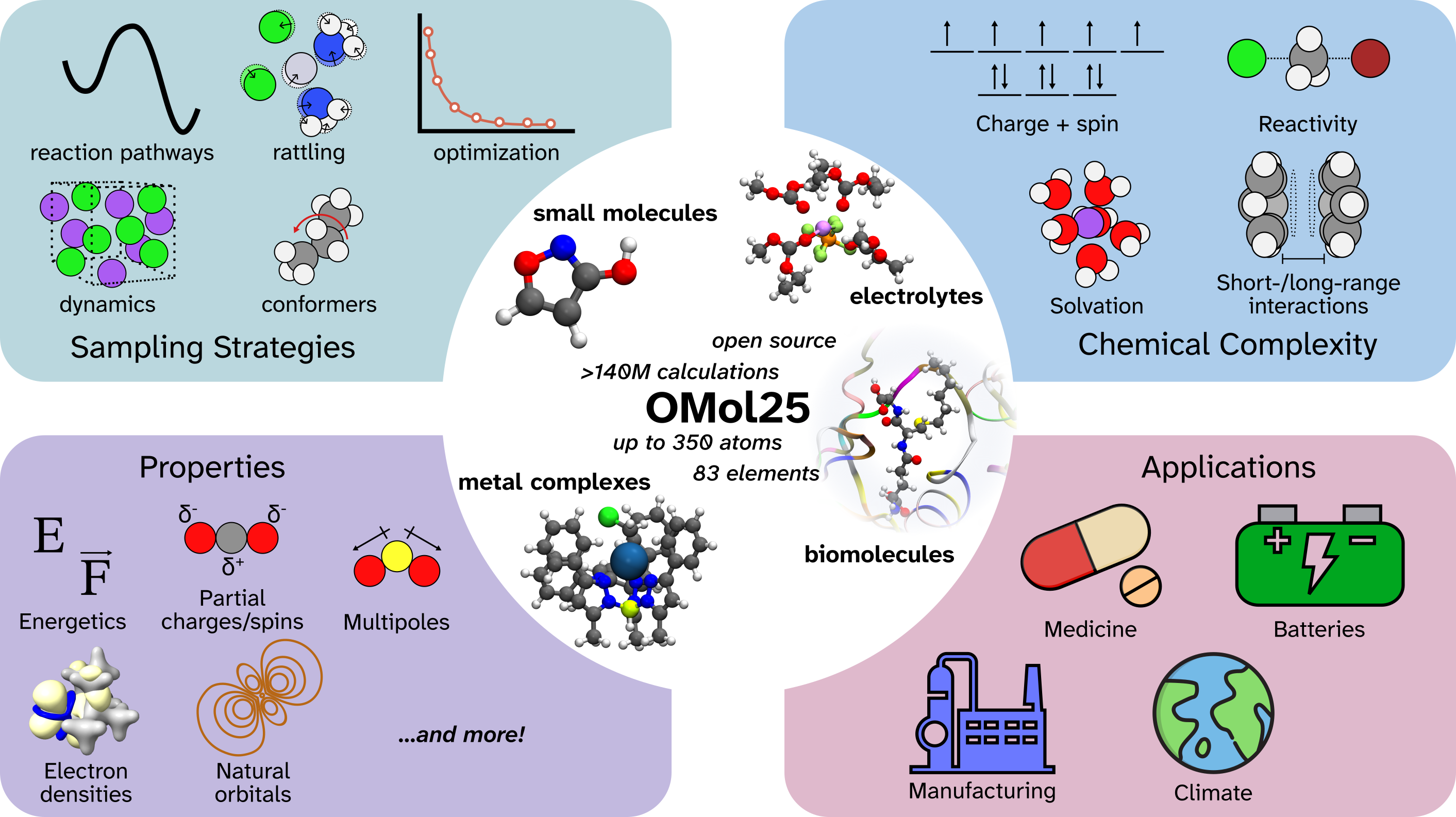}
\caption{Overview of \omol, including chemical scope, sampling strategies used to construct structures, chemical phenomena we seek to capture, properties available for each datapoint, and envisioned application areas.}
\label{fig:overview}
\end{figure}
\section{Open Molecules 2025 Dataset}

For ML models to act as DFT surrogates across the broad field of molecular chemistry, we must ensure the training dataset encapsulates the behavior of atoms across the field's numerous individual subdisciplines. To accomplish this, we explore three major domains within the Open Molecule 2025 dataset: \textbf{biomolecules}, \textbf{metal complexes}, \textbf{electrolytes}, and \textbf{main-group molecules}; we additionally re-evaluate existing \textbf{community} datasets and derivatives thereof. Biomolecules are focused on proteins, DNA, and RNA. Metal complexes feature monometallic transition metal (TM), main group metal, and lanthanide systems with diverse ligands. The electrolytes domain includes collections of multiple molecules, often charged, and their interactions with solvent molecules. The main-group domain includes heavy $p$-block molecules, molecular clusters, and organic molecules undergoing reactions and in reduced, oxidized, protonated, and deprotonated states. Finally, the community domain includes various datasets and derivatives that are generally focused on organic chemistry. 

Within each of these domains, we sought to ensure coverage of several important chemical concepts: the effect of charge and spin, the role of molecular conformations, and the propensity for reactivity. As a result of these different domains and cross-cutting considerations, there are no fewer than 14 conceptually distinct methods used to generate input data. A thorough explanation of all of the methods employed is given in the Appendix and the code used for generation will be made available on GitHub. In this section, we shall only summarize the methods and describe the resulting dataset.

\subsection{Biomolecules}

Accurately capturing the interactions of biomolecules is critical for applications in drug design, both of small molecules and biologics, and biochemistry research generally (\textit{e.g.}, computationally studying a protein's mechanism of action). To enable these diverse applications, we include a broad set of protein--ligand, protein--protein, nucleic acid--nucleic acid, and protein--nucleic acid interactions.  

Annotations for ligands (both small molecules and metal ions), DNA, and RNA bound to proteins were taken from the BioLiP2 database of protein-ligand interactions~\cite{zhang_biolip2_2023}. To obtain inputs of suitable size for DFT, fragments of these large macromolecular systems are extracted. This is accomplished by pulling out the immediate protein pocket environment of the experimental PDB structure around a small-molecule ligand, metal, or nucleic acid residue (and potentially additional nearby residues). To these extracted pockets, hydrogens are added to generate various protonation/tautomeric states of the protein and ligands, and protein and nucleic acid residues are appropriately capped. Molecular dynamics (MD) simulation of the extracted system is then performed with restraints on the protein and nucleic acid backbones to prevent the fragment from falling apart now that they are not held together by the rest of the protein or nucleic acid structure. This workflow utilizes the Schr\"odinger Software Suite~\cite{SDGR_2024-2} for manipulating and elaborating these biomolecular systems.

Additional protein-ligand structures were generated by docking random drug-like molecules from the GEOM~\cite{axelrod_geom_2022}, ChEMBL~\cite{mendez_chembl_2019}, and ZINC20~\cite{irwin_zinc20free_2020} databases into the above-extracted protein pockets using smina~\cite{koes_lessons_2013} and simulating them by the same MD procedure as above. Only the two closest chains to the ligand are retained, as opposed to the entire pocket.

Protein--protein interactions are sampled by extracting the environments around buried residues (as determined by the per residue relative solvent accessible surface area~\cite{miller_interior_1987}) and residues at the interfaces of protein subunits (according to the DIPS-Plus database~\cite{morehead_dips-plus_2023}). These extracted protein--protein fragments were similarly prepared, capped, and run through the MD protocol described above. Nucleic acid--nucleic acid interactions were probed by extracting collections of nearby residues in various topologies with an analogous procedure. In addition to the protein--nucleic acid structures we obtain from BioLiP2, the Nucleic Acid Knowledge Base~\cite{lawson_nucleic_2024} was mined for non-traditional nucleic acid structures, such as A-form and Z-form DNA, triplex systems, and Holliday junctions~\cite{ortiz-lombardia_crystal_1999}.

\subsection{Metal Complexes}

Metal complexes, including organometallic species, Werner coordination complexes, and any well-defined molecular system in which a set of ligands stabilizes one or more metal centers, are critically important for catalysis, energy harvesting, and manufacturing~\cite{nandy_computational_2021,*kinzel_transition_2021,*larsen_photoredox_2018,*sun_chemical_2021,*george_atomic_2010}. The metals in these structures, which span the periodic table, present several challenges relative to organic chemistry. Metals have a much wider array of configurational flexibility, from strongly enforced bond angles to loose, complexly-varying arrangements around the metal center. Metal--ligand bonds tend to break more easily than main-group bonds, which allows for variability in the number of bonded ligands. Finally, metals have much more flexibility when it comes to electronic structure. Many metals support multiple oxidation states with each potentially having profoundly different behaviors and chemistries.

\subsubsection{Ground State Structures}

In order to sample the diverse space of metal complexes, we employed the \texttt{Architector} package~\cite{taylor_architector_2023}. Architector takes in a specification of the metal center (\textit{e.g.}, element, oxidation state, coordination number) and a set of ligands and generates 3D structures of the requested metal center coordinated by the ligands in multiple geometries and conformations. We leveraged a curated list of experimentally used ligands~\cite{architector_data} and metal-oxidation state pairs from the \texttt{mendeleev} package~\cite{mendeleev}. In this way, an extremely large number of complexes can be generated by randomly assembling collections of ligands to attach to randomly chosen metal centers. The structures which are output by Architector have been optimized by xTB~\cite{bannwarth_gfn2-xtbaccurate_2019} during and after generation and can be used directly in DFT calculations. The spin state for each system was set to the maximum that would be reasonable for the given metal oxidation state (e.g., a $d^8$ center would be a triplet). Optimizations with no more than five steps were carried out on these complexes. The TM complex inputs were also run as single points in the lowest possible spin state (singlets for even-number electron systems, doublets of odd-number) and a subset of these were also optimized with a maximum of five steps. For complexes with spin multiplicity greater than 4, we also ran these complexes at relevant intermediate spin states as single points.

\subsubsection{Metal Reactivity}\label{sec:dataset:mc:react}
Diverse samples of reactive metal complex structures were generated by taking existing datasets of metal complex reactivity MOR41~\cite{Dohm2018}, ROST61~\cite{Maurer2021}, and MOBH35~\cite{Semidalas2022} and swapping the identity of metals and ligands using Architector. The atoms of these reactant and product complexes were renumbered to bring them into correspondence and a reaction path was generated using the Artificial Force Induced Reaction (AFIR) scheme~\cite{maeda_artificial_2016,levine_large_2023}. In AFIR, the atoms in bonds that are breaking are pushed apart from each other while the atoms in the bonds that are forming are pushed toward each other with a fictitious, constant force. A geometry optimization is run with progressively higher force constants until the reaction occurs. The points along this optimization path form a reasonable guess for the minimum energy path from reactant to product. This path was subsampled to generate DFT inputs of metal complexes in reactive geometries using the highest spin configuration.

\subsection{Electrolytes}

Electrolytes, solutions containing ions and other additives, are vital components of batteries, play a central role in biological and geochemical processes, and are essential to electrochemical manufacturing. In this work, we consider electrolytes broadly defined to include aqueous and non-aqueous solutions, ionic liquids, and molten salts. The solvation of molecular and ionic electrolytes, and the presence of other electrolyte species, profoundly affects molecular stability and structure, particularly by stabilizing highly charged groups. This creates a challenging problem for MLIPs seeking to predict the subtle forces governing intermolecular interactions, particularly due to the presence of varying local charge. 

\subsubsection{Molecular Dynamics-Based Sampling}

We employ MD to simulate a diverse sampling of electrolyte structures in large periodic boxes. The initial structures are created using the Desmond MD package~\cite{10.1145/1188455} along with Schr\"odinger's Disordered System Builder. After equilibration, the structures were simulated with NPT MD using the OPLS4 \cite{lu_opls4_2021} force field at a range of temperatures and concentrations. From these simulations, a set of frames were collected and the environment around every ion was extracted. The ion's environment included all molecules where any atom of that molecule was within some fixed cutoff radius of the atom(s) of the central ion. In this work, we used both 3\angs{} and 5\angs{} for the cutoffs, which corresponds roughly to the primary and secondary coordination shells, respectively. These extracted clusters are subsampled to obtain a set of clusters that vary significantly in composition and geometry. A similar procedure is employed around solvent molecules, except that we require no solute to be present in the extracted solvent-centered conformers, i.e. they are solvent-only clusters. Clusters from the 3\angs~cutoff were optimized for up to five steps, while clusters from the 5\angs~cutoff were run as single-points.

As electrolytes are used in batteries and subject to high electric fields, electrons are frequently pulled out of or pushed onto clusters of the kinds described here. We therefore also compute a random sample of systems with an either an electron added or removed from the cluster.

Different intermolecular interactions have different scaling behavior with distance and these effects must be captured by an MLIP. A random sample of clusters were dilated by random amounts (changing all intermolecular distances, but keeping intramolecular distances fixed), spreading the molecules apart or contracting them together.

In purely classical MD, many possible configurations may not be sampled. To account for this, we utilize Ring Polymer Molecular Dynamics (RPMD) that includes the use of nuclear quantum effects (NQE) to increase the diversity of the configurations. We used OpenMM's~\cite{Eastman_openmm_2023} RPMD implementation to simulate another batch of electrolytes which were skewed to contain more light atoms that are more affected by NQE. The same extraction procedure as above was employed to obtain DFT input structures for single point calculations.

At interfaces, solvents can form different structures than in bulk systems due to the absence of, for example, hydrogen bonding partners on the other side of the interface. We apply a spherical restraining potential around a large droplet of electrolyte to create a gas-solvent interface. This droplet is sampled via the same strategy as above to create a set of interfacial structures for DFT single points.

Given water's importance as a solvent, additional large clusters of up to 70 water molecules were also included. Water was simulated with the AMOEBA~\cite{Shi2013, Ponder2010, Ren2003} force field and snapshots were used as DFT single point inputs.

\subsubsection{Small Molecule Generation}

Electrolyte research often involves the development of new ions and additives that may stabilize battery chemistry or prevent degradation. In order to sample from diverse electrolyte-like molecules, an array of 77 electrolyte ``core'' structures and 240 functional groups were curated and these cores and functional groups were randomly chained together to create novel small molecules. Ions and solvent molecules were placed around these new species with the Architector package~\cite{taylor_architector_2023}. These systems were optimized for up to five steps.

\subsubsection{Electrolyte Reactivity}

Electrolyte reactions were taken from previous work on reaction networks and mechanistic studies of electrolyte decomposition and solid-electrolyte interphase formation \cite{Xie2021, Barter2023, SpotteSmith2022, SpotteSmith2022a, SpotteSmith2023}. Coordinates for all of these reaction templates were generated if they did not already exist and random metals were swapped with the metals present in the templates to generate a library of reactants and products undergoing electrolyte-type reactions in the presence of various ions. The Popcornn~\cite{Khegazy2025} method was used to generate reaction paths between reactant and product; these paths were subsampled to generate DFT inputs.

\subsection{Main-group Molecules}

While main-group compounds are found throughout the other domains, there are several undersampled classes of molecules and clusters that are practically important to more fully cover the domain of molecular DFT. These include reactive trajectories, heavy main-group containing-compounds and clusters, unusual protonation and ionization states of main-group compounds, larger clusters of molecules, and noble gas compounds and clusters.

To increase the number of reactivity samples within \omol, we utilize several reactivity datasets of reactant-transition state-product triples~\cite{zhao_comprehensive_2023} and elementary reaction steps~\cite{tavakoli_pmechdb_2024,tavakoli_rmechdb_2023}. For the former, we carry out an interpolation in internal coordinate space from reactant to transition state to product, and, for the latter, we generate 3D structures and use the AFIR procedure as described in Appendix \ref{sec:app:react:mechdb}.

The Crystallographic Open Database was mined for non-metal containing structures of clusters (such as boron hydride clusters, partially condensed fullerenes, and polynuclear cages). Cluster-type structures of the mixtures of main-group elements were also generated with Architector. Architector was also employed, treating heavy main-group elements (Si, Ge, Se, Te, Sb, As) as "metal centers" around which to attach random organic "ligands". Compounds from the ANI-2x dataset were randomly substituted with heavy analogues of the corresponding light atoms (and bond lengths adjusted) to create further coverage of heavy-main group chemistry.

Molecules from the ANI-2X and SPICE2 datasets were also randomly ionized (between -1 and +2) and protonated or deprotonated, including at more extreme pKa sites. Clusters of molecules taken from the Open Molecular Crystals 2025 dataset\cite{gharakhanyan_open_2025} were also computed. Finally, collections of noble gases and other molecules were simulated with molecular dynamics to sample the phase space of these weakly interacting elements.

\subsection{Community}

\subsubsection{Existing Community Datasets}

Numerous molecular datasets have been previously released~\cite{devereux_extending_2020,schreiner_transition1x_2022,zhang_ani-1xbb_2025,christensen_orbnet_2021,Eastman_spice_2023,eastman_nutmeg_2024,unke_physnet_2019, Chmiela_2017_md17, Ram_2014_qm9} with varying levels of theory. As part of Open Molecules 2025, we have recomputed several of the most widely used datasets to upgrade them to a consistent and higher level of theory and to fill in missing data, such as forces. The datasets calculated were ANI-2X~\cite{devereux_extending_2020}, Transition-1X~\cite{schreiner_transition1x_2022}, ANI-1xBB~\cite{zhang_ani-1xbb_2025}, Orbnet Denali~\cite{christensen_orbnet_2021}, SPICE2~\cite{Eastman_spice_2023,eastman_nutmeg_2024}, and Solvated Protein Fragments~\cite{unke_physnet_2019, gems}. We also recomputed approximately 30\% of the GEOM~\cite{axelrod_geom_2022} dataset. The GEOM systems were optimized, with a fraction having their initial positions randomly perturbed. The Transition-1X dataset is a database of reactive trajectories and was recomputed in the UKS formalism.

\subsection{ML-Based MD}

In order to increase the structural diversity of the dataset and discover areas where the ML models may need additional data to avoid pathological behavior, we undertook ML-based MD of the types of inputs used in the first three domains above. Three EquiformerV2 models~\cite{liao_equiformerv2_2024} were trained on approximately half of the data from each of the three domains. Additional Architector metal complexes, periodic boxes of electrolytes, and protein--protein interface clusters were prepared as described above. The EqV2 models and the MACE-MP0 model~\cite{batatia2024foundationmodelatomisticmaterials} were used to simulate short MD trajectories which were randomly subsampled to create ML-MD-based DFT inputs. For metal complexes, a small amount of data was obtained by rattling the atomic positions according to a Boltzmann distribution as was done in Open Materials 2024~\cite{barroso-luque_open_2024}.

\subsection{Calculation Details}\label{sec:dataset:lot}

In selecting the level of theory for Open Molecules 2025, we sought to create a long-lasting dataset using the highest quality settings that were possible given the computational resources available. The DFT functional selected, the range-separated hybrid meta-GGA $\omega$B97M-V~\cite{mardirossian_omegab97m}, has consistently been shown by various authors to be one of, if not the most, accurate functional for a broad array of quantum chemistry tasks~\cite{mardirossian_thirty_2017,goerigk_look_2017,*najibi_nonlocal_2018,*santra_minimally_2019,*garrison2023_jcim}. Only double-hybrid functionals have been consistently shown to outperform it, at prohibitive computational cost. Because the dataset contains anions, a basis set with diffuse functions was required; the triple-zeta def2-TZVPD basis set~\cite{weigend2005a, *rappoport2010a, *rappoport_property-optimized_2021} was thus selected as the Ahlrichs basis sets are well-optimized for use with DFT, with effective core potentials (ECPs)~\cite{dolg1989a, *andrae1990a, *kaupp1991a, *dolg1993b, *leininger1996a, *metz2000a, *metz2000b, *peterson2003b} that allow support for elements 1--83. We computed all singlet systems which contain transition metal and lanthanide complexes or where bonds are expected to be breaking in the Unrestricted Kohn-Sham (UKS) formalism and rotate by 20$^\circ$ between the HOMO and LUMO in the $\beta$ space in order to break spin symmetry in the initial guess. Non-singlet systems were also run in UKS.

Calculations were carried out with the ORCA 6.0.0 DFT package~\cite{RN204,neese_software_2025}, using both RI-J~\cite{RN29} and COSX~\cite{RN192} integral acceleration techniques and \texttt{tight} convergence settings. The LibXC\cite{lehtola_recent_2018} implementation of the $\omega$B97M-V functional was employed. Benchmarking of grid settings (both the exchange-correlation grid and the COSX grid) indicated that typical grids led to small numerical inconsistencies between energy and forces due to grid incompleteness. These errors were significant on the scale of errors with state-of-the-art MLIPs. In order to achieve sufficiently tight consistency, ORCA's \texttt{DEFGRID3} offered the best trade-off of convergence and cost. Calculations were only considered in \omol{} if they met several quality control and error checks; details are described in Appendix \ref{sec:app:quality}.

The released dataset required 6.6 billion CPU core-hours in total. Nearly all calculations were run on Elastic Compute within Meta's private cloud \cite{Gupta_2024}, utilizing a heterogeneous pool of servers that can be preempted at any time to accommodate increased demand on higher-priority workloads. Although this presents challenges at the software level, especially around handling of partially complete calculations when hosts are reclaimed, it allowed for the creation of this dataset on servers that would otherwise have been sitting idle.

\begin{figure}
\includegraphics[width=0.9\linewidth]{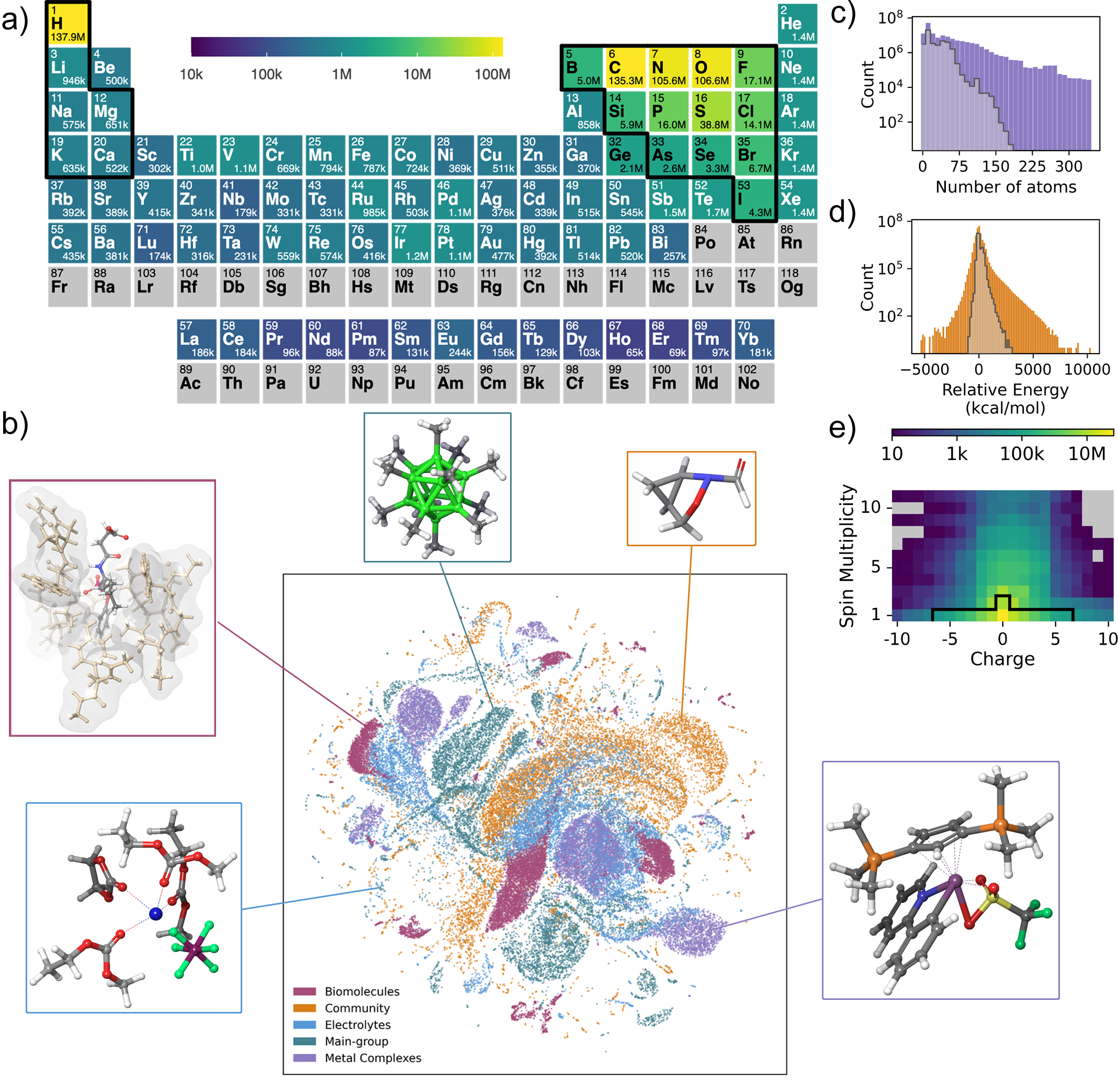}
\caption{OMol25 dataset composition. a) Periodic table heat map, showing the number of snapshots in the training set containing a given element. Elements which were included in previous widely used large-scale molecular DFT dataset
are noted with a black border. b) UMAP showing the distribution of the various domains in \omol, with specific dataset examples highlighted. c) Histogram in log scale for the number of training set snapshots with a given number of atoms. d) Histogram in log scale of training set snapshots with a given energy value relative to atomic references. e) Heat map for number of training set snapshots of different charge:spin. Charge/spin combinations included in the pre-existing datasets which comprise the Community domain are boxed in black.}
\label{fig:dataset}
\end{figure}

\subsection{Dataset Profile}

\begin{figure}[ht!]
\centering
\includegraphics[width=1.0\linewidth]{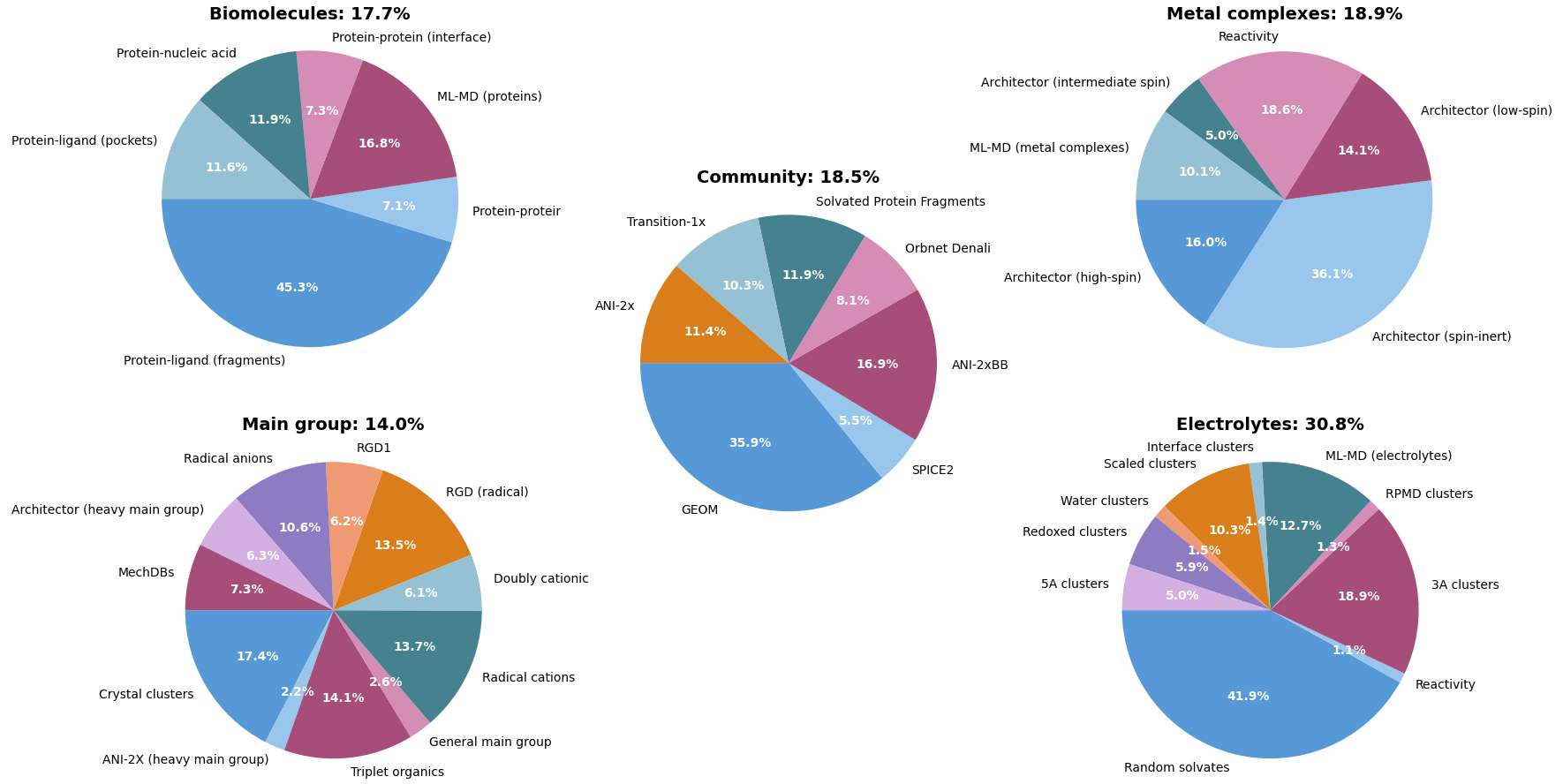}
\caption{OMol25 breakdown by domain and sampling strategy by number of atoms. The percentage next to each domain indicates its  share of the total dataset's number of atoms.}
\label{fig:splits}
\end{figure}

The final Open Molecules 2025 dataset contains more than 100 million DFT calculations. The number of atoms ranges from 2 to 350, with 50 on average. Charges vary from -10 to +10 and spin multiplicity varies from 1 to 11. Owing to the wide range of system sizes in the dataset, we report the breakdowns of each domain by the total number of atoms rather than the total number of systems. There are approximately 1.2B atoms in each of Biomolecules and Metal Complexes, 1.4B in Community, and 2.0B in Electrolytes. The breakdown of each domain into the subdomains discussed above is given in Figure \ref{fig:splits}. In addition to energy and force data, the dataset includes a variety of partial charge and spin schemes, orbital energies, Fock matrices, densities, and more as described in Appendix \ref{sec:app:calc}.

\begin{table}[ht!]
\caption{Size of the \omol~train, validation, and test splits.}
\centering
\resizebox{0.6\textwidth}{!}{%
\begin{tabular}{ccrl}
\toprule
\multicolumn{1}{l}{}  & Split   & Size      & \multicolumn{1}{c}{Description}                   \\ \midrule
\multirow{3}{*}{Train} & All     & 140,641,161 & Full training set \\
& 4M      & 3,986,754   & Uniform $\sim$4M subset \\
& Neutral & 34,335,828  & Charge neutral, singlet subset \\ \midrule
Val             & Comp     & 2,762,021   & Out-of-distribution compositions \\ \midrule

\multirow{6}{*}{Test}  & Comp    & 2,805,046   & Out-of-distribution compositions \\
& M-Lig   & 42,028     & Unique metal-ligand pairs                         \\
& PDB-TM  & 26,614     & Metal-containing protein structures               \\
& Reactivity & 64,898 & Held-out organic and metal-complex reactions \\
& COD     & 84,807     & Experimental crystal structures                       \\
& Anions  & 16,488     & Unique anion structures                                     \\ \bottomrule
\end{tabular}%
}
\label{tab:dataset}
\end{table}

\subsection{\omol~Training, Validation, and Test Splits}
The \omol~dataset is divided into training, validation, and test splits to ensure consistent evaluations within the community. Splits are created based on compositions (\textit{i.e.}, molecular formula). After enumerating all unique compositions among the computed data, we hold out $\sim$2.5\% each for a validation and test set. We create three dataset sizes - the full \omol~dataset (``All''), a much smaller $\sim$4M split (``4M''), and a charge-neutral, singlet split (``Neutral''). The 4M split is sampled uniformly across the electrolytes, metal complexes, biomolecules, reactivity (RMechDB, PMechDB, and ANI-1xBB), and community domains. The Neutral split corresponds to only the charge-neutral singlets from a subset of the community domain - ANI-2X, Orbnet Denali, SPICE2, GEOM, Transition-1X, and RGD1. This split is intended to measure the performance of models on datasets the community is familiar with, without worrying about the complexity of charge and spin. All training and validations splits are made publicly available.

To test generalizability beyond just out-of-distribution (OOD) compositions, we also generate several explicit OOD splits. Dataset splits are summarized in Table \ref{tab:dataset}.

\subsubsection{Out-of-Distribution Splits}

\textbf{Metal-Ligand Bonds} As a test of the generalizability of metal-ligand interactions, we randomly sampled 50 metal-ligand (M-L) bond combination (e.g. Pt-OAc) to hold out from training. This validation set consists of 39,615 complexes which contain one or more of the OOD M-L bonds.

\textbf{PDB-TM} \omol{} explicitly excludes metal-containing protein structures from the main training split so that it can be used to test the learning transfer from metal complex and electrolytes data. These include single ions coordinated to amino acids (e.g. the Zn of a zinc finger), metal complexes (e.g. the Fe of a heme group), and multimetallic clusters (e.g. 4Fe-4S clusters).

\textbf{Reactivity} Out-of-distribution organic reactivity data is taken from the work of Grambow, Pattanaik, and Green~\cite{grambow_reactants_2020}. More specifically, 1782 of their reactions are not found in either Transition-1x or RGD1. These reactant, transition state, product triples were elaborated into reaction paths as was done in Section \ref{sec:methods:interp_rxn}. Additionally, several of the reaction templates in Section \ref{sec:dataset:mc:react} were incompatible with the automated ligand swapping methodology, and thus were held out of training data generation. After metal swapping, these templates yielded 1436 reactions, which were then subjected to the AFIR procedure described in Section \ref{sec:dataset:mc:react}.

\textbf{COD} To assess the sufficiency of \omol's generated metal complexes for describing real-world systems, we draw from the Crystallography Open Database (COD)~\cite{vaitkus_workflow_2023, *grazulis_crystallography_2012, *merkys_graph_2023, *vaitkus_validation_2021, *merkys_codcifparser_2016, *grazulis_computing_2015} of experimental crystal structures of metal complexes. This replicates the typical starting point of real-world computational projects involving metal complexes. In this case, both the highest and lowest reasonable spin states were used for first-row TMs, only low-spin states for second and third row TMs, and only high-spin states for lanthanide complexes.

\textbf{Anions}\label{sec:methods:ood_anions} Battery research often involves the development of new electrolytes. To assess the ability of models to describe new types of electrolyte systems, two additional electrolyte anions were simulated according the same procedure described in Section \ref{app:elyte:desmond}. While additional non-OOD anions, cations, and solvents appear in the resulting clusters, all of them also contain one or more of these new anions. 

Additional splits including out-of-distribution cations/solvents, TorsionNet500~\cite{rai_torsionnet_2022}, and the Wiggle150~\cite{brew_wiggle150_2025} benchmark are described in the Appendix \ref{sec:app:addl_res}.
\section{Evaluations}

\begin{figure}
\includegraphics[width=1.0\linewidth]{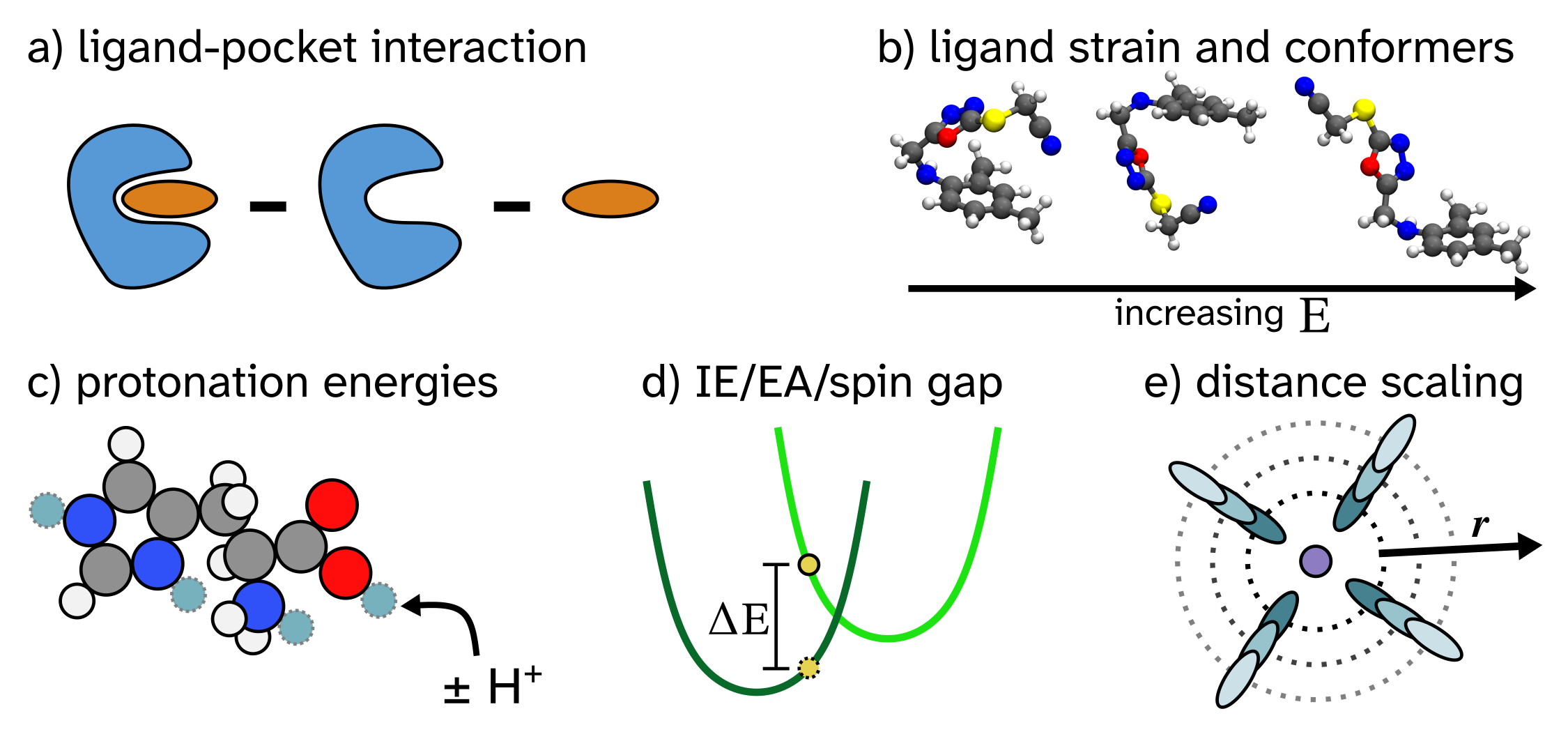}
\caption{\omol{} evaluations. a) Ligand-pocket interaction energy/force, as defined by the energy/force difference between the ligand-pocket complex and the isolated ligand and isolated pocket. b) Ligand strain energy and conformer optimization/ordering, both of which involve a global optimization where many conformers of a structure are all subjected to a tightly converged geometry optimization. c) Relative protonation energies, as defined by the energy difference between optimized structures of different protonation states. d) Unoptimized ionization energy (IE) / electron affinity (EA) / spin gap, as defined by the energy difference between static structures of varying charge and/or spin multiplicity. e) Distance scaling, which seeks to predict the energy difference between instances of the same structure containing multiple molecular components, with the inter-component distance scaled by some factor.}
\label{fig:evals}
\end{figure}

In order to evaluate the capacity for models trained on \omol{} to capture the breadth of chemistry that the training data seeks to cover, we developed a range of evaluation tasks, as summarized in Figure \ref{fig:evals}. Each task is carried out on at least 1,000 different structures to achieve robust statistics.

\subsection{Protein-ligand Interaction Energy and Forces}

Protein-ligand binding is central to biological processes and is the mechanistic underpinning of many life-saving treatments. It is highly desirable that models trained on \omol{} be able to accurately capture the physics of protein-ligand binding in realistic environments. Since calculating a true binding free energy is beyond the scope of an MLIP evaluation, requiring an appropriate sampling procedure, we use ligand-pocket interaction energy ($E_{ligand+pocket} - E_{ligand} - E_{pocket}$) MAE and ligand-pocket interaction forces ($\vec{F}_{ligand+pocket} - \vec{F}_{ligand} - \vec{F}_{pocket}$) MAE as a proxy for the binding free energy task. A full free energy procedure would be required to accurately predict interaction energies at each time step (Fig. \ref{fig:evals}a). Further details are provided in Appendix \ref{sec:app:eval:protein_ligand}.

\subsection{Ligand Strain}

Another recently established evaluation task for atomistic simulation of protein-ligand binding is the ligand strain energy \cite{wallace2025strainproblemsgottwist}, defined as the energy difference between the local minimum corresponding to the bioactive conformation of a ligand and the ligand's global minimum energy. Details of how we obtain the local minimum of the bioactive geometry are provided in Appendix \ref{sec:app:eval:lig_strain}. The global minimum is identified via tightly converged geometry optimizations on structures generated by CREST \cite{10.1063/5.0197592}, RDKit~\cite{landrum_rdkitrdkit_2025}, and MacroModel~\cite{Mohamadi1990, Watts2014, macromodel} run on the original bioactive conformation. Evaluation metrics are then the strain energy MAE and the RMSD between the DFT global minimum structure and the MLIP global minimum structure.

\subsection{Conformers}

Molecules with many torsional degrees of freedom and configurationally flexible metal complexes can have a large number of local minima on their potential energy surface, and it is frequently of critical importance to recover the lowest energy conformer(s) whose properties dominate the thermal ensemble (Fig. \ref{fig:evals}b). To evaluate model capacity in this context, families of conformers are optimized to very tight convergence with both DFT and the MLIP. We then map the DFT-optimized to the MLIP-optimized conformers via RMSD-based linear sum assignment, which finds the correspondence between DFT and MLIP structures with lowest total RMSD. Using this mapping, the total RMSD are used as evaluation metrics (see Appendix \ref{sec:app:eval:conformers} for more details). We further compare the $\Delta E$ between each conformer and the global minimum conformer versus the equivalent value from the MLIP, yielding a $\Delta E$ MAE evaluation metric. 

\subsection{Protonation Energies}

Protonation state plays a central role in many chemical transformations in biological, environmental, and industrial processes. Since a rigorous treatment of acidity/basicity in solution (\textit{i.e.}, pKa prediction) is beyond the scope of MLIP evaluation, we opt for a proxy evaluation task of protonation energies (Fig. \ref{fig:evals}c). These protonation energies are used as part of DFT-based pKa prediction workflows~\cite{cao_quantum_2024,*tang_discovery_2022,johnston_epik_2023}. We perform geometry optimization on pairs of structures that differ by one proton and then calculate the $\Delta E$ between the two structures, yielding $\Delta E$ MAE, and the RMSD between MLIP-optimized and ORCA-optimized structures as evaluation metrics. Note that this evaluation is only possible for MLIPs which can simulate species of different total charge. Further details are provided in Appendix \ref{sec:app:eval:protonation}.

\subsection{Unoptimized IE/EA and Spin Gap}

The addition, removal, or transfer of electrons are central to enzymatic, catalytic, electrochemical, and other redox processes. An MLIP which faithfully describes such processes would be able to predict energy and force differences between two electronic states of the same atomic configuration (Fig. \ref{fig:evals}d). The $\Delta E$ between two such states are the well-known vertical ionization energy (IE) and electron affinity (EA), while $\Delta F$ captures the differences in the potential energy surface between the two states. We therefore evaluate $\Delta E$ MAE and $\Delta \vec{F}$ MAE between principal and charge-modified structures. Further details are provided in Appendix \ref{sec:app:eval:ip_ea}.

Of similar importance is a model's ability to accurately predict the energy and force on different spin surfaces. Accurately simulating systems containing open-shell d-block metals often requires determining which is the lowest energy spin state, which may change in the course of a reaction, and the energy gap between spin states can play a critical role in properties of molecular optical devices and photoactive catalysts~\cite{sco_spintronics,spin_catalysis,sco_materials}. Our evaluation metrics are thus $\Delta E$ MAE and $\Delta \vec{F}$ MAE between structures simulated at their highest viable spin multiplicity and each possible lower spin multiplicity. Further details are provided in Appendix \ref{sec:app:eval:ip_ea}.

\subsection{Distance Scaling: Short-range and Long-range Interactions}

Different intermolecular interactions have distinct scaling behavior with distance (\textit{e.g.}, $1/r$ for charge-charge interactions, $1/r^3$ for dipole-dipole interactions, $1/r^6$ for dispersion forces). It is important that an MLIP smoothly and accurately captures both the correct magnitude and scaling behavior to make valid predictions on observables such as phase changes, ion/mass transport, density, viscosity, etc. We probe this feature by scaling the distance between components of molecular clusters and complexes. Many MLIP models employ a distance cutoff beyond which two atoms do not communicate directly. We therefore split this evaluation into a ``short-range'' and ``long-range'' regime around 6\angs{} of separation (typical of many model's cutoff radius). We thus compute a distance scaling scan and evaluate $\Delta E$ and $\Delta \vec{F}$ between the points on that scan and the lowest energy structure along the scan that lies in the short-range regime. Further details are provided in Appendix \ref{sec:app:eval:scaling}.

\begin{table}[ht!]
\caption{Summary of metrics for each evaluation task.}
\centering
\resizebox{.4\textwidth}{!}{%
\begin{tabular}{lr}
\toprule
\textbf{Evaluation Task} & \textbf{Metrics} \\ \midrule
\multirow{2}{*}{Protein-ligand} & Interaction Energy MAE \\
 & Interaction Forces MAE \\ \midrule
\multirow{2}{*}{Ligand Strain} & Strain Energy MAE \\
 & Global Minimum RMSD \\ \midrule
\multirow{2}{*}{Conformers} & Ensemble RMSD \\
 & $\Delta E$ MAE \\ \midrule
\multirow{2}{*}{Protonation} & RMSD \\
 & $\Delta E$ MAE \\ \midrule
\multirow{2}{*}{IE/EA} & $\Delta E$ MAE \\
 & $\Delta \vec{F}$ MAE \\ \midrule
\multirow{2}{*}{Spin Gap} & $\Delta E$ MAE \\
 & $\Delta \vec{F}$ MAE \\ \midrule
\multirow{2}{*}{Distance Scaling} & SR and LR $\Delta E$ \\
 & SR and LR $\Delta \vec{F}$ \\
 \bottomrule
\end{tabular}
}
\end{table}

\section{Baseline Models}

We present a set of baseline results to aid the community in evaluating the performance of current state-of-the-art models on the \omol~validation and test sets as well as the \omol~evaluations. The current\footnote{Additional baseline models will be included in the future} set of baseline models includes eSEN~\cite{fu_2025_esen}, GemNet-OC~\cite{gasteiger2022gemnetocdevelopinggraphneural}, MACE~\cite{batatia2022_mace}, and UMA~\cite{wood_uma_2025}. While these models are by no means an exhaustive survey of the field, they serve to provide a representative overview of the capabilities of current equivariant and invariant models.

All the baseline MLIPs are message-passing graph neural networks where nodes represent atoms and edges represent interactions with neighboring atoms. Typically, these models take atom types (\textit{i.e.}, elements) and 3D atom positions as input and return energy as an output. Conserving models compute per-atom forces by taking the gradient of the energy with respect to the positions, while direct models predict the per-atom forces as an additional output of the model.

One necessary modification to baseline models is enabling the use of total charge and spin as inputs. Given the extent to which molecules with different change and spin exist in \omol~(see Figure \ref{fig:dataset}) charge- and spin-aware models are a requirement when training on the full dataset. For example, there are numerous examples in the dataset where the same 3D structure has a different charge or spin. If the model is not given this information, it would incorrectly predict these structures to have the same energy. To accomplish this in baseline models, we introduce a simple combined charge and spin embedding based on the input total charge and spin, which is added to the node embeddings (more details can be found in Appendix \ref{sec:app:models}). We recognize that this may not be sufficient to describe how charge and spin might localize in a system and leave more complex schemes for incorporating charge and spin for future work.

Additionally, we train eSEN models of different sizes to understand how performance changes with increasing model capacity.  Due to the scale of \omol, we utilize a multi-step training procedure to improve the efficiency of medium and large models. The procedure involves first pre-training direct force models with lower precision~\cite{fu_2025_esen}, followed by a second stage at FP32. Small models are trained in both a direct and conserving configuration using full precision. GemNet-OC was trained using its default cutoff radius of 12\angs, but also a configuration with 6\angs, consistent with eSEN. For training, energies are referenced using a modified heat of formation followed by a linear reference. Full training details and model hyperparameters can be found in Appendix \ref{sec:app:models}.
\section{Results}

\subsection{Test}
For all splits, we evaluate our models' performance on predicting a structure's energy and per-atom forces. The per-atom mean absolute error (MAE) is used as the primary evaluation metric, similar to previous work~\cite{oc20,barroso-luque_open_2024,odac}.

\begin{table}[h!]
\caption{\textbf{Structure to energy and force results across the different test splits}. Total energy and force mean absolute error metrics are reported across the two different training splits - All and 4M.}
\begin{threeparttable}
\centering
\resizebox{\textwidth}{!}{%
\begin{tabular}{llr SS SS SS SS SS SS}
\multicolumn{15}{c}{Test} \\
\hline
 & & &
\multicolumn{2}{c}{Comp} &
\multicolumn{2}{c}{M-Lig} &
\multicolumn{2}{c}{PDB-TM} &
\multicolumn{2}{c}{Reactivity} &
\multicolumn{2}{c}{COD} &
\multicolumn{2}{c}{Anions}
\\
\cmidrule(l){4-15} 
 Dataset & Model & \# of params & \mcc{Energy}   & \mcc{Forces}    
 & \mcc{Energy}   & \mcc{Forces}    
 & \mcc{Energy}   & \mcc{Forces}     
 & \mcc{Energy}   & \mcc{Forces}  
 & \mcc{Energy}   & \mcc{Forces}   
 & \mcc{Energy}   & \mcc{Forces}  
 \\ 
 \midrule[1pt]
\multirow{2}{*}{OMol-1}
 & eSEN-sm-cons. & 6.3M & 3.35 & 0.19 & 1.97 & 0.36 & 2.56 & 0.31 & 2.59 & 1.04 & 3.09 & 0.56 & 1.44 & 0.24 \\
 & UMA-S-1.2 (OMol) & 290M$^1$ & 1.07 & 0.15 & 1.55 & 0.34 & 1.82 & 0.29 & 2.20 & 1.00 & 2.78 & 0.57 & 1.06 & 0.21 \\
\midrule
\multirow{8}{*}{OMol-0} 
& eSEN-sm-d. & 6.3M & 2.43 & 0.210 & 2.03 & 0.395 & 2.63 & 0.353 & 2.84 & 1.204 & 4.22 & 0.541 & 2.41 & 0.262 \\
& eSEN-sm-cons. & 6.3M & 2.15 & 0.170 & 1.77 & 0.338 & 2.23 & 0.293 & 2.38 & 1.020 & 3.06 & 0.474 & 1.32 & 0.213\\
& eSEN-md-d. & 50.7M & 1.35 & 0.099 & 1.16 & 0.203 & 1.53 & 0.209 & 1.96 & 0.743 & 2.56 & 0.330 & 1.34 & 0.125 \\
& GemNet-OC-r6 & 39.1M & 1.15 & 0.150 & 1.31 & 0.297 & 1.71 & 0.272 & 2.23 & 0.985 & 3.07 & 0.435 & 1.39 & 0.186 \\
& GemNet-OC & 39.1M & 0.80 & 0.135 & 1.26 & 0.289 & 1.56 & 0.263 & 2.18 & 0.968 & 3.02 & 0.425 & 1.45 & 0.166 \\
& UMA-S-1.1 (OMol) & 150M$^1$ & 1.70 & 0.199 & 1.58 & 0.378 & 2.14 & 0.357 & 2.33 & 1.088 & 2.95 & 0.582 & 1.44 & 0.263 \\
& UMA-M-1.1 (OMol) & 1.4B$^1$ & 1.38 & 0.126 & 1.13 & 0.240 & 1.27 & 0.234 & 1.77 & 0.770 & 2.09 & 0.421 & 1.06 & 0.169 \\
\midrule
\multirow{5}{*}{4M}
& eSEN-sm-d. & 6.3M & 3.45 & 0.273 & 2.75 & 0.499 & 4.04 & 0.431 & 3.70 & 1.489 & 6.01 & 0.696 & 3.98 & 0.365 \\
& eSEN-sm-cons. & 6.3M & 3.25 & 0.228 & 2.21 & 0.435 & 3.00 & 0.367 & 2.88 & 1.249 & 4.20 & 0.625 & 2.33 & 0.294  \\
& eSEN-md-d. & 50.7M & 2.05 & 0.138 & 1.69 & 0.277 & 2.69 & 0.282 & 2.78 & 1.012 & 3.75 & 0.438 & 2.36 & 0.190  \\
& GemNet-OC-r6 & 39.1M & 2.05 & 0.204 & 1.96 & 0.396 & 3.28 & 0.360 & 3.14 & 1.316 & 5.90 & 0.605 & 2.46 & 0.270  \\
& GemNet-OC & 39.1M & 1.38 & 0.184 & 1.84 & 0.379 & 2.88 & 0.362 & 3.08 & 1.281 & 5.31 & 0.562 & 2.07 & 0.253  \\
\bottomrule
\end{tabular}%
}
\begin{tablenotes}
  \item \kindatiny Energy (kcal/mol), Forces (kcal/mol/\angs)
  \item $^1$UMA models are mixture of experts models so while the models have many parameters (e.g. 150M, 290M, 1.4B), only 6M, 6M, and 50M,\\ respectively, are active when making inference calls.
\end{tablenotes}
\end{threeparttable}
\label{tab:test_ood_total}
\end{table}

Results for eSEN and GemNet-OC are given in Table \ref{tab:test_ood_total} across the different test splits. Conserving models (cons.) outperform their direct (d.) counterparts across all splits and metrics. Larger models, such as eSEN-md, also outperform their smaller variants. GemNet-OC outperforms eSEN-sm across all splits, and is comparable to eSEN-md across most metrics, outperforming it only on the out-of-distribution compositions (``Comp'') split. When we compare a GemNet-OC model trained with a smaller cutoff radius of 6\angs{} (the same as used in eSEN models), we generally observe diminishing performance. Models trained on the full dataset exhibit 50\% to 100\% improved performance versus models trained on the 4M subset. Model trends between the All and 4M datasets, however, correlate very well, suggesting that model development can be confidently done on the subset. Of the test splits, Comp and metal ligand pairs (``M-Lig'') have the lowest errors; with reactivity and COD the highest. Averaged across all splits, UMA-M-1.1 performs the best with an energy MAE of 1.38 kcal/mol and 0.13 kcal/mol/\angs~forces MAE on the OMol-0 set.

To measure model performance across the different domains, we evaluate the respective subsets within the out-of-distribution Comp test split. Results are given in Table \ref{tab:test_breakdown_total}. As a general statement, the neutral organics and biomolecules tend to have lower errors than electrolytes and metal complexes.

\begin{table}[ht!]
\caption{\textbf{Out-of-distribution composition \textbf{test} results}. Results broken across biomolecules, electrolytes, metal complexes, and neutral organics. Total energy and force mean absolute error metrics are reported across the two different training splits - All and 4M.}
\begin{threeparttable}
\centering
\resizebox{\textwidth}{!}{%
\begin{tabular}{ll SS SS SS SS}
\multicolumn{10}{c}{Test-Comp} \\ \midrule
&
&
\multicolumn{2}{c}{Biomolecules} &
\multicolumn{2}{c}{Electrolytes} &
\multicolumn{2}{c}{Metal Complexes} &
\multicolumn{2}{c}{Neutral Organics} \\
\cmidrule(l){3-10} 
Dataset & Model & \mcc{Energy}   & \mcc{Forces}    
& \mcc{Energy}   & \mcc{Forces}     
& \mcc{Energy}   & \mcc{Forces}  
& \mcc{Energy}   & \mcc{Forces}  \\
\midrule
\multirow{2}{*}{OMol-1}
& eSEN-sm-cons. & 3.67 & 0.13 & 3.64 & 0.24 & 3.01 & 0.65 & 0.78 & 0.39 \\
& UMA-S-1.2 (OMol) & 0.71 & 0.09 & 1.40 & 0.19 & 2.51 & 0.61 & 0.62 & 0.37 \\
\midrule
\multirow{8}{*}{OMol-0}
 & eSEN-sm-d. & 2.28 & 0.147 & 2.84 & 0.240 & 3.20 & 0.731 & 1.04 & 0.431 \\
 & eSEN-sm-cons. & 2.01 & 0.107 & 2.56 & 0.210 & 2.84 & 0.628 & 0.73 & 0.378 \\
 & eSEN-md-d. & 1.18 & 0.060 & 1.64 & 0.119 & 2.03 & 0.431 & 0.52 & 0.177 \\
 & GemNet-OC-r6 & 0.94 & 0.104 & 1.37 & 0.170 & 2.24 & 0.562 & 0.77 & 0.346 \\
 & GemNet-OC & 0.57 & 0.091 & 0.92 & 0.151 & 2.20 & 0.553 & 0.67 & 0.335 \\
 & MACE-OMOL-L-0 & 6.08 & 0.155 & 6.53 & 0.285 & 3.59 & 0.704 & 1.54 & 0.518 \\
 & UMA-S-1.1 (OMol) & 1.55 & 0.131 & 1.99 & 0.242 & 2.57 & 0.669 & 0.75 & 0.472 \\
 & UMA-M-1.1 (OMol) & 1.35 & 0.075 & 1.56 & 0.160 & 1.94 & 0.472 & 0.44 & 0.219 \\
\midrule
\multirow{5}{*}{4M}  
 & eSEN-sm-d. & 3.00 & 0.190 & 4.31 & 0.319 & 4.29 & 0.894 & 1.79 & 0.633 \\
 & eSEN-sm-cons. & 2.90 & 0.144 & 4.10 & 0.285 & 3.43 & 0.768 & 1.35 & 0.587 \\
 & eSEN-md-d. & 1.62 & 0.078 & 2.65 & 0.174 & 3.01 & 0.587 & 1.06 & 0.289 \\
 & GemNet-OC-r6 & 1.49 & 0.137 & 2.69 & 0.237 & 3.35 & 0.737 & 1.66 & 0.550 \\
 & GemNet-OC & 0.94 & 0.122 & 1.72 & 0.211 & 3.18 & 0.718 & 1.36 & 0.523 \\
\bottomrule
\end{tabular}%
}
\begin{tablenotes}
  \item \kindatiny Energy (kcal/mol), Forces (kcal/mol/\angs)
\end{tablenotes}
\end{threeparttable}
\label{tab:test_breakdown_total}
\end{table}

\subsection{Evaluations}
Evaluations are broken into two categories: optimization and single-point based tasks. Optimization results (ligand-strain, conformers, and protonation) are reported in Table \ref{tab:opts_evals}. Single-point results (protein-ligand interaction, IE/EA, spin gap, and distance scaling) are reported in Table \ref{tab:single_evals}. Evaluations requiring an optimization used Sella~\cite{hermes2022sella} to ensure consistency with the underlying DFT calculated.

\begin{table*}[ht!]
\caption{\textbf{Optimization Evaluations} including ligand-strain, conformer prediction, and protonation states. Results are reported across a variety of energy and structure based metrics for each task.}
\centering
\resizebox{0.7\textwidth}{!}{%
\tablestyle{1.5pt}{1.05}

\begin{tabular}{y{25}y{55} ww  ww  ww}
\multirow{2}{*}{\vspace{-2.9cm} Dataset} 
& \multirow{2}{*}{\vspace{-2.9cm} Model} 
& \multicolumn{2}{c}{\ct[c3]{\it Ligand strain}} 
& \multicolumn{2}{c}{\ct[c4]{\it Conformers}} 
& \multicolumn{2}{c}{\ct[c5]{\it Protonation}} \\
& 
& \cb[c3]{RMSD}{min. [\angs~]$\downarrow$} 
& \cb[c3]{Strain energy }{MAE [kcal/mol]$\downarrow$}
& \cb[c4]{RMSD}{ensemble [\angs~]$\downarrow$} 
& \cb[c4]{$\Delta$ Energy}{MAE [kcal/mol]$\downarrow$}
& \cb[c5]{RMSD}{[\angs~]$\downarrow$} 
& \cb[c5]{$\Delta$ Energy}{MAE [kcal/mol]$\downarrow$}
\\
\hline
\addpadding
\multirow{2}{*}{OMol-1} 
    & esen-sm-cons.    & 0.24 & 0.11 & 0.04 & 0.12 & 0.06 & 0.54 \\
    & UMA-S-1.2 (OMol) & 0.17 & 0.10 & 0.03 & 0.09 & 0.05 & 0.53 \\
\hline
\addpadding
\multirow{7}{*}{OMol-0} 
    & esen-sm-cons.    & 0.19 & 0.11 & 0.03 & 0.10 & 0.05 & 0.67 \\
    & esen-md-d.       & 0.19 & 0.07 & 0.03 & 0.08 & 0.05 & 0.49 \\
    & GemNet-OC-r6     & 0.42 & 0.22 & 0.17 & 0.26 & 0.21 & 1.16 \\
    & GemNet-OC        & 0.44 & 0.19 & 0.18 & 0.28 & 0.18 & 0.95 \\
    & MACE-OMOL-L-0    & 0.24 & 0.18 & 0.05 & 0.16 & 0.06 & 0.82 \\
    & UMA-S-1.1 (OMol) & 0.21 & 0.11 & 0.04 & 0.12 & 0.06 & 0.83 \\
    & UMA-M-1.1 (OMol) & 0.12 & 0.07 & 0.02 & 0.06 & 0.04 & 0.55 \\
\hline
\addpadding
\multirow{4}{*}{4M}
    & esen-sm-cons.    & 0.26 & 0.13 & 0.05 & 0.16 & 0.07 & 1.12 \\
    & esen-md-d.       & 0.23 & 0.10 & 0.05 & 0.12 & 0.08 & 0.74 \\
    & GemNet-OC-r6     & 0.65 & 0.31 & 0.31 & 0.54 & 0.33 & 1.58 \\
    & GemNet-OC        & 0.65 & 0.29 & 0.28 & 0.46 & 0.29 & 1.50 \\
\shline
\end{tabular}
}
\label{tab:opts_evals}
\end{table*}

\textbf{Ligand strain} evaluation are in very good agreement with DFT for all of the methods investigated here, with the largest errors of only 0.19 kcal/mol for GemNet-OC down to 0.07 kcal/mol for UMA-M-1.1. Energy differences were well within chemical accuracy ($\sim$1 kcal/mol). Models trained on the 4M dataset were only worse by a hundredths of kcal/mol. RMSDs are also very close to DFT, indicating very good structural matching. The eSEN, UMA, and MACE models performed similarly, with an RMSD of between 0.12 and 0.24\angs, respectively.

\textbf{Conformers} energy differences were also all well within chemical accuracy for conformer ranking; the best performing models had errors less than 0.1 kcal/mol. While there was erosion in performance for models trained on the 4M dataset, they also were well within chemical accuracy.

\textbf{Protonation} proved to be the hardest of the optimization-based evaluations by energy. Energy differences ranged from 0.55 kcal/mol to about 1 kcal/mol. Though this is still quite good, it is 5-10x worse than for conformers. RMSD errors remained quite small, indicating that model have no trouble accounting for bond length and angle changes in (de)protonated molecules. It is interesting to note that models trained on the full OMol dataset which feature greater amounts of protonated/deprotonated data do tend to perform better on this evaluation.

\begin{table*}[ht!]
\caption{\textbf{Singlepoint Evaluations} including protein-ligand, ionization energies/electron affinity (IE/EA), spin gap, and distance scaling. Results are reported across a variety of energy and force based metrics for each task.}
\centering
\resizebox{\textwidth}{!}{%
\tablestyle{1.5pt}{1.05}

\begin{tabular}{y{25}y{55}w ww ww ww wwww}
%\shline
\multirow{2}{*}{\vspace{-2.9cm} Dataset} 
& \multirow{2}{*}{\vspace{-2.9cm} Model} 
& \multicolumn{2}{c}{\ct[c3]{\it Protein-ligand}} 
& \multicolumn{2}{c}{\ct[c4]{\it IE/EA}} 
& \multicolumn{2}{c}{\ct[c5]{\it Spin gap}} 
& \multicolumn{4}{c}{\ct[c6]{\it Distance scaling}} \\
& 
& \cb[c3]{Ixn Energy}{MAE} 
& \cb[c3]{Ixn Forces}{MAE} 
& \cb[c4]{$\Delta$ Energy}{MAE} 
& \cb[c4]{$\Delta$ Forces}{MAE} 
& \cb[c5]{$\Delta$ Energy}{MAE} 
& \cb[c5]{$\Delta$ Forces}{MAE} 
& \cb[c6]{$\Delta$ Energy (SR)}{MAE} 
& \cb[c6]{$\Delta$ Forces (SR)}{MAE} 
& \cb[c6]{$\Delta$ Energy (LR)}{MAE} 
& \cb[c6]{$\Delta$ Forces (LR)}{MAE} 
\\
\hline
\addpadding
\multirow{2}{*}{OMol-1} 
    & esen-sm-cons.    & 4.83 & 0.11 & 4.68 & 1.26 & 7.24 & 1.30 & 0.56 & 0.22 & 9.70 & 0.36 \\
    & UMA-S-1.2 (OMol) & 1.54 & 0.09 & 4.08 & 1.24 & 5.69 & 1.30 & 0.31 & 0.18 & 0.74 & 0.12 \\
\hline
\addpadding
\multirow{7}{*}{OMol-0} 
    & esen-sm-cons.    & 3.40 & 0.10 & 5.14 & 1.34 & 9.03  & 1.34 & 0.50 & 0.20 & 4.54 & 0.26 \\
    & esen-md-d.       & 1.48 & 0.05 & 3.36 & 0.99 & 7.01  & 1.02 & 0.34 & 0.13 & 2.53 & 0.14 \\
    & GemNet-OC-r6     & 1.08 & 0.09 & 4.61 & 1.17 & 8.55  & 1.23 & 0.32 & 0.16 & 1.86 & 0.14 \\
    & GemNet-OC        & 0.45 & 0.06 & 4.09 & 1.12 & 8.62  & 1.22 & 0.23 & 0.14 & 1.78 & 0.09 \\
    & MACE-OMOL-L-0    & 6.86 & 0.15 & 7.81 & 1.68 & 10.12 & 1.50 & 0.72 & 0.24 & 5.65 & 0.23 \\
    & UMA-S-1.1 (OMol) & 2.95 & 0.12 & 4.77 & 1.42 & 8.51  & 1.42 & 0.47 & 0.22 & 4.49 & 0.28 \\
    & UMA-M-1.1 (OMol) & 1.77 & 0.09 & 3.15 & 1.12 & 7.73  & 1.19 & 0.36 & 0.18 & 3.18 & 0.29 \\
\hline
\addpadding
\multirow{4}{*}{4M} 
    & esen-sm-cons.    & 5.62 & 0.13 & 6.76 & 1.59 & 11.03 & 1.58 & 0.63 & 0.25 & 5.37 & 0.27 \\
    & esen-md-d.       & 2.27 & 0.06 & 5.54 & 1.33 & 9.93  & 1.34 & 0.49 & 0.18 & 2.56 & 0.16 \\
    & GemNet-OC-r6     & 2.21 & 0.10 & 7.61 & 1.52 & 12.89 & 1.61 & 0.46 & 0.20 & 1.92 & 0.16 \\
    & GemNet-OC        & 0.84 & 0.08 & 5.48 & 1.44 & 11.31 & 1.51 & 0.33 & 0.18 & 2.31 & 0.10 \\
\shline
\multicolumn{12}{l}{All energies in kcal/mol, all forces in kcal/mol/\AA{}}
\end{tabular}
}
\vspace{2pt}
\label{tab:single_evals}
\end{table*}

\textbf{Protein-ligand Interaction} evaluation metrics improve dramatically between MACE-OMOL-L (6.86 kcal/mol interaction energy MAE, 0.15 kcal/mol/\angs{} interaction forces MAE) and UMA-S-1.2 (1.54 kcal/mol interaction energy MAE, 0.09 kcal/mol/\angs{} interaction forces MAE). Still further improvements in interaction energy are seen with GemNet-OC (0.45 kcal/mol). These improvements may imply that higher angular momentum representations are important for accurately capturing the interactions. While interaction forces MAE is comparable with the biomolecules portion of the Test force MAE, the interaction energy MAE is significantly greater than biomolecules Test energy MAE, demonstrating the additional challenge inherent in specifically capturing intermolecular protein-ligand interactions. 

\textbf{IE/EA} and \textbf{Spin Gap} evaluation metrics demonstrate the severe challenge presented by energy and force differences of metal-containing complexes with different charge and/or spin states. UMA-M-1.1 achieves 3.15 kcal/mol error on IE/EA and 7.73 kcal/mol $\Delta E$ MAE on spin gaps. Force errors were also notably higher by a large margin than any test set or evaluation metric. While $\Delta E$ and $\Delta \vec{F}$ errors do improve by nearly 50\% on IE/EA for larger models, they only improve by around 25\% on spin gap, suggesting that more than just model scaling will be required to achieve sufficiently low error for scientific utility on these tasks.

\textbf{Distance Scaling} evaluation metrics demonstrate that capturing non-covalent interactions over different length scales was a significant challenge for current baseline models. While $\Delta E$ and $\Delta F$ errors are fairly small inside the cutoff distance (SR), they dramatically increase once systems are passed that cutoff (LR), though this is significantly amelioriated in the case of UMA-S-1.2. While this is to be expected since none of these models contain any correction for long range interactions as are sometimes included in other models~\cite{gong_predictive_2025,anstine_aimnet2_2025}, the fact that these models can display significant discontinuities in the PES at the cutoff is a serious issue that further work should address.
\section{Outlook and future directions}
\omol~is the first dataset of its kind to span the breadth of major chemistry domains (inorganic, organic, biochemistry) at a high level of theory. However, we recognize that gaps in the coverage of chemical space still exist. For example, the \omol~dataset does not include any of the radioactive elements after bismuth (\textit{e.g.}, actinides) or structures which might arise in polymer materials. Furthermore, the coverage of certain classes of materials such as lanthanides complexes, multimetallic structures, and solvated protonated organic molecules and metal complexes are relatively limited, although baseline models trained on \omol~may still be suitable for these applications. We invite the community to explore its capabilities to help inform future dataset development.

Baseline models trained on \omol~have demonstrated great potential for MLIPs to be highly accurate across a wide range of chemical tasks. With models like UMA-M-1.1 and UMA-S-1.2 achieving average energy and force errors as low as 1.38 kcal/mol and 1.07 kcal/mol, respectively, across all test splits; \omol’s combination of scale, diversity, quality, and complexity represents a significant leap in molecular DFT datasets for training MLIPs.

While baseline models have shown strong performance, approaching or surpassing chemical accuracy ($\sim$ 1kcal/mol) especially in domains such as biomolecules and small-molecule organics; \omol’s evaluation tasks reveal significant gaps that need to be addressed. Notably, ionization energies/electron affinity, spin-gap, and long range scaling have errors as high as 4-9 kcal/mol. Architectural improvements around charge, spin, and long-range interactions are especially critical here. Partial charge and spin data were included in \omol~to assist in these efforts. In the future, we hope to extend our evaluations to include free energy and reactivity tasks (for example, transition state or reaction path optimization). Such evaluations would require accurate geometry second derivatives (\textit{i.e.}, Hessians), tests of which are currently absent from our MLIP evaluations, and may reveal new challenges to be overcome. Other properties computed in this dataset, such as multipole moments and electron densities, may enable new models which can also predict spectroscopic observables.

In order to foster community engagement and motivate rapid method development, we released a public leaderboard to track the community’s efforts on \omol’s evaluation tasks. We hope the \omol~dataset will be a valuable resource to the community as we all seek to advance the state of the art. Whether \omol~is leveraged for pre-training, specialized domain training (\textit{e.g.} biomolecules), or fine-tuning, we are eager to see what challenges the community will address with this dataset.

\clearpage
\section{Acknowledgments}

M.G.T. acknowledges the support of the U.S. Department of Energy (DOE), Office of Science, Office of Basic Energy Sciences (BES), Heavy Element Chemistry Program (KC0302031) under contract number E3M2. M.G.T. acknowledges support from the Laboratory Directed Research and Development program through the Institute for Materials Science (IMS) of Los Alamos National Laboratory (LANL) for his work on development of sampling methods for metal complexes. LANL is operated by Triad National Security, LLC, for the National Nuclear Security Administration of the U.S. Department of Energy (contract no. 89233218CNA000001).

M.R.H. acknowledges support by a grant from the Simons Foundation (Grant 839534, MET) and the NYU IT High Performance Computing resources, services, and staff expertise.

I.B. and G.C. acknowledge the use of the AI computing and storage resources by GENCI at IDRIS thanks to the grant 2024-GC010815458 on the supercomputer Jean Zay's  H100 partition.

P.E. acknowledges support from the National Institute of General Medical Sciences of the National Institutes of Health under award number R01GM140090.

A. S. K. and S. R. acknowledge support from the U.S. Department of Energy, Office of Science, Energy Earthshot initiatives as part of the Center for Ionomer-based Water Electrolysis at Lawrence Berkeley National Laboratory under Award Number DE-AC02-05CH11231. 

A.S.R. acknowledges support from the Wilke 1989 School of Engineering and Applied Science Innovation Fund at Princeton University.

S.V. was supported by the DOE's National Nuclear Security Administration's Office of Defense Nuclear Nonproliferation Research and Development (NA-22) as part of the NextGen Nonproliferation Leadership Development Program.

S.M.B. acknowledges support from the Laboratory Directed Research and Development program of Lawrence Berkeley National Laboratory (LBNL), supported by the Office of Science, Office of Basic Energy Sciences (BES), of the U.S. Department of Energy (DOE) under Contract No. DE-AC02-05CH11231. For his contributions to the electrolyte modeling portion of the dataset, S.M.B. acknowledges support from the Energy Storage Research Alliance (ESRA) (DE-AC02-06CH11357), an Energy Innovation Hub funded by the U.S. Department of Energy, Office of Science, Basic Energy Sciences.

\clearpage
\newpage
\bibliographystyle{achemso}
\bibliography{paper}

\clearpage
\newpage

\let\addcontentsline\oldaddcontentsline
\appendix
\renewcommand{\contentsname}{Appendix Table of Contents}
\section*{Appendix}
\tableofcontents
\clearpage
\section{Calculation Details}

All calculations in OMol25 were carried out with the ORCA 6.0.0 DFT package~\cite{RN204,neese_software_2025}. ORCA supports various integral acceleration techniques, including RI-J and COSX, which dramatically improve the computational cost of these calculations at a very small cost in error, and these were used here. Experimentation with integral threshold settings indicated that the best trade-off between robust convergence and computational cost was to set the integral threshold (\texttt{thresh} in ORCA) to 1e-12 and the primitive batch threshold (\texttt{tcut} in ORCA) to 1e-13; these values were subsequently adopted by the ORCA package to be defaults in future versions of ORCA. ORCA's \texttt{tight} convergence settings were employed. Extensive benchmarking with various combinations of grid settings (both the exchange-correlation grid and the COSX grid) indicated that typical grids led to small numerical inconsistencies between energy and forces. In other words, there were discrepancies between the derivative of the energy with respect to coordinates and the computed forces due to grid incompleteness, and these errors were significant on the scale of errors with state-of-the-art MLIPs). In order to achieve sufficiently tight consistency, ORCA's \texttt{DEFGRID3} offered the best trade-off of convergence and cost. This corresponds to a pruned grid with 590 angular points for exchange-correlation and 302 for the final COSX grid.

As described in Section \ref{sec:dataset:lot}, we typically compute systems where bonds may be breaking in UKS. We could not find mention of whether Transition-1X was originally run in RKS or UKS and so have computed the dataset both ways, though we advocate for the use of the UKS version. In the end, this is a small consideration as <5\% of the Transition-1X data in UKS had an expectation value of $S^2$ greater than 0.001, indicating that, for more than 95\% of the dataset, the RKS and UKS SCF solutions are the same.

\subsection{Calculation Quality Filters}\label{sec:app:quality}
Given the scale and diversity of \omol{}, adequate quality and error checks are required before considering a calculation valid, particularly for ML training. We enforce several quality checks on the resulting DFT calculations before considering them in the final dataset:

\begin{itemize}
    \item Referenced energies shall not exceed $\pm$150 eV. This removes highly unreasonable configurations; 
    \item Max per atom force shall not exceed 50eV/\angs. This also removes highly unreasonable configurations;
    \item $S^2$ < 0.5 for open-shell metal-containing systems, $S^2$ < 1.1 otherwise. We enforce tighter constraints on metal-systems to avoid incorrect SCF solutions but allow full spin-unpairing which may occur in organic reactivity;
    \item Enforce alpha, beta, and total electron consistency with the integrated densities. This indicates insufficient grid density;
    \item ORCA errors where the final COSX exchange deviates considerably. This indicate convergence errors;
    \item Non-negative HOMO-LUMO gaps.
\end{itemize}

Furthermore, using a trained model on an early snapshot of \omol{}, we perform an error analysis similar to the work of MACE-OFF on SPICE~\cite{kovacs2023mace}. Energy errors were evaluated for a held-out test set of \omol{}, followed by manual inspection on some of the worst offenders. This allowed us to identify and filter systematically problematic inputs, such as isolated metal centers far away from the rest of the structure resulting in convergence to excited SCF minima.

\subsection{Computed Properties}\label{sec:app:calc}

The properties computed for each point in \omol{} are the following:
\begin{enumerate}
    \item Total energy (in eV)
    \item Forces (in eV/\angs)
    \item charge
    \item spin
    \item Number of atoms
    \item Number of electrons
    \item Number of ECP electrons
    \item Number of basis functions
    \item Unrestricted vs. Restricted
    \item Number of SCF steps
    \item Energy computed by VV10
    \item S$^2$ expectation value
    \item Deviation of S$^2$ from ideal
    \item Integrated density (should be very close to the total number of electrons)
    \item HOMO energy (in eV), $\alpha$ and $\beta$ for unrestricted
    \item HOMO-LUMO gap (in eV), $\alpha$ and $\beta$ for unrestricted
    \item Maximum force magnitude for a given atom in a given direction (fmax)
    \item Mulliken charges (and spins if unrestricted)
    \item Loewdin charges (and spins if unrestricted)
    \item NBO charges (and spins if unrestricted) if the total number of atom <= 70
    \item Any ORCA warnings that are generated
    \item ORCA .gbw files and densities will be made available in the near future

Additional properties which have been computed, such as multipole moments, Fock matrices, and orbital energies, will be released in future versions of the dataset.
\end{enumerate}

\section{OMol Dataset Versions}

A pre-release version of the \omol{} dataset was initially published on arXiv on May 13, 2025 along with models. That data, which we term OMol-0, contained about 100M DFT datapoints. After feedback from the community was obtained on the performance of models and areas for improvements of \omol, an additional 40M DFT datapoints were added to produce the final version of the dataset contained in this manuscript. Intermediate spin metal complexes and additional oxidation states were added to the Metal Complex domain; low-spin (i.e. antiferromagnetically coupled versions) of high-spin electrolytes and larger electrolyte systems were added to the Electrolytes domains. The remaining additions comprise most of the Main-group Molecules domain, described in detail below (section \ref{sec:methods:maingroup}), with the exception of non-radical RGD1 data and the MechDBs which were allocated in the preprint to the Community domain (it was pointed out that that classification was improper as none of the geometries therein had actually appeared in a previous community dataset).

Models trained on OMol-0 have already been utilized in numerous works by other authors. As such, we felt it was important to include OMol-0-trained models in this manuscript. Those models are: eSEN-sm-cons., UMA-S-1.1, UMA-M-1.1, GemNet-OC, GemNet-OC-r6, and MACE-OMOL.

The expanded OMol-1 provides notable improvements in performance of models for ionization energies, electron affinities, spin-gaps (particularly for otherwise closed-shell organic molecules), noble gases, and the heavy $p$-block elements. The OMol-1 models are: UMA-S-1.2. No significant regressions were noted by expanding the dataset to include these additional data.
\section{Biomolecules}

Biological application represent one of the largest current uses of computational chemistry. Annual spending by pharmaceutical companies on structure-based drug design, both of small molecules and biologics is on the order of a billion dollars. Methods such as free energy perturbation require extremely high accuracy in the potential used. Due to the large size of biological systems and the need to simulate often for tens of nanoseconds, atomistic simulations nearly universally employ classical molecular dynamics force field. To achieve \textemdash and surpass\textemdash{} state-of-the-art classical forcefields, high quality quantum chemistry data which covers the breadth of biological interactions is needed.

To this end, we developed several focused subdatasets intended to probe a variety of applications which would reasonably be expected to arise in biochemistry research. All structures taken from the Protein Data Bank (PDB) were first prepared with Schr\"{o}dinger's Protein Preparation Wizard~\cite{madhavi_sastry_protein_2013}. Molecular Dynamics (MD) calculations were performed with Schr\"{o}dinger's Desmond molecular dynamics engine with the OPLS4 forcefield~\cite{lu_opls4_2021}. Desmond/OPLS4 automatically assigns forcefield parameters to an extremely broad amount of chemical matter automatically, which facilitated this high-throughout process. Protonation and tautomeric states of ligands were sampled with Schr\"{o}dinger's Epik program~\cite{johnston_epik_2023,madhavi_sastry_protein_2013}.

\subsection{Protein Pocket + Ligand}\label{sec:app:bio:prot-lig}

Ligand-protein interactions are some of the most important for computational chemistry, and it is critical to obtain a broad and diverse sampling of ligand chemistries and geometries. Moreover, often neglected from ligand-protein interaction is the effect of protonation state and tautomerism on both ligands and proteins. The BioLiP2 database contains nearly 9.5M entries detailing ligands (broadly defined, including small molecules, metal ions, peptides, DNA, and RNA) and their corresponding receptor residues. All non-peptide, non-nucleic acid ligand structures in the BioLiP2 database as of May 31, 2024 were extracted and processed. In order to obtain unique ligand binding environments, we removed the following entries as duplicates:
\begin{enumerate}
    \item if the same ligand id appeared with identical receptor residues (including residue numbering); we do not require PDB IDs to be identical as closely related proteins often have conserved binding sites
    \item if the same ligand ID appeared in the same PDB structure with receptor residues that were largely the same (one being a proper subset of the other), we discard the larger binding site, as those marginal residues to not represent key interactions
\end{enumerate}

This resulted in 258,467 unique ligand-pocket pairs. We further elaborated those pairs using Schr\"{o}dinger's Epik in conjunction with the Protein Preparation Wizard to add up to the 10 lowest energy protonation/tautomer states and optimize the protonation of receptor residues in their presence. The ligand and the residues identified by the BioLiP2 entry were extracted along with any residues with any atom within 2.5\angs{} of the ligand (these residues include such things as waters coordinated to metal centers and ligands which may consist of multiple covalently attached positions). All protein residues were capped with ACE or NMA at the N- and C-termini, respectively, and these caps were rotated to match the C$\alpha$ positions of the native protein structure. If two residues were extracted that had a single residue in between that was not extracted, then these capping groups would overlap and instead of capping, a glycine is placed between the two residues with C$\alpha$ at the same location as the residue in the native protein.

Various cleanup steps were applied to ensure that these extracted pockets were suitable for further study:
\begin{enumerate}
    \item Hydrogen atoms were added or removed as needed to match the expected Lewis structure
    \item Ions were adjusted to the expected oxidation states: Li, Na, K, Rb, Cs, Ag to +1, Mg, Ca, Sr, Zn, Cd, Hg to +2, Al, Eu to +3, Fe (if an oxidation state was not specified in the PDB entry) to +2, Cu (if an oxidation state was not specified) to +2, Os (if not specified) to +3, Sn (if not specified) to +4 unless it was part of Sn-C or Sn-S bonding.
    \item Formal charges on various functional groups were adjusted
    \item Visual inspection of many systems suggest that Epik does not always do a good job with the protonation state of sulfates, phosphates, and ligands bound to metals, especially heme groups. As such, if a fully deprotonated state of these functional groups were not present, then one was added.
    \item Disulfide bonds that were disrupted by the extraction procedure were capped with hydrogen
    \item Several minor adjustments to bonding and Lewis structure were made to allow the system to work properly with Schr\"{o}dinger's Desmond MD engine.

\end{enumerate} 

Charge and spin states were assigned to these extracted pockets on the basis of the formal charge of the atoms and any metal oxidation states that were determined by this process. After confirming that the charge and spin were physically consistent and restricting the number of atoms to 350 or fewer, this left 416,324 systems.

Since PDB structures are often not suitable for use directly with DFT, owing to, among other things, imprecisely determined atomic bond lengths, and to increase the structural diversity present, the extracted pockets were used as inputs to an MD simulation. Per the recommended procedure from Desmond, the systems were run with Brownian dynamics for 100ps at 10K with all heavy atoms restrained, then Langevin dynamics for 12ps at 10K with all heavy atoms restrained. The restraints were then removed and the system energy was minimized. We then applied restraints to the C$\alpha$ positions of all present residues and 1.02 nanoseconds of Langevin dynamics were simulated. The systems were split randomly into two groups: 85\% were run at 300K and 15\% were run at 400K. The C$\alpha$ constraints were included to prevent the pocket from disintegrating due to the absence of the rest of the protein structure. We discarded frames where the ligand centroid moved more than 5\angs{} from the starting centroid position in the final stage, with the idea being that this represented the ligand falling out of the extracted pocket. Frames with an RMSD less than 1\angs{} from the initial frame were discarded (this occurred in some cases were the combination of the C$\alpha$ constraints and strong ligand-pocket interactions (\textit{e.g.}, with some metals) led to little movement by MD). The remaining frames were evenly subsampled into 10 frames. The first and last, being farthest in time and therefore most likely to be decorrelated, were selected as inputs for DFT. This results in 622,360 DFT inputs from the 300K MD and 122,906 inputs from 400K MD.

\subsection{Fragmented Protein Pockets + Ligand}

Having exhausted experimental ligand-protein interaction structures in BioLip2 with ~750k datapoints, synthetic ligand-protein structures were required both to increase the sampling of diverse interactions and especially to improve coverage of interactions as they might pertain to docking simulations with novel chemical matter. Receptor pockets obtained in the previous step were identified as "drug-like" pockets (in contrast to \textit{e.g.}, metal-binding pockets) by noting if any binding affinity data (\textit{e.g.}, from PDBbind) was associated with the BioLiP2 entry for the pocket. This procedure identified 32,833 pockets which are likely to be druggable. Drug-like systems were randomly sampled from Zinc20's "clean leads" set (molecular weight between 250 and 350, predicted LogP $\leq$ 3.5, and 7 or fewer rotatable bonds), ChemBL, and GEOM drugs and docked into these pockets with smina. Up to three protonation/tautomer states were computed for each ligand with Epik before docking. These systems were randomly split 50/50 between 300K and 400K MD using the procedure described above. However, in this case, after MD, instead of the entire pocket being used as input for DFT, only the two receptor fragments that were closest to the ligand are retained (as long as that is less than 5 residues in total). Only the last frame of the MD was used. These fragmented pockets are smaller, which allow a larger number to be run for the same computational cost. This procedure resulted in 2,323,228 inputs from 300K MD and 2,326,639 inputs from 400K MD.

\subsection{Protein + Protein: Core}

While there is some degree of protein-protein interaction in the Pocket + Ligand and Fragment Pocket + Ligand data, we sought to increase coverage of protein-protein interaction data. To ensure broad coverage, we focus separately on the cores of proteins and on the interfaces between protein subunits.

We randomly sample proteins in the PDB and, for each structure, randomly sample up to 20 residues whose relative solvent accessible surface areas is 0.2 or less (indicating that the residue was part of the protein core). All residues where any heavy atom of the side-chain was within 4\angs{} (in 80\% of cases) or 4.5\angs{} (in 20\% of cases) of the heavy atoms in the randomly selected core side-chain was extracted as the residue's protein environment. These residues were treated as for ligand-pockets: adding hydrogens with Protein Preparation Wizard, fixing Lewis structures, capping termini of residues, and/or replacing single-amino acid gaps with glycine. To avoid being too similar to the SPICE dataset of amino acid dimers, we also exclude any system that consists of fewer than three residues. The resulting prepared systems where run with the same MD procedure as described above, 50\% at 300K and 50\% at 400K. Both the first and last frame of the MD were used for DFT, resulting in 1,432,225 inputs. Additionally, 1,209,514 samples were submitted without running them through the MD procedure first in order to assess whether the DFT machinery could handle the sort of raw geometries typical of PDB structures.

\subsection{Protein + Protein: Interface}\label{app:bio:ppi}

Protein-protein interfaces were generated by randomly sampling residues from the DIPS-plus database to identify residues at protein interfaces. Any residue that had a heavy atom of its side-chain within 4\angs{} of the selected residue's side-chain heavy atoms and was on the same protein chain or within 6\angs{} if it was on a different protein chain were extracted as the environment of that interface residue. Preparing, capping, and MD was performed in the same manner as protein-core systems above, giving 929,656 DFT inputs. An additional <n> were prepared and capped but instead of running with classical MD, these inputs were run with MD using an MLIP, as described in Section \ref{sec:methods:mlmd}.

\subsection{DNA/RNA}

DNA and especially RNA often are also part of protein structures, and proper coverage of protein-nucleic acid (and nucleic acid-nucleic acid) interactions are critical to ensuring ML models can accurately treat the breadth of biological systems. Fortunately, the BioLiP2 also features extensive coverage of these kinds of interactions. Due to the structural differences of nucleic acid vs. proteins and ligand, we look for specific motifs rather than simple radius cutoffs. Starting from a randomly selected nucleic acid base residue, which we denote as the ``core'', we classify the types of motifs we will extract as follows. Note that some of these are subsets of each other, but we consider them separate types to obtain a range of sizes of extracted systems.
\begin{enumerate}
    \item Residues with any atom within 2.5\angs{} of the core and not directly bound to the core. For DNA, this is typically the complementary base pair and any nearby protein residues.\label{S--}
    \item Residues with any atom within 2.5\angs{} of the core and not DNA or RNA. This is generally the base itself and any protein nearby. \label{S-}
    \item Residues directly attached to the core. This is generally three-some stacks of bases.\label{E}
    \item Residues with any atom within 2.5\angs{} of the core that aren't directly bound to it and not protein. For DNA, this is typically the complementary base.\label{--}
    \item Residues with any atom within 2.5\angs{} of the core and not protein. For DNA, this is typically the complementary base and the residues immediately before and after the core on the same chain.\label{E-}
    \item Residues with any atom within 2.5\angs{} of the core or the residue immediately after it in the chain. For DNA, this is typically two consecutive base pairs.\label{==}
\end{enumerate}

For all nucleic acid ligands in BioLiP2 (which are typically at least 8 nucleic acid bases (note that this implies 16 DNA residues total per chain)), we randomly select three of type \ref{S--}, two of type \ref{S-}, one of type \ref{E}, one of type \ref{--}, and one of either type \ref{E-} or type \ref{==} with 50\% probability of either. After extracting these residues, proteins are capped as described above and nucleic acids are capped with an O3' OH group. MD was run on these systems as well; in addition to protein C$\alpha$s being restrained, the phosphate P and C4' positions of nucleic acids were restrained to keep the nucleic acid backbone intact. For the 400K MD, after discretizing into 10 equally spaced frames, rather than the first and last frame being used, the first two are used. This is because significant melting of nucleic acid structures occurs at this temperature. In our simulations, a large number of base pairs had moved apart to distances greater than 10\angs{} even with the backbones frozen. The first two frames do not feature significant melting yet. For 300K MD, the first two and final frames are taken. This resulted in 770,537 RNA inputs and 570,087 DNA inputs.

Additionally, the Nucleic Acid Knowledge Base (NAKB) was mined for PDB structures of RNA and DNA with unusual structural motifs. Structures had been annotated as being: A-form DNA, Z-form DNA, two-strands forming a parallel helix, a helix composed of base triple steps, a helix composed of base quadruple steps, helix composed of steps with 5 or more hydrogen-bonded bases in plane, or a Holliday junction. The same selection procedure as above was used to sample the various unique environments present in these structures. This yielded 129,029 additional DFT inputs.
\section{Metal Complexes}

\subsection{Complexes from Architector}\label{app:mc:arch}
Metal complex chemistry spans the periodic table and includes unique features in terms of molecular symmetry and ligand-metal-ligand interactions. To generate complexes with diverse metal centers and ligand chemistry, the open-source Architector package \cite{taylor_architector_2023} was used. As a brief summary, Architector takes as input a metal center, metal oxidation state, metal spin, metal coordination number, and ligands to surround the metal center. Ligands are specified by both a SMILES string \cite{smiles} and a list of indices of the atoms of the SMILES string that are bound to the metal center. The combination of this information leads to a full molecular graph encoding a mononuclear molecular complex. 

Possible metal oxidation states and coordination numbers were mined from a combination of the mendeleev python package \cite{mendeleev} and a previously-published dataset derived from the Cambridge Structural Database (CSD summary) \cite{taylor_architector_2023,architector_data}. The mendeleev package contains common oxidation states for each metal center, and the CSD summary dataset contains the frequency of coordination numbers (CNs) observed for each metal center in mononuclear metal complexes. Only CNs with more than 1\% frequency in the CSD summary dataset and CNs less than 13 were selected. Since promethium complexes do not exist in the CSD summary dataset due to its radioactivity, relatively short half-life, and abundance, the coordination numbers and oxidation states known for neighboring samarium were applied to promethium. All oxidation states and coordination numbers sampled are found in Table \ref{tab:metals}. When sampling metal complexes from this space, metal and oxidation state were uniformly sampled followed by a uniform sampling from possible coordination numbers. Metal spin states were assigned from the mendeleev electron configuration for the given oxidation state, which tends to give the highest reasonable spin for metal at a given oxidation state.

\begin{table}[ht!]
\centering
\caption{Metals, oxidation states, and coordination numbers sampled for OMol25 metal complexes.}
\label{tab:metals}
\noindent
\begin{minipage}[t]{0.45\textwidth}
  \begin{tabular}{lrl}
    \toprule
    metal & oxidation state & coordination numbers \\
    \midrule
Li & 1 & 2,3,4,5,6 \\
Be & 2 & 4,5 \\
Na & 1 & 2,4,5,6,7,8,9 \\
Mg & 2 & 4,5,6,7 \\
Al & 2 & 3,4,5,6,7,8 \\
Al & 3 & 4,5,6 \\
K & 1 & 2,3,4,5,6,7,8,9,10 \\
Ca & 2 & 4,5,6,7,8,9 \\
Sc & 3 & 4,5,6,7,8 \\
Ti & 2 & 5,6,7,8 \\
Ti & 3 & 4,5,6,7,8 \\
Ti & 4 & 4,5,6,7,8 \\
V & 2 & 4,5,6,7,8 \\
V & 3 & 4,5,6,7 \\
V & 4 & 4,5,6,8 \\
V & 5 & 4,5,6 \\
Cr & 2 & 4,5,6,7 \\
Cr & 3 & 5,6,7 \\
Cr & 6 & 4,5,6,7 \\
Mn & 1 & 3,4,5,6,7,8 \\
Mn & 2 & 4,5,6,7 \\
Mn & 3 & 4,5,6 \\
Mn & 4 & 4,5,6 \\
Mn & 6 & 4,5,6,7 \\
Mn & 7 & 4,5,6,7 \\
Fe & 2 & 4,5,6,7 \\
Fe & 3 & 4,5,6,7 \\
Co & 1 & 3,4,5,6,7,8 \\
Co & 2 & 4,5,6,7 \\
Co & 3 & 4,5,6 \\
Ni & 2 & 4,5,6 \\
Ni & 3 & 2,3,4,5,6,7,8 \\
Cu & 1 & 2,3,4,5 \\
Cu & 2 & 4,5,6 \\
Zn & 2 & 4,5,6 \\
Ga & 1 & 2,3,4,5,6,7 \\
Ga & 3 & 4,5,6 \\
Rb & 1 & 5,6,7,8,9,10,11,12 \\
Sr & 2 & 4,5,6,7,8,9,10 \\
Y & 3 & 4,5,6,7,8,9 \\
Zr & 4 & 4,5,6,7,8,9 \\
Nb & 5 & 4,5,6,7,8 \\
Mo & 4 & 4,5,6,7,8 \\
Mo & 6 & 4,5,6 \\
Tc & 4 & 4,6,7 \\
Tc & 7 & 4,6,7 \\
Ru & 2 & 4,5,6,7,8 \\
Ru & 3 & 5,6,7 \\
Ru & 4 & 4,5,6 \\
Rh & 1 & 2,3,4,5,6\\
Rh & 2 & 2,3,4,5,6\\
Rh & 3 & 4,5,6 \\
Pd & 0 & 4,5 \\
Pd & 2 & 4,5 \\
    \bottomrule
  \end{tabular}
\end{minipage}
\hfill
\begin{minipage}[t]{0.45\textwidth}
  \begin{tabular}{lrl}
    \toprule
    metal & oxidation state & coordination numbers \\
    \midrule
Pd & 4 & 4,5,6 \\
Ag & 1 & 2,3,4,5,6 \\
Cd & 2 & 4,5,6,7,8 \\
In & 3 & 4,5,6,7,8 \\
Sn & 2 & 3,4,5,6,7,8,9 \\
Sn & 4 & 4,5,6,7 \\
Cs & 1 & 4,5,6,7,8,9,10,11,12 \\
Ba & 2 & 4,5,6,7,8,9,10,11,12 \\
La & 3 & 4,5,6,7,8,9,10,11 \\
Ce & 3 & 4,5,6,7,8,9,10 \\
Ce & 4 & 4,5,6,7,8,9,10 \\
Pr & 3 & 5,6,7,8,9,10 \\
Pr & 4 & 4,5,6,7,8,9,10,11,12 \\
Nd & 3 & 4,5,6,7,8,9,10 \\
Pm & 3 & 4,5,6,7,8,9,10 \\
Sm & 2 & 4,5,6,7,8,9,10,11,12 \\
Sm & 3 & 4,5,6,7,8,9,10 \\
Eu & 2 & 5,6,7,8,9,10 \\
Eu & 3 & 6,7,8,9,10 \\
Gd & 3 & 5,6,7,8,9,10 \\
Tb & 3 & 4,6,7,8,9,10 \\
Tb & 4 & 4,5,6,7,8,9,10 \\
Dy & 3 & 4,5,6,7,8,9,10 \\
Dy & 4 & 4,5,6,7,8,9,10 \\
Ho & 3 & 4,5,6,7,8,9,10 \\
Er & 3 & 4,5,6,7,8,9,10 \\
Tm & 3 & 5,6,7,8,9,10 \\
Yb & 2 & 4,5,6,7,8,9,10 \\
Yb & 3 & 4,5,6,7,8,9,10 \\
Lu & 3 & 4,5,6,7,8,9,10 \\
Hf & 4 & 4,5,6,7,8 \\
Ta & 5 & 4,5,6,7,8 \\
W & 4 & 4,5,6,7,8 \\
W & 6 & 4,5,6,7,8 \\
Re & 3 & 4,5,6,7 \\
Re & 4 & 4,5,6,7 \\
Re & 5 & 4,5,6,7 \\
Re & 7 & 4,5,6 \\
Os & 2 & 4,5,6 \\
Os & 4 & 4,5,6 \\
Ir & 2 & 3,4,5,6,7 \\
Ir & 3 & 4,5,6 \\
Ir & 4 & 4,5,6,7 \\
Pt & 2 & 4,5 \\
Pt & 4 & 4,6 \\
Au & 1 & 2,3,4 \\
Au & 2 & 2,3,4,5,6 \\
Au & 3 & 4,5 \\
Hg & 1 & 2,3,4,5,6,7,8 \\
Tl & 1 & 2,3,4,5,6,7,8,10 \\
Tl & 3 & 2,4,5,6,7,8 \\
Pb & 2 & 4,5,6,7,8,9,10 \\
Pb & 4 & 4,5,6,7,8 \\
Bi & 3 & 4,5,6,7,8,9 \\
    \bottomrule
  \end{tabular}
\end{minipage}
\end{table}

For ligand sampling, sets of ligands have previously been curated by several groups from the CSD, but these focused on accurately extracting as many different ligands as possible for computational design efforts, rather than for large-scale computation \cite{Kneiding2024,Chen2023}. Further, such datasets contain reported accuracies of only $\approx$90, so manual curation is regardless required. As the CSD represents a set of experimentally derived structures and it is much easier to synthesize ligands from around known chemistries, these ligand datasets tend to consist of many repeated chemical motifs \cite{Nandy2023}. To overcome these limitations, here, a set of 723 base ligands was manually curated from the CSD summary dataset based on coordinating atom type diversity and frequency of occurrence in the CSD summary dataset.

\begin{figure}
\centering\includegraphics[width=0.7\linewidth]{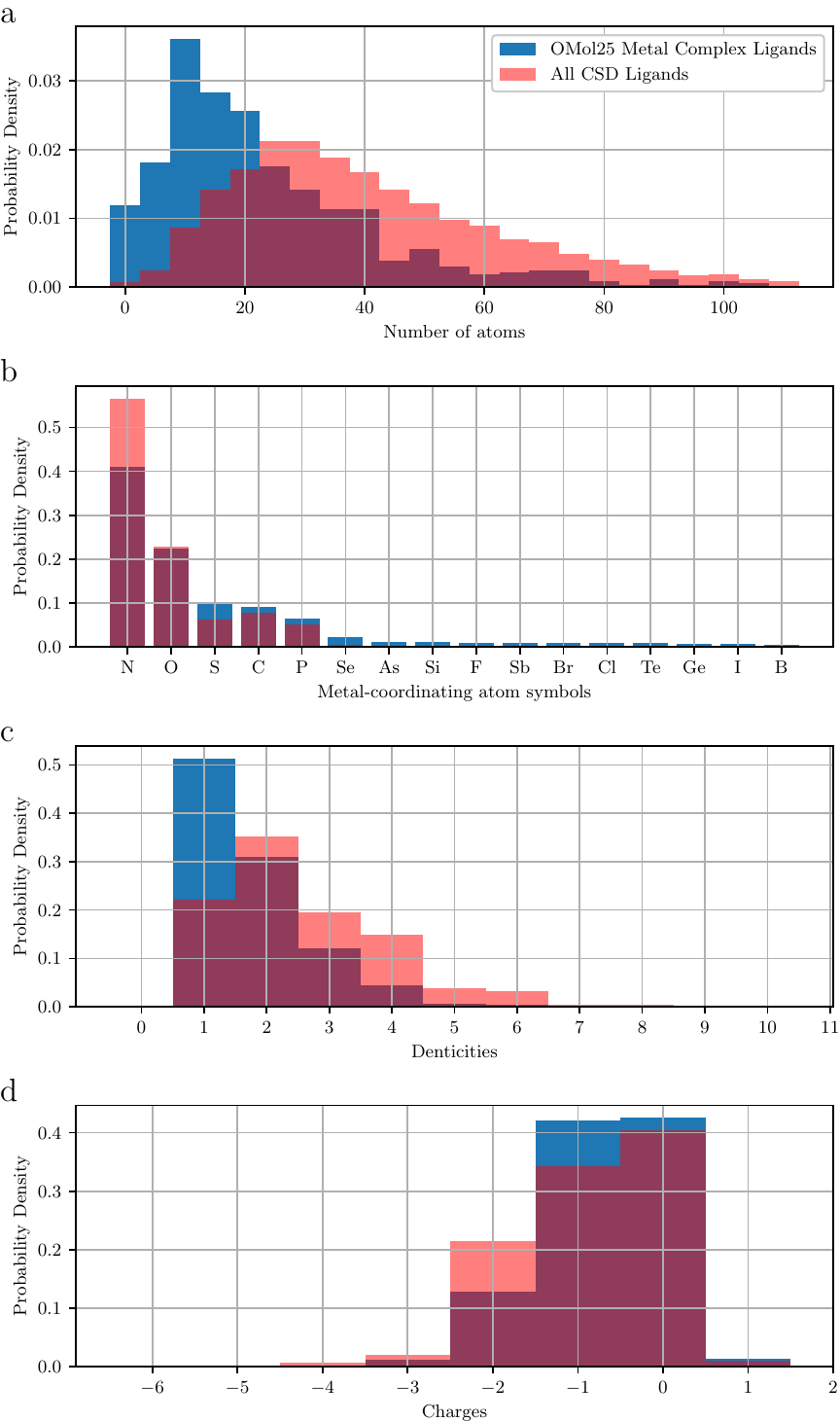}
\caption{Histogram cross-sections of ligands sampled for metal complex datasets. a) total number of atoms in the ligands b) metal-coordinating atoms from the ligands c) denticity of the ligands and d) the total charges on the ligands.}
\label{fig:lig_hists}
\end{figure}

The curation process which generated these 723 ligands proceeded as follows. Ligands were selected based on a total of 32 unique SYBYL atom types (\textit{e.g.}, C.2, N.3, N.ar ...) of both metal-coordinated and uncoordinated atoms. For each unique SYBYL atom type the 10 most frequent ligands along with 10 randomized ligands from the CSD summary dataset containing either the metal-bound atom type or non-metal-bound atom type were selected. After removing duplicates, a total of 719 ligands were produced, followed by manual inspection of all ligands resulting in the addition of 3 ligands encoding additional organometallic bonds (cyclopentadienyl, pentamethylcyclopentadienyl, and bulky cyclopentadienyl) and hydride for 723 ligands from which to sample. Basic statistics of the OMol25 sampling ligands and CSD ligands are highlighted in Figure \ref{fig:lig_hists}. Note the resulting curated ligand set spans a wide range of number of atoms (1-120), metal-coordinating atom types (32), denticities (1-8), and charges (-4 to +1). Additionally, note that metal-coordinating atom type diversity is higher, denticities and number of atoms are lower, and distributions of charges are similar for the OMol25 ligand sampling compared to the unique ligands in the CSD.

A ligand sampling algorithm was created to build the required Architector inputs. First, a metal, oxidation state, and coordination number were randomly selected. Ligands were randomly and iteratively selected until no remaining coordination sites on the metal remained. With each selection, only ligands that could fit at the remaining coordination sites, did not move the charge of the overall complex outside of the range -2 and +4, and did not make the complex exceed a set number of atoms (both 120 and 250 were used in this dataset) were permitted. From these remaining ligands, a weighted sampling was performed by denticity to ensure the per-site likelihood of all possible binding sites was equal. This was done to avoid under sampling high-denticity ligands. The result of the sampling pipeline is a massively chemically diverse, largely high-spin, set of Architector inputs for metal complexes of a size and charge likely to converge well in DFT calculations.

From here, Architector was run on samples of up to 1M metal complexes. For the most part, Architector was run with default parameters with a few exceptions: to obtain a variety of metal coordination environments and symmetries, up to 10 initial ligand symmetries were screened and up to 3 conformers were requested to be relaxed for every core geometry (\textit{e.g.}, octahedral, trigonal prismatic, hexagonal planar). For initial screening and final relaxation of the generated complexes, GFN2-xTB was used \cite{bannwarth_gfn2-xtbaccurate_2019}. An additional Architector parameter tuned for this work was allowing a looser graph distance tolerance of 2, which allows generated molecules to have bond distances up to twice as long as the defined molecular graph sum of covalent radii, which was done to allow for structures with more dissociated ligands when sampling at higher coordination numbers. The code to reproduce the sampling workflow and Architector generation are found in the OMol repository, while the curation of metal and ligand datasets can be found in a separate repository \cite{coordcomplexsampling2025}.

\subsection{Complexes from the Crystallography Open Database (COD)}

In order to evaluate the performance of models trained on computationally-generated metal complexes, a test set of experimentally determined metal complexes was obtained. The Crystallography Open Database (COD) contains over 500,000 experimental crystal structures extracted from various journals and publications under a permissive license and a convenient SQL interface. We extracted all crystal structures that were marked as not containing disorder and whose unit cell contents matched the indicated contents and which contained both carbon (as we are interested especially in complexes which contain organic ligands) and hydrogen (as a proxy for checking if the crystal structure has been decorated with hydrogens, as is not always the case, and which would require manual intervention to obtain molecules suitable for DFT). This yields 305,535 structures. These structures were then processed by a workflow to contract molecules into valid structures (\textit{i.e.}, without being split by periodic boundary conditions, crystallographic symmetry, etc.), and assign Lewis structures. This Lewis structure information was used to determine the metal oxidation state and, from there, the spin of the system.

Systems which did not contain exactly one metal center, where the absolute value of the charge on any given molecule was greater than 2, which contained actinides, which had more 250 atoms, or which had colliding atoms (defined as having a bond distance of less than 0.55 times the sum of covalent radii for the two atoms) were discarded. While multi-metallic centers are of considerable interest, substantial manual intervention is required to determine accurate oxidation states in multimetallic complexes. Given the large scale of the \omol~dataset and the significant effort that would be required, we elected not to include multimetallic complexes at this time.Additionally, systems which contained carbon atoms with more than 4 bonds to non-metals, hydrogen atoms with more than 1 bond to non-metals, and boron atoms forming more than 4 bonds to non-metals were also rejected. Such situations very frequently indicated undocumented disorder in the crystal structure, or, in the case of boron, a cluster structure. Boron clusters are not well described by a 2-electron per bond Lewis structure, and so the automated procedure for assigning charges to these systems often produces nonsensical results for such structural motifs. Lastly, if the assigned metal oxidation states was not one of the common oxidation states for a given metal, the assignment was considered suspect and rejected. We note that rejections based on uncommon oxidation states made up fewer than 10\% of the remaining structures at this point. These filters resulted in 70,004 structures with reasonably trustworthy structure and electronic assignment. Second and third row complexes were assigned low-spin configurations and first row complexes were calculated both in their highest and lowest reasonable spin configurations, resulting in 90,359 DFT inputs.
\section{Electrolytes}

Electrolytes are collections of ions and solvent molecules which feature complex intermolecular interactions, dynamic short-range solvation and in some cases long-range network structures. They are also important for a range of practical applications, from biochemistry, biotechnology, and medicine (the cellular medium is essentially an aqueous electrolyte solution) to battery energy storage, chemical synthesis, and catalysis. In this thrust, we aim to capture solvation structures, the chemical diversity of the small molecules used in electrolyte solutions, and electrolyte reactivity.

\subsection{Bulk Electrolytes}
\label{sec:methods:md}

Many electrolyte properties of interest, including transport (\textit{e.g.}, conductivity, transference number, viscosity), (de)solvation, and wetting, are inherently dynamic in nature. These are properties that static calculations of isolated molecules cannot capture. To simulate electrolyte dynamics well, ML models need diverse structures in their training set; for electrolytes specifically, these structures must include clusters of noncovalently bound ions and solvent molecules. To this end, we employ classical molecular dynamics (MD) as a means to sample clusters of solvated ions. The data obtained from MD simulations have been initiated in two generations. Below, we describe each generation of MD-based data:

\begin{table}[p]
\centering
\caption{List of cations used for MD simulations of bulk electrolytes and the force field parameters, sorted by valency. Abbreviations: BMPyrr$^+$ (1-butyl-1-methylpyrrolidinium), BPPyrr$^+$ (1-butyl-1-propylpyrrolidin-1-ium), EMIM$^+$ (1-ethyl-3-methylimidazolium), BMIM$^+$ (1-butyl-3-methylimidazolium), MPA$^+$ (1-Methoxypropan-2-ylazanium), AMA$^+$ (acetyl(methyl)azanium), BMEA$^+$ (bis(2-methoxyethyl)azanium), MPPip$^+$ (N-methyl-N-propylpiperidinium), PMPyrr$^+$ (N-propyl-N-methylpyrrolidinium), DEMA$^+$ (N,N-Diethyl-2-methoxy-N-methylethanaminium), PQ$^{2+}$ (paraquat), TBP$^+$ (tetrabutyl-phosphonium).}
\label{tab:cations_ff}

\begin{minipage}[t]{0.48\textwidth}
\centering
\small
\textbf{\underline{Monovalent Cations}}
  \begin{tabular}{lcc}
    \toprule
    \multirow{2}{*}{Name} & \multicolumn{2}{c}{FF Parameters} \\
    \cmidrule{2-3}
    & \textit{1st Gen} & \textit{2nd Gen} \\
    \midrule
    H$_3$O$^+$ & OPLS4 \cite{lu_opls4_2021} & Jang, at al. (2004) \cite{jang2004nanophase} \\
    Li$^+$ & OPLS4 \cite{lu_opls4_2021} & Sengupta, et al. (2021) \cite{sengupta_parameterization_2021} \\ 
    Na$^+$ & OPLS4 \cite{lu_opls4_2021} & Sengupta, et al. (2021) \cite{sengupta_parameterization_2021} \\
    NH$_4$$^+$ & OPLS4 \cite{lu_opls4_2021} & Sengupta, et al. (2021) \cite{sengupta_parameterization_2021} \\
    K$^+$ & OPLS4 \cite{lu_opls4_2021} & Sengupta, et al. (2021) \cite{sengupta_parameterization_2021} \\
    Cu$^+$ & OPLS4 \cite{lu_opls4_2021} & Sengupta, et al. (2021) \cite{sengupta_parameterization_2021} \\
    Rb$^+$ & OPLS4 \cite{lu_opls4_2021} & Sengupta, et al. (2021) \cite{sengupta_parameterization_2021} \\
    Ag$^+$ & OPLS4 \cite{lu_opls4_2021} & Sengupta, et al. (2021) \cite{sengupta_parameterization_2021} \\
    Cs$^+$ & OPLS4 \cite{lu_opls4_2021} & Sengupta, et al. (2021) \cite{sengupta_parameterization_2021} \\
    Ti$^+$ & OPLS4 \cite{lu_opls4_2021} & Sengupta, et al. (2021) \cite{sengupta_parameterization_2021} \\
    (VO$_2$)$^+$ & OPLS4 \cite{lu_opls4_2021} & Gupta, et al. (2016) \cite{gupta2016force} \\
    TEMPO$^+$ & OPLS4 \cite{lu_opls4_2021} & OPLS-AA \cite{jorgensen2005potential,dodda20171,dodda2017ligpargen} \\
    \bottomrule
  \end{tabular}

\vspace{0.4cm}

\textbf{\underline{Trivalent Cations}}
  \begin{tabular}{lcc}
    \toprule
    \multirow{2}{*}{Name} & \multicolumn{2}{c}{FF Parameters} \\
    \cmidrule{2-3}
    & \textit{1st Gen} & \textit{2nd Gen} \\
    \midrule
    Al$^{3+}$ & OPLS4 \cite{lu_opls4_2021} & Li, et al. (2021) \cite{li2021parametrization} \\
    V$^{3+}$ & OPLS4 \cite{lu_opls4_2021} & Li, et al. (2021) \cite{li2021parametrization} \\
    Cr$^{3+}$ & OPLS4 \cite{lu_opls4_2021} & Li, et al. (2021) \cite{li2021parametrization} \\
    Fe$^{3+}$ & OPLS4 \cite{lu_opls4_2021} & Li, et al. (2021) \cite{li2021parametrization} \\
    In$^{3+}$ & OPLS4 \cite{lu_opls4_2021} & Li, et al. (2021) \cite{li2021parametrization} \\
    Y$^{3+}$ & OPLS4 \cite{lu_opls4_2021} & Li, et al. (2021) \cite{li2021parametrization} \\
    Tl$^{3+}$ & OPLS4 \cite{lu_opls4_2021} & Li, et al. (2021) \cite{li2021parametrization} \\
    Ti$^{3+}$ & OPLS4 \cite{lu_opls4_2021} & Li, et al. (2021) \cite{li2021parametrization} \\
    \bottomrule
  \end{tabular}

\vspace{0.4cm}

\textbf{\underline{Ionic Liquid Cations}}
  \begin{tabular}{lcc}
    \toprule
    \multirow{2}{*}{Name} & \multicolumn{2}{c}{FF Parameters} \\
    \cmidrule{2-3}
    & \textit{1st Gen} & \textit{2nd Gen} \\
    \midrule
    EMIM$^+$ & OPLS4 \cite{lu_opls4_2021} & OPLS-AA \cite{jorgensen2005potential,dodda20171,dodda2017ligpargen} \\
    BMIM$^+$ & OPLS4 \cite{lu_opls4_2021} & OPLS-AA \cite{jorgensen2005potential,dodda20171,dodda2017ligpargen} \\
    BMPyrr$^+$ & OPLS4 \cite{lu_opls4_2021} & OPLS-AA \cite{jorgensen2005potential,dodda20171,dodda2017ligpargen} \\
    BPPyrr$^+$ & OPLS4 \cite{lu_opls4_2021} & OPLS-AA \cite{jorgensen2005potential,dodda20171,dodda2017ligpargen} \\
    PMPyrr$^+$ & OPLS4 \cite{lu_opls4_2021} & OPLS-AA \cite{jorgensen2005potential,dodda20171,dodda2017ligpargen} \\
    MPPip$^+$ & OPLS4 \cite{lu_opls4_2021} & OPLS-AA \cite{jorgensen2005potential,dodda20171,dodda2017ligpargen} \\
    DEMA$^+$ & OPLS4 \cite{lu_opls4_2021} & OPLS-AA \cite{jorgensen2005potential,dodda20171,dodda2017ligpargen} \\
    \bottomrule
  \end{tabular}
\end{minipage}
\hfill
\begin{minipage}[t]{0.48\textwidth}
\centering
\small
\textbf{\underline{Divalent Cations}}
  \begin{tabular}{lcc}
    \toprule
    \multirow{2}{*}{Name} & \multicolumn{2}{c}{FF Parameters} \\
    \cmidrule{2-3}
    & \textit{1st Gen} & \textit{2nd Gen} \\
    \midrule
    Be$^{2+}$ & OPLS4 \cite{lu_opls4_2021} & Li, et al. (2020) \cite{li2020systematic} \\
    Mg$^{2+}$ & OPLS4 \cite{lu_opls4_2021} & Li, et al. (2020) \cite{li2020systematic} \\
    Ca$^{2+}$ & OPLS4 \cite{lu_opls4_2021} & Li, et al. (2020) \cite{li2020systematic} \\
    V$^{2+}$ & OPLS4 \cite{lu_opls4_2021} & Li, et al. (2020) \cite{li2020systematic} \\
    Cr$^{2+}$ & OPLS4 \cite{lu_opls4_2021} & Li, et al. (2020) \cite{li2020systematic} \\
    Mn$^{2+}$ & OPLS4 \cite{lu_opls4_2021} & Li, et al. (2020) \cite{li2020systematic} \\
    Fe$^{2+}$ & OPLS4 \cite{lu_opls4_2021} & Li, et al. (2020) \cite{li2020systematic} \\
    Co$^{2+}$ & OPLS4 \cite{lu_opls4_2021} & Li, et al. (2020) \cite{li2020systematic} \\
    Ni$^{2+}$ & OPLS4 \cite{lu_opls4_2021} & Li, et al. (2020) \cite{li2020systematic} \\
    Cu$^{2+}$ & OPLS4 \cite{lu_opls4_2021} & Li, et al. (2020) \cite{li2020systematic} \\
    Zn$^{2+}$ & OPLS4 \cite{lu_opls4_2021} & Li, et al. (2020) \cite{li2020systematic} \\
    Sr$^{2+}$ & OPLS4 \cite{lu_opls4_2021} & Li, et al. (2020) \cite{li2020systematic} \\
    Cd$^{2+}$ & OPLS4 \cite{lu_opls4_2021} & Li, et al. (2020) \cite{li2020systematic} \\
    Sn$^{2+}$ & OPLS4 \cite{lu_opls4_2021} & Li, et al. (2020) \cite{li2020systematic} \\
    Ba$^{2+}$ & OPLS4 \cite{lu_opls4_2021} & Li, et al. (2020) \cite{li2020systematic} \\
    VO$^{2+}$ & OPLS4 \cite{lu_opls4_2021} & Gupta, et al. (2016) \cite{gupta2016force} \\
    Pd$^{2+}$ & OPLS4 \cite{lu_opls4_2021} & Li, et al. (2020) \cite{li2020systematic} \\
    Ag$^{2+}$ & OPLS4 \cite{lu_opls4_2021} & Li, et al. (2020) \cite{li2020systematic} \\
    Hg$^{2+}$ & OPLS4 \cite{lu_opls4_2021} & Li, et al. (2020) \cite{li2020systematic} \\
    Pb$^{2+}$ & OPLS4 \cite{lu_opls4_2021} & Li, et al. (2020) \cite{li2020systematic} \\
    PQ$^{2+}$ & OPLS4 \cite{lu_opls4_2021} & OPLS-AA \cite{jorgensen2005potential,dodda20171,dodda2017ligpargen} \\
    Pt$^{2+}$ & OPLS4 \cite{lu_opls4_2021} & Li, et al. (2020) \cite{li2020systematic} \\
    \bottomrule
  \end{tabular}

\vspace{0.4cm}

\textbf{\underline{Tetravalent Cations}}
  \begin{tabular}{lcc}
    \toprule
    \multirow{2}{*}{Name} & \multicolumn{2}{c}{FF Parameters} \\
    \cmidrule{2-3}
    & \textit{1st Gen} & \textit{2nd Gen} \\
    \midrule
    Ti$^{4+}$ & OPLS4 \cite{lu_opls4_2021} & Li, et al. (2021) \cite{li2021parametrization} \\
    Mn$^{4+}$ & OPLS4 \cite{lu_opls4_2021} & Li, et al. (2021) \cite{li2021parametrization} \\
    Zr$^{4+}$ & OPLS4 \cite{lu_opls4_2021} & Li, et al. (2021) \cite{li2021parametrization} \\
    \bottomrule
  \end{tabular}

\vspace{0.4cm}

\textbf{\underline{Ionic Liquid Cations (continued)}}
  \begin{tabular}{lcc}
    \toprule
    \multirow{2}{*}{Name} & \multicolumn{2}{c}{FF Parameters} \\
    \cmidrule{2-3}
    & \textit{1st Gen} & \textit{2nd Gen} \\
    \midrule
    AMA$^+$ & OPLS4 \cite{lu_opls4_2021} & OPLS-AA \cite{jorgensen2005potential,dodda20171,dodda2017ligpargen} \\
    MPA$^+$ & OPLS4 \cite{lu_opls4_2021} & OPLS-AA \cite{jorgensen2005potential,dodda20171,dodda2017ligpargen} \\
    BMEA$^+$ & OPLS4 \cite{lu_opls4_2021} & OPLS-AA \cite{jorgensen2005potential,dodda20171,dodda2017ligpargen} \\
    Choline & OPLS4 \cite{lu_opls4_2021} & OPLS-AA \cite{jorgensen2005potential,dodda20171,dodda2017ligpargen} \\
    TBP$^+$ & OPLS4 \cite{lu_opls4_2021} & OPLS-AA \cite{jorgensen2005potential,dodda20171,dodda2017ligpargen} \\
    PQ$^{2+}$ & OPLS4 \cite{lu_opls4_2021} & OPLS-AA \cite{jorgensen2005potential,dodda20171,dodda2017ligpargen} \\
    \bottomrule
  \end{tabular}
\end{minipage}

% Add some space at the bottom to avoid page number overlap
\vspace{0.5cm}
\end{table}

\begin{table}[p]
\centering
\caption{Anionic species with their associated force field parameters, sorted by valency and grouped by parameter types. Abbreviations: TFSI$^-$ (bis(trifluoromethylsulfonyl)imide), BOB$^-$ (bisoxalatoborate), HMDS$^-$ (hexamethyldisilazide), TEMPO$^-$ (2,2,6,6-tetramethylpiperidin-1-yl)oxide), FSI$^-$ (bis(fluorosulfonyl)imide), TFA$^-$ (triflate), Ac$^-$ (acetate), MeO$^-$ (methoxide), EtO$^-$ (ethoxide), iPrO$^-$ (isopropoxide), ETFA$^-$ (ethyl trifluoroacetate), DFP$^-$ (difluorophosphate), DHB$^{2-}$ (2,5-dihydroxy-[1,4]-benzoquinone dianion), EDO$^{2-}$ (ethane-1,2-diolate), BOB$^-$ (bisoxalatoborate).}
\label{tab:anions_ff}

\begin{minipage}[t]{0.48\textwidth}
\centering
\small
\textbf{\underline{Monovalent Anions}}
  \begin{tabular}{lcc}
    \toprule
    \multirow{2}{*}{Name} & \multicolumn{2}{c}{FF Parameters} \\
    \cmidrule{2-3}
    & \textit{1st Gen} & \textit{2nd Gen} \\
    \midrule
    F$^-$ & OPLS4 \cite{lu_opls4_2021} & Sengupta, et al. (2021) \cite{sengupta_parameterization_2021} \\
    Cl$^-$ & OPLS4 \cite{lu_opls4_2021} & Sengupta, et al. (2021) \cite{sengupta_parameterization_2021} \\
    Br$^-$ & OPLS4 \cite{lu_opls4_2021} & Sengupta, et al. (2021) \cite{sengupta_parameterization_2021} \\
    I$^-$ & OPLS4 \cite{lu_opls4_2021} & Sengupta, et al. (2021) \cite{sengupta_parameterization_2021} \\
    OH$^-$ & OPLS4 \cite{lu_opls4_2021} & GAFF2 \cite{wang2004gaff,wang2005gaff_erratum} \\
    BH$_4$$^-$ & OPLS4 \cite{lu_opls4_2021} & Mamatlukov, at al. \cite{mamatkulov2024unveiling} \\
    AsF$_6$$^-$ & OPLS4 \cite{lu_opls4_2021} & Ishida, et al. \cite{ishida2009atom} \\
    NO$_3$$^-$ & OPLS4 \cite{lu_opls4_2021} & OPLS-2009IL \cite{sambasivarao2009development,doherty2017revisiting} \\
    ClO$_4$$^-$ & OPLS4 \cite{lu_opls4_2021} & OPLS-2009IL \cite{sambasivarao2009development,doherty2017revisiting} \\
    BF$_4$$^-$ & OPLS4 \cite{lu_opls4_2021} & OPLS-2009IL \cite{sambasivarao2009development,doherty2017revisiting} \\
    PF$_6$$^-$ & OPLS4 \cite{lu_opls4_2021} & OPLS-2009IL \cite{sambasivarao2009development,doherty2017revisiting} \\
    TFSI$^-$ & OPLS4 \cite{lu_opls4_2021} & OPLS-2009IL \cite{sambasivarao2009development,doherty2017revisiting} \\
    TFA$^-$ & OPLS4 \cite{lu_opls4_2021} & OPLS-2009IL \cite{sambasivarao2009development,doherty2017revisiting} \\
    HCO$_3$$^-$ & OPLS4 \cite{lu_opls4_2021} &  OPLS-AA \cite{jorgensen2005potential,dodda20171,dodda2017ligpargen} \\
    Ac$^-$ & OPLS4 \cite{lu_opls4_2021} &  OPLS-AA \cite{jorgensen2005potential,dodda20171,dodda2017ligpargen} \\
    MeO$^-$ & OPLS4 \cite{lu_opls4_2021} &  OPLS-AA \cite{jorgensen2005potential,dodda20171,dodda2017ligpargen} \\
    EtO$^-$ & OPLS4 \cite{lu_opls4_2021} &  OPLS-AA \cite{jorgensen2005potential,dodda20171,dodda2017ligpargen} \\
    iPrO$^-$ & OPLS4 \cite{lu_opls4_2021} &  OPLS-AA \cite{jorgensen2005potential,dodda20171,dodda2017ligpargen} \\
    HMDS$^-$ & OPLS4 \cite{lu_opls4_2021} &  OPLS-AA \cite{jorgensen2005potential,dodda20171,dodda2017ligpargen} \\
    TEMPO$^-$ & OPLS4 \cite{lu_opls4_2021} &  OPLS-AA \cite{jorgensen2005potential,dodda20171,dodda2017ligpargen} \\
    ETFA$^-$ & OPLS4 \cite{lu_opls4_2021} &  OPLS-AA \cite{jorgensen2005potential,dodda20171,dodda2017ligpargen} \\
    BOB$^-$ & OPLS4 \cite{lu_opls4_2021} & Wang, et al. (2014) \cite{wang2014atomistic} \\
    FSI$^-$ & OPLS4 \cite{lu_opls4_2021} & GAFF2 \cite{wang2004gaff,wang2005gaff_erratum} \\
    DFP$^-$ & OPLS4 \cite{lu_opls4_2021} & GAFF2 \cite{wang2004gaff,wang2005gaff_erratum} \\
    \bottomrule
  \end{tabular}
\end{minipage}
\hfill
\begin{minipage}[t]{0.48\textwidth}
\centering
\small
\textbf{\underline{Divalent Anions}}
  \begin{tabular}{lcc}
    \toprule
    \multirow{2}{*}{Name} & \multicolumn{2}{c}{FF Parameters} \\
    \cmidrule{2-3}
    & \textit{1st Gen} & \textit{2nd Gen} \\
    \midrule
    CO$_3$$^{2-}$ & OPLS4 \cite{lu_opls4_2021} & GAFF2 \cite{wang2004gaff,wang2005gaff_erratum} \\
    SO$_4$$^{2-}$ & OPLS4 \cite{lu_opls4_2021} & GAFF2 \cite{wang2004gaff,wang2005gaff_erratum} \\
    EDO$^{2-}$ & OPLS4 \cite{lu_opls4_2021} & OPLS-AA \cite{jorgensen2005potential,dodda20171,dodda2017ligpargen} \\
    DHB$^{2-}$ & OPLS4 \cite{lu_opls4_2021} & OPLS-AA \cite{jorgensen2005potential,dodda20171,dodda2017ligpargen} \\
    \bottomrule
  \end{tabular}

\vspace{0.4cm}

\textbf{\underline{Trivalent Anions}}
  \begin{tabular}{lcc}
    \toprule
    \multirow{2}{*}{Name} & \multicolumn{2}{c}{FF Parameters} \\
    \cmidrule{2-3}
    & \textit{1st Gen} & \textit{2nd Gen} \\
    \midrule
    PO$_4$$^{3-}$ & OPLS4 \cite{lu_opls4_2021} & GAFF2 \cite{wang2004gaff,wang2005gaff_erratum} \\
    \bottomrule
  \end{tabular}
\end{minipage}
\end{table}

\begin{table}[p]
\centering
\caption{Solvent species with their associated force field parameters, organized by solvent type. Abbreviations: ACN (acetonitrile), BuCN (butyronitrile), DMSO (dimethylsulfoxide), DMF (N,N-dimethylformamide), NMA (N-methylacetamide), MPA (1-methoxy-2-propylamine), DHQ (dihydroquinone), TMPOH (2,2,6,6-tetramethylpiperidin-1-ol), BMEA (bis(2-methoxyethyl)amine), DCM (dichloromethane), DBE (1,2-dibromoethane), DCE (1,2-dichloroethane), NM (nitromethane), NB (nitrobenzene), PYR (pyridine), TEA (triethylamine), THF (tetrahydrofuran), DEE (diethyl ether), DME (dimethoxyethane), DTD (ethylene sulfate), EG (ethylene glycol), EC (ethylene carbonate), FEC (fluoroethylene carbonate), PC (propylene carbonate), DMC (dimethyl carbonate), EMC (ethyl methyl carbonate), EtAc (ethyl acetate), ETFA (ethyl trifluoroacetate), GBL ($\gamma$-butyrolactone), EMS (ethyl methyl sulfonate), TMP (trimethylphosphate), TEP (triethylphosphate), HMPA (hexamethylphosphoramide), AQ (anthraquinone), BQ (benzoquinone), DHBQ (2,5-dihydroxy-1,4-benzoquinone), TEMPO (2,2,6,6-tetramethylpiperidine-1-oxyl), TTE (tris(2,2,2-trifluoroethyl) borate).}
\label{tab:solvents_ff}

\begin{minipage}[t]{0.48\textwidth}
\centering
\small
\textbf{\underline{Protic Solvents}}
  \begin{tabular}{lcc}
    \toprule
    \multirow{2}{*}{Name} & \multicolumn{2}{c}{FF Parameters} \\
    \cmidrule{2-3}
    & \textit{1st Gen} & \textit{2nd Gen} \\
    \midrule
    Water & TIP3P \cite{lu_opls4_2021} & TIP4P/2005f \cite{gonzalez2011flexible} \\
    MeOH & OPLS4 \cite{lu_opls4_2021} & OPLS-AA \cite{jorgensen2005potential,dodda20171,dodda2017ligpargen} \\
    EtOH & OPLS4 \cite{lu_opls4_2021} & OPLS-AA \cite{jorgensen2005potential,dodda20171,dodda2017ligpargen} \\
    iPrOH & OPLS4 \cite{lu_opls4_2021} & OPLS-AA \cite{jorgensen2005potential,dodda20171,dodda2017ligpargen} \\
    EG & OPLS4 \cite{lu_opls4_2021} & OPLS-AA \cite{jorgensen2005potential,dodda20171,dodda2017ligpargen} \\
    Urea & OPLS4 \cite{lu_opls4_2021} & OPLS-AA \cite{jorgensen2005potential,dodda20171,dodda2017ligpargen} \\
    AcNH$_2$ & OPLS4 \cite{lu_opls4_2021} & OPLS-AA \cite{jorgensen2005potential,dodda20171,dodda2017ligpargen} \\
    NMA & OPLS4 \cite{lu_opls4_2021} & OPLS-AA \cite{jorgensen2005potential,dodda20171,dodda2017ligpargen} \\
    Pyrrole & OPLS4 \cite{lu_opls4_2021} & OPLS-AA \cite{jorgensen2005potential,dodda20171,dodda2017ligpargen} \\
    MPA & OPLS4 \cite{lu_opls4_2021} & OPLS-AA \cite{jorgensen2005potential,dodda20171,dodda2017ligpargen} \\
    DHQ & OPLS4 \cite{lu_opls4_2021} & OPLS-AA \cite{jorgensen2005potential,dodda20171,dodda2017ligpargen} \\
    TMPOH & OPLS4 \cite{lu_opls4_2021} & OPLS-AA \cite{jorgensen2005potential,dodda20171,dodda2017ligpargen} \\
    BMEA & OPLS4 \cite{lu_opls4_2021} & OPLS-AA \cite{jorgensen2005potential,dodda20171,dodda2017ligpargen} \\
    \bottomrule
  \end{tabular}

\vspace{0.4cm}

\textbf{\underline{Nonpolar Aprotic Solvents}}
  \begin{tabular}{lcc}
    \toprule
    \multirow{2}{*}{Name} & \multicolumn{2}{c}{FF Parameters} \\
    \cmidrule{2-3}
    & \textit{1st Gen} & \textit{2nd Gen} \\
    \midrule
    Pentane & OPLS4 \cite{lu_opls4_2021} & OPLS-AA \cite{jorgensen2005potential,dodda20171,dodda2017ligpargen} \\
    Hexane & OPLS4 \cite{lu_opls4_2021} & OPLS-AA \cite{jorgensen2005potential,dodda20171,dodda2017ligpargen} \\
    Benzene & OPLS4 \cite{lu_opls4_2021} & OPLS-AA \cite{jorgensen2005potential,dodda20171,dodda2017ligpargen} \\
    Toluene & OPLS4 \cite{lu_opls4_2021} & OPLS-AA \cite{jorgensen2005potential,dodda20171,dodda2017ligpargen} \\
    PhF & OPLS4 \cite{lu_opls4_2021} & OPLS-AA \cite{jorgensen2005potential,dodda20171,dodda2017ligpargen} \\
    CCl$_4$ & OPLS4 \cite{lu_opls4_2021} & OPLS-AA \cite{jorgensen2005potential,dodda20171,dodda2017ligpargen} \\
    \bottomrule
  \end{tabular}

\vspace{0.4cm}

\textbf{\underline{Halogenated Solvents}}
  \begin{tabular}{lcc}
    \toprule
    \multirow{2}{*}{Name} & \multicolumn{2}{c}{FF Parameters} \\
    \cmidrule{2-3}
    & \textit{1st Gen} & \textit{2nd Gen} \\
    \midrule
    CHCl$_3$ & OPLS4 \cite{lu_opls4_2021} & OPLS-AA \cite{jorgensen2005potential,dodda20171,dodda2017ligpargen} \\
    CHBr$_3$ & OPLS4 \cite{lu_opls4_2021} & OPLS-AA \cite{jorgensen2005potential,dodda20171,dodda2017ligpargen} \\
    DCM & OPLS4 \cite{lu_opls4_2021} & OPLS-AA \cite{jorgensen2005potential,dodda20171,dodda2017ligpargen} \\
    DBE & OPLS4 \cite{lu_opls4_2021} & OPLS-AA \cite{jorgensen2005potential,dodda20171,dodda2017ligpargen} \\
    DCE & OPLS4 \cite{lu_opls4_2021} & OPLS-AA \cite{jorgensen2005potential,dodda20171,dodda2017ligpargen} \\
    \bottomrule
  \end{tabular}

\vspace{0.4cm}

\textbf{\underline{Redox-Active Compounds}}
  \begin{tabular}{lcc}
    \toprule
    \multirow{2}{*}{Name} & \multicolumn{2}{c}{FF Parameters} \\
    \cmidrule{2-3}
    & \textit{1st Gen} & \textit{2nd Gen} \\
    \midrule
    AQ & OPLS4 \cite{lu_opls4_2021} & OPLS-AA \cite{jorgensen2005potential,dodda20171,dodda2017ligpargen} \\
    BQ & OPLS4 \cite{lu_opls4_2021} & OPLS-AA \cite{jorgensen2005potential,dodda20171,dodda2017ligpargen} \\
    DHBQ & OPLS4 \cite{lu_opls4_2021} & OPLS-AA \cite{jorgensen2005potential,dodda20171,dodda2017ligpargen} \\
    TEMPO & OPLS4 \cite{lu_opls4_2021} & OPLS-AA \cite{jorgensen2005potential,dodda20171,dodda2017ligpargen} \\
    TTE & OPLS4 \cite{lu_opls4_2021} & OPLS-AA \cite{jorgensen2005potential,dodda20171,dodda2017ligpargen} \\
    \bottomrule
  \end{tabular}
\end{minipage}
\hfill
\begin{minipage}[t]{0.48\textwidth}
\centering
\small
\textbf{\underline{Ethers, Carbonates, Esters \& Ketones}}
  \begin{tabular}{lcc}
    \toprule
    \multirow{2}{*}{Name} & \multicolumn{2}{c}{FF Parameters} \\
    \cmidrule{2-3}
    & \textit{1st Gen} & \textit{2nd Gen} \\
    \midrule
    THF & OPLS4 \cite{lu_opls4_2021} & OPLS-AA \cite{jorgensen2005potential,dodda20171,dodda2017ligpargen} \\
    DEE & OPLS4 \cite{lu_opls4_2021} & OPLS-AA \cite{jorgensen2005potential,dodda20171,dodda2017ligpargen} \\
    DME & OPLS4 \cite{lu_opls4_2021} & OPLS-AA \cite{jorgensen2005potential,dodda20171,dodda2017ligpargen} \\
    Diglyme & OPLS4 \cite{lu_opls4_2021} & OPLS-AA \cite{jorgensen2005potential,dodda20171,dodda2017ligpargen} \\
    EC & OPLS4 \cite{lu_opls4_2021} & OPLS-AA \cite{jorgensen2005potential,dodda20171,dodda2017ligpargen} \\
    FEC & OPLS4 \cite{lu_opls4_2021} & OPLS-AA \cite{jorgensen2005potential,dodda20171,dodda2017ligpargen} \\
    PC & OPLS4 \cite{lu_opls4_2021} & OPLS-AA \cite{jorgensen2005potential,dodda20171,dodda2017ligpargen} \\
    DMC & OPLS4 \cite{lu_opls4_2021} & OPLS-AA \cite{jorgensen2005potential,dodda20171,dodda2017ligpargen} \\
    EMC & OPLS4 \cite{lu_opls4_2021} & OPLS-AA \cite{jorgensen2005potential,dodda20171,dodda2017ligpargen} \\
    EtAc & OPLS4 \cite{lu_opls4_2021} & OPLS-AA \cite{jorgensen2005potential,dodda20171,dodda2017ligpargen} \\
    DTD & OPLS4 \cite{lu_opls4_2021} & OPLS-AA \cite{jorgensen2005potential,dodda20171,dodda2017ligpargen} \\
    ETFA & OPLS4 \cite{lu_opls4_2021} & OPLS-AA \cite{jorgensen2005potential,dodda20171,dodda2017ligpargen} \\
    GBL & OPLS4 \cite{lu_opls4_2021} & OPLS-AA \cite{jorgensen2005potential,dodda20171,dodda2017ligpargen} \\
    Acetone & OPLS4 \cite{lu_opls4_2021} & OPLS-AA \cite{jorgensen2005potential,dodda20171,dodda2017ligpargen} \\
    \bottomrule
  \end{tabular}

\vspace{0.4cm}

\textbf{\underline{Nitrogen-Containing Solvents}}
  \begin{tabular}{lcc}
    \toprule
    \multirow{2}{*}{Name} & \multicolumn{2}{c}{FF Parameters} \\
    \cmidrule{2-3}
    & \textit{1st Gen} & \textit{2nd Gen} \\
    \midrule
    ACN & OPLS4 \cite{lu_opls4_2021} & OPLS-AA \cite{jorgensen2005potential,dodda20171,dodda2017ligpargen} \\
    BuCN & OPLS4 \cite{lu_opls4_2021} & OPLS-AA \cite{jorgensen2005potential,dodda20171,dodda2017ligpargen} \\
    DMF & OPLS4 \cite{lu_opls4_2021} & OPLS-AA \cite{jorgensen2005potential,dodda20171,dodda2017ligpargen} \\
    NM & OPLS4 \cite{lu_opls4_2021} & OPLS-AA \cite{jorgensen2005potential,dodda20171,dodda2017ligpargen} \\
    NB & OPLS4 \cite{lu_opls4_2021} & OPLS-AA \cite{jorgensen2005potential,dodda20171,dodda2017ligpargen} \\
    PYR & OPLS4 \cite{lu_opls4_2021} & OPLS-AA \cite{jorgensen2005potential,dodda20171,dodda2017ligpargen} \\
    TEA & OPLS4 \cite{lu_opls4_2021} & OPLS-AA \cite{jorgensen2005potential,dodda20171,dodda2017ligpargen} \\
    \bottomrule
  \end{tabular}

\vspace{0.4cm}

\textbf{\underline{Sulfur \& Phosphorus Solvents}}
  \begin{tabular}{lcc}
    \toprule
    \multirow{2}{*}{Name} & \multicolumn{2}{c}{FF Parameters} \\
    \cmidrule{2-3}
    & \textit{1st Gen} & \textit{2nd Gen} \\
    \midrule
    DMSO & OPLS4 \cite{lu_opls4_2021} & OPLS-AA \cite{jorgensen2005potential,dodda20171,dodda2017ligpargen} \\
    Sulfolane & OPLS4 \cite{lu_opls4_2021} & OPLS-AA \cite{jorgensen2005potential,dodda20171,dodda2017ligpargen} \\
    EMS & OPLS4 \cite{lu_opls4_2021} & OPLS-AA \cite{jorgensen2005potential,dodda20171,dodda2017ligpargen} \\
    TMP & OPLS4 \cite{lu_opls4_2021} & OPLS-AA \cite{jorgensen2005potential,dodda20171,dodda2017ligpargen} \\
    TEP & OPLS4 \cite{lu_opls4_2021} & OPLS-AA \cite{jorgensen2005potential,dodda20171,dodda2017ligpargen} \\
    HMPA & OPLS4 \cite{lu_opls4_2021} & OPLS-AA \cite{jorgensen2005potential,dodda20171,dodda2017ligpargen} \\
    \bottomrule
  \end{tabular}
\end{minipage}

% Add some space at the bottom to avoid page number overlap
\vspace{0.5cm}
\end{table}

\begin{figure}[t]
    \centering  \includegraphics[width=\linewidth]{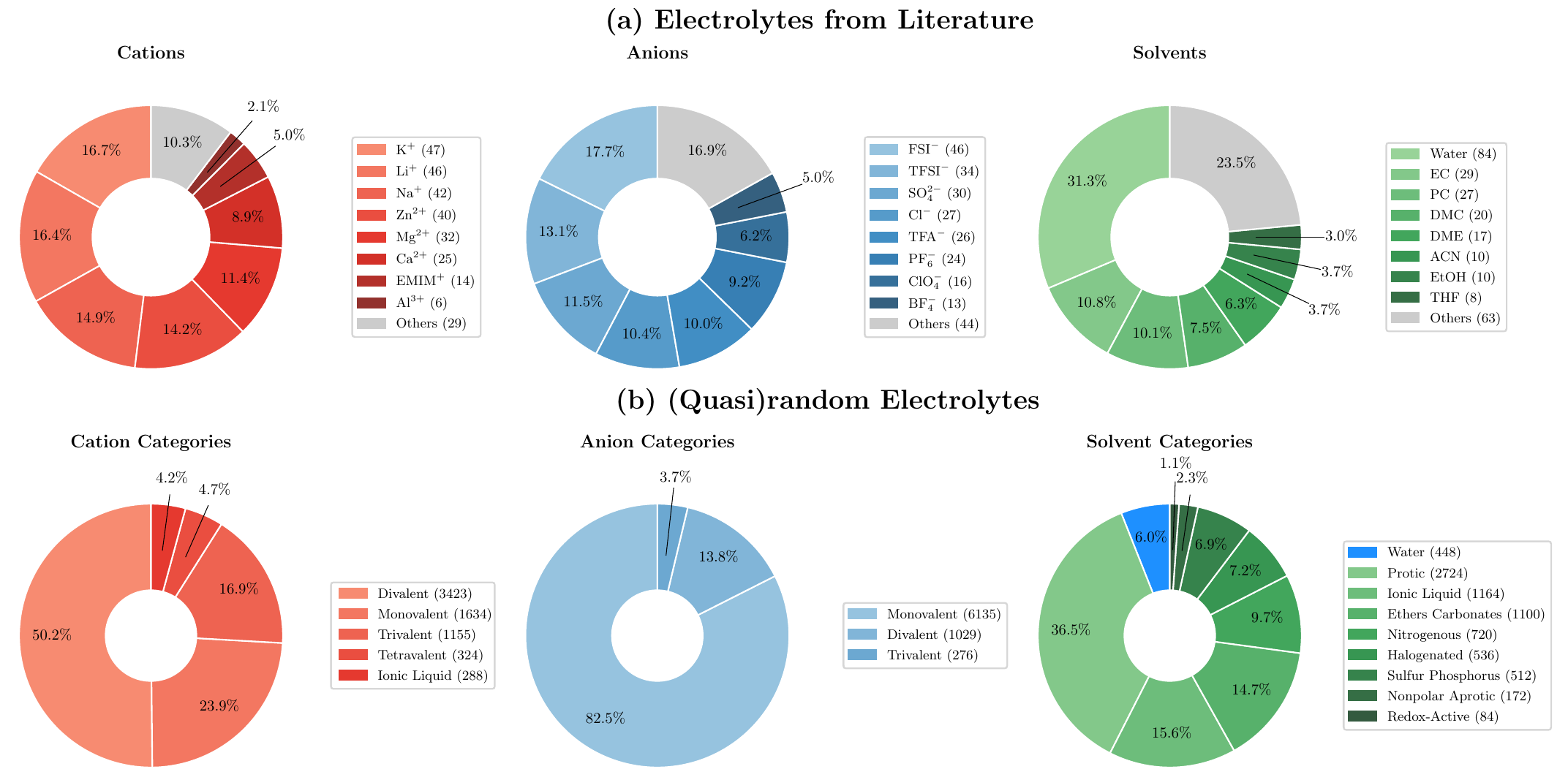}
    \caption{(a) Statistics of electrolyte systems obtained from the literature. Alkali ions ($\rm Li^+$, $\rm K^+$, and $\rm Na^+$) make up the top three cations, while bis(fluorosulfonyl)imide (FSI$^-$),bis(trifluoromethylsulfonyl)imide (TFSI$^-$), and sulfate ($\rm SO_4^{2-}$ make up the top three anions. Water stands out as the most common solvent, with various carbonates like ethylene carbonate (EC) and propylene carbonate (PC) forming the second tier of common solvents. (b) Statistics of (quasi)random electrolyte systems. The final set includes 54 unique cation types, dominated by monovalent ions, 28 anion types, and 79 different solvents with compositions averaging approximately 2 cations, 2 anions, and 2 solvents per unique formulation.}
    \label{fig:1stgen-electrolyte}
\end{figure}

\paragraph{First generation.}\label{app:elyte:desmond} In the first generation, we aimed to compile a list of electrolyte mixtures that are both relevant for practical applications and chemically diverse. To this end, we used the literature as a reference point and supplemented it with a workflow that generates random mixtures \cite{Adil_2020, Adil_2022, Aiken_2022, Aurbach_1991, B_rresen_1997, Bhide_2014, Bialik_2008, Biria_2020, Carbone_2017, Casteel_1972, Chang_2019, Chen_2017, Chen_2018, Cheng_2015, Deng_2019, Ding_2013, Doi_2017, Eiberweiser_2015, Etman_2020, Geng_2022, Hosaka_2018, Hou_2021, Hou_2021a, Jiang_2019, Jiao_2017, K_hnel_2017, Kao_2009, Keyzer_2016, Ko_2020, Komaba_2011, Kubota_2008, Kubota_2012, Lazouski_2019, Leonard_2018, Li_2022, Lin_2015, Liu_2020, Liu_2020a, Ma_2020, Marczewski_2014, Morishita_1997, Naveed_2019, Nguyen_2020, NuLi_2005, Pan_2016, Pluha_ov_2013, Ponrouch_2015, Qian_2015, Ren_2018, Shi_2017, Shterenberg_2015, Son_2018, Song_2016, Soundharrajan_2018, Su_2016, Suo_2015, Suo_2017, Tsuneto_1994, Vidal_Abarca_2012, Wang_2016, Wang_2017, Wang_2017a, Wang_2018, Wang_2019, Wang_2023, Watanabe_2023, Watarai_2008, Weng_2017, Xiao_2017, Xie_2019, Yamamoto_2017, Yang_2021, Yoo_2022, Yoon_2014, Zhang_2016, Zhang_2019, Zhang_2019a, Zhao_2016, Zhou_2018, Zhu_2021}. From the literature, we have included systems of technological relevance such as batteries (including Li-, Na-, K-, Mg-, Ca-, Zn-, and Al-ion batteries as well as Pb-acid batteries). We have also included examples from redox flow batteries, catalytic systems (for instance Li-mediated and Ca-mediated electrochemical ammonia synthesis), ionic liquids, and molten salt electrolytes, as well as several common aqueous and non-aqueous solutions. In total, we obtained 223 unique electrolyte systems, encompassing different types of solvents (\textit{e.g.}, protic, aprotic, ionic liquids), concentration (\textit{e.g.}, dilute and high-concentration), electrolyte components (\textit{e.g.}, solvents and salts), and number of components.

Next, we generated random electrolytes in five classes: (1) salts dissolved in protic solvents, (2) salts dissolved in aprotic solvents , (3) ionic liquid mixtures, (4) aqueous solutions, and (5) molten salts.
Aqueous solutions, despite being a subset of protic solutions, received special attention due to their importance and the challenges in capturing their dynamic behavior.
Each class was chosen with a fixed probability (40\% protic, 40\% aprotic, 10\% ionic liquid, 5\% molten salt, 5\% aqueous).
These main categories are then complemented by fully randomized formulations (110 unique systems), creating a comprehensive exploration of electrolyte chemical space. 
Every random electrolyte mixture is built from a curated list of electrolyte components based on our initial literature review, where components are labeled to ensure compatibility between cations, anions, and solvents; see \Cref{tab:cations_ff,tab:anions_ff,tab:solvents_ff} for the final list.
Charge neutrality is maintained by solving charge balance equations, which determines the stoichiometric coefficients for the selected ionic species. 
For each base composition, we produce four variant simulations, adjusting both temperature and concentration to reasonable high and low values. 
For temperatures, we ensure that the lowest and highest temperatures are 5\% above and below the freezing and boiling/decomposition temperatures, respectively. 
The concentrations are chosen such that, if the electrolyte system behaves as an ideal solution, the inter-solute separation distances are 2.25 and 1.75 nm, respectively. Note that molten salt systems have no solutes, by definition. For these systems, concentration was not varied, and temperatures were set to 1000 K and 1300 K. 
Overall, the random electrolyte dataset we produced includes 884 unique compositions and 3,536 simulations in total. 

We set up classical MD simulations using Desmond \cite{bowers2006scalable} and Schr\"{o}dinger's disordered system builder \cite{desmond2025}. Simulations were performed using the OPLS4 force field \cite{lu_opls4_2021} and the TIP3P water model with a van der Waals scaling factor of 0.5. Custom charges were generated for [AsF$_6$]$^-$, TEMPO, [VO]$^{2+}$, and [VO$_2$]$^+$. After system generation, we prepared a Desmond multisim script (.msj file) that defines a simulation protocol with multiple stages. The protocol begins with an initial relaxation using Brownian dynamics at 10 K, followed by a series of molecular dynamics simulations with gradually increasing time steps. The system then transitions from NVT to NPT ensemble, undergoes a high-temperature annealing phase at 700 K, and concludes with a production run in the NPT ensemble at the specified temperature using a 2 fs time step for 250 ns and snapshots were taken every 2.5 ns.

\begin{figure}[t]
    \centering
    \includegraphics[width=\linewidth]{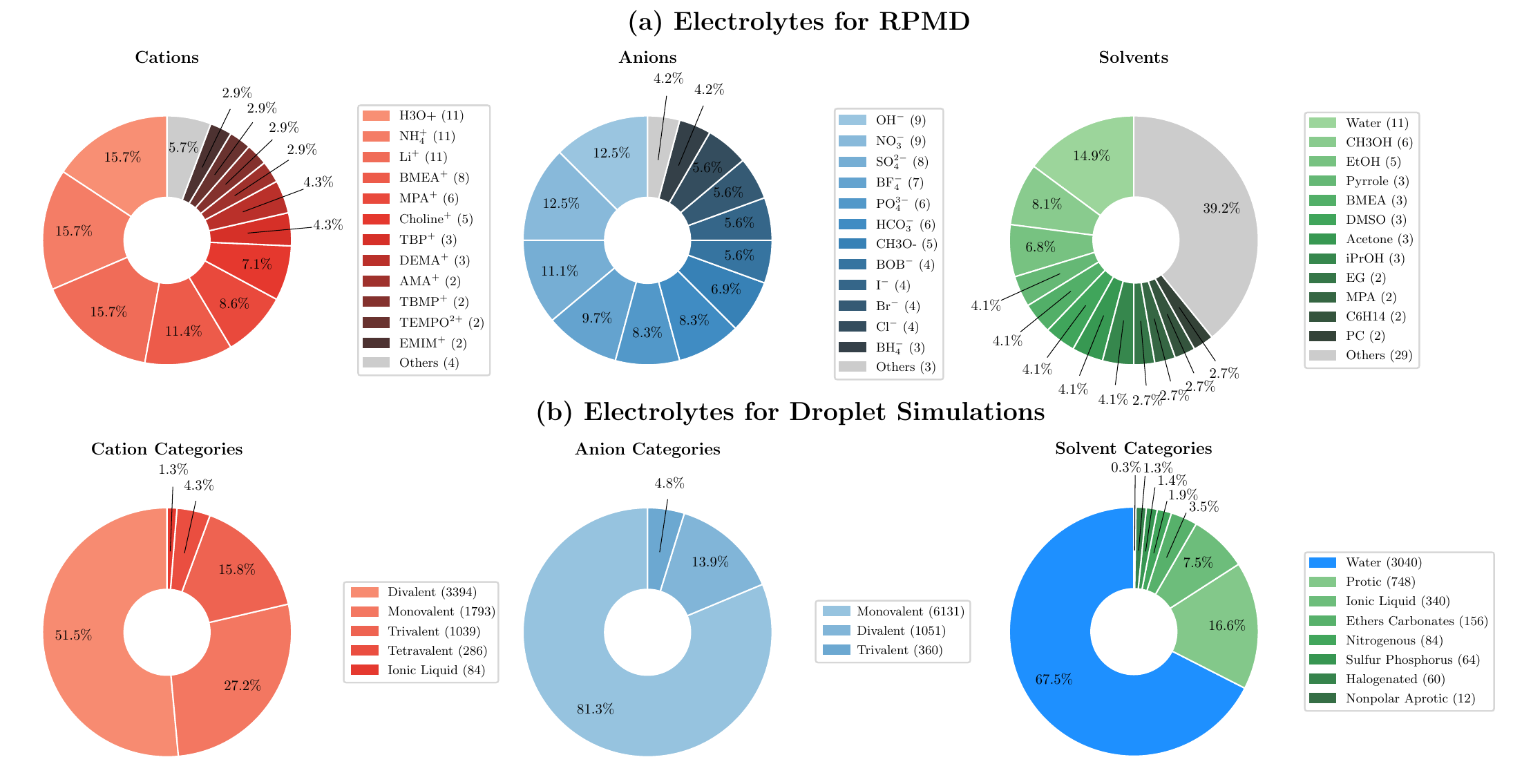}
    \caption{(a) Statistics of electrolyte systems obtained for RPMD simulations. Overall, the final set contains 16 unique cations, 13 anions, and 38 different solvents. The cation set includes both simple ions like H$_3$O$^+$, NH$_4^+$ (ammonium), and Li$^+$, as well as more complex organic cations derived from ionic liquids. Anions include simple species like OH$^-$ and NO$_3^-$, as well as  boron-containing anions such as BF$_4^-$, BH$_4^-$, and $BOB^-$. The diversity of solvents is particularly broad, including water (H$_2$O), alcohols (methanol, ethanol), and organic carbonates. (b) Statistics of random electrolyte systems obtained for spherical droplet simulations. Each electrolyte system allows up to 4 cations, 4 anions, and 4 solvents per system. The resulting dataset for spherical droplet simulations maintains substantial chemical diversity with 50 unique cations, 29 anions, and 60 different solvents.}
    \label{fig:2ndgen-electrolyte}
\end{figure}

We also prepared additional clusters which contained an OOD anion, cation, or solvent, via the methods described here. The OOD anions were difluorooxalatoborate (DFOB) and hexamethyldisilazide (HMDS), the OOD cations were tetramethylammonium (TMA) and tetraethylphosphonium, and the OOD solvents were toluene and 1,4-dioxane. These clusters contained at least one of the OOD species and any number of ID species. We also prepared a set of clusters that consists only of OOD species. The set containing OOD anions, ID anions, ID cation, and ID solvents was labeled ``OOD Anions'' and was discussed in the main text Section \ref{sec:methods:ood_anions}. Results for the similarly defined sets of cations and solvents, as well as the all OOD set is described in Appendix \ref{sec:app:addl_res}.

\paragraph{Second Generation.} The second generation of electrolyte MD simulations sought to capture specific nuclear quantum (using ring-polymer molecular dynamics, or RPMD) and interfacial effects (using droplet simulations). RPMD approximates quantum nuclei as classical ring polymers to incorporate nuclear quantum effects, which are crucial for accurate descriptions of zero-point energy, tunneling, and hydrogen bonding \cite{habershon2013ring}. 
RPMD represents each atom as a ring of harmonically bonded beads to capture a wider range of conformations than classical MD \cite{habershon2013ring}, making such configurations valuable for our dataset.
For RPMD, we selected 50 electrolyte systems containing light atoms (\textit{e.g.}, \ce{H}, \ce{Li}, and \ce{B}) where nuclear quantum effects are expected to be most important. 
For all systems, we chose relatively low temperatures, ranging from 164.85 to 325.71 K (averaging 247 K or $-26^\circ$C), where quantum effects of the light atoms are expected to be dominant. 
Meanwhile, spherical droplet simulations provide an opportunity to sample solvation clusters affected by interfacial effects.
To this end, we generated 3,734 electrolyte compositions where the compositions are generated in the same way as the first generation with 20\% salt in protic solvents, 15\% salt in aprotic solvents, and 65\% aqueous electrolytes.
Note that the second generation focuses heavily on aqueous systems to compensate for their under-representation in the first generation, where we have elected to use rigid water models.

For the second generation, we used OpenMM version 8.1.1 \cite{Eastman_openmm_2023}, which has the capabilities to perform RPMD and droplet simulations. Force-field parameters were obtained from multiple sources \cite{gonzalez2011flexible,sengupta_parameterization_2021,li2020systematic,li2021parametrization,jorgensen2005potential,dodda20171,dodda2017ligpargen,wang2004gaff,wang2005gaff_erratum,sambasivarao2009development,doherty2017revisiting} and we have set up an additional workflow to translate a curated set of parameters into OpenMM XML force field files that could be readily used and mixed for different electrolyte compositions. Water parameters were derived from the TIP4P/2005f model \cite{gonzalez2011flexible}. Metal cations and some small anions (e.g., I$^-$) used Lennard-Jones parameterization benchmarked with the TIP3P-FB water model \cite{sengupta_parameterization_2021,li2020systematic,li2021parametrization}. Neutral solvents and many ions were parameterized using Optimized Potentials for Liquid Simulations -- All-Atom (OPLS-AA) force-field parameters with 1.14*CM1A atomic partial charges obtained from the LigParGen online server \cite{jorgensen2005potential,dodda20171,dodda2017ligpargen}. For selected anions, we employed complementary parameters from various sources such as general AMBER force field (GAFF) \cite{wang2004gaff,wang2005gaff_erratum}, OPLS-2009IL, which is a variant of OPLS-AA designed for ionic liquids \cite{sambasivarao2009development,doherty2017revisiting}, and customized paramaters from the literature \cite{jang2004nanophase,ishida2009atom,wang2014atomistic,gupta2016force,mamatkulov2024unveiling}. While mixing parameters from different forcefields is generally discouraged due to potential incompatibilities in energy terms and molecular topologies, our aim is not to simulate accurate dynamics but to sample diverse electrolyte configurations. Chemically implausible or high-energy structures are not a drawback; in fact, they are crucial for training ML models to recognize and avoid unstable regions of the potential energy surface. 

For RPMD, we set up an OpenMM simulation using a Langevin integrator with Monte Carlo barostat for NPT ensemble simulations. We first perform energy minimization to relax any high-energy configurations followed by production runs. The simulation operates at a specified temperature and pressure (1.0 bar), with a default timestep of 1 fs, though these parameters were adjusted depending on the system requirements. For light elements, we used 32 beads and employed a ring-polymer contraction \cite{markland2008efficient} to reduce computational overhead, where direct space (short-range) forces use 3 beads and reciprocal space (long-range) forces use just 1 bead, while the rest of intramolecular forces use the full 32 beads. For droplet simulations, we configured a nonperiodic simulation and added a spherical harmonic container force. This container uses a harmonic potential with a force constant of 100 kJ/mol/nm$^2$ that activates when particles move beyond a defined radius. Both RPMD and droplet simulations were run for 20 ns and snapshots were taken every 0.2 ns.

\begin{figure}
    \centering
    \includegraphics[width=0.5\linewidth]{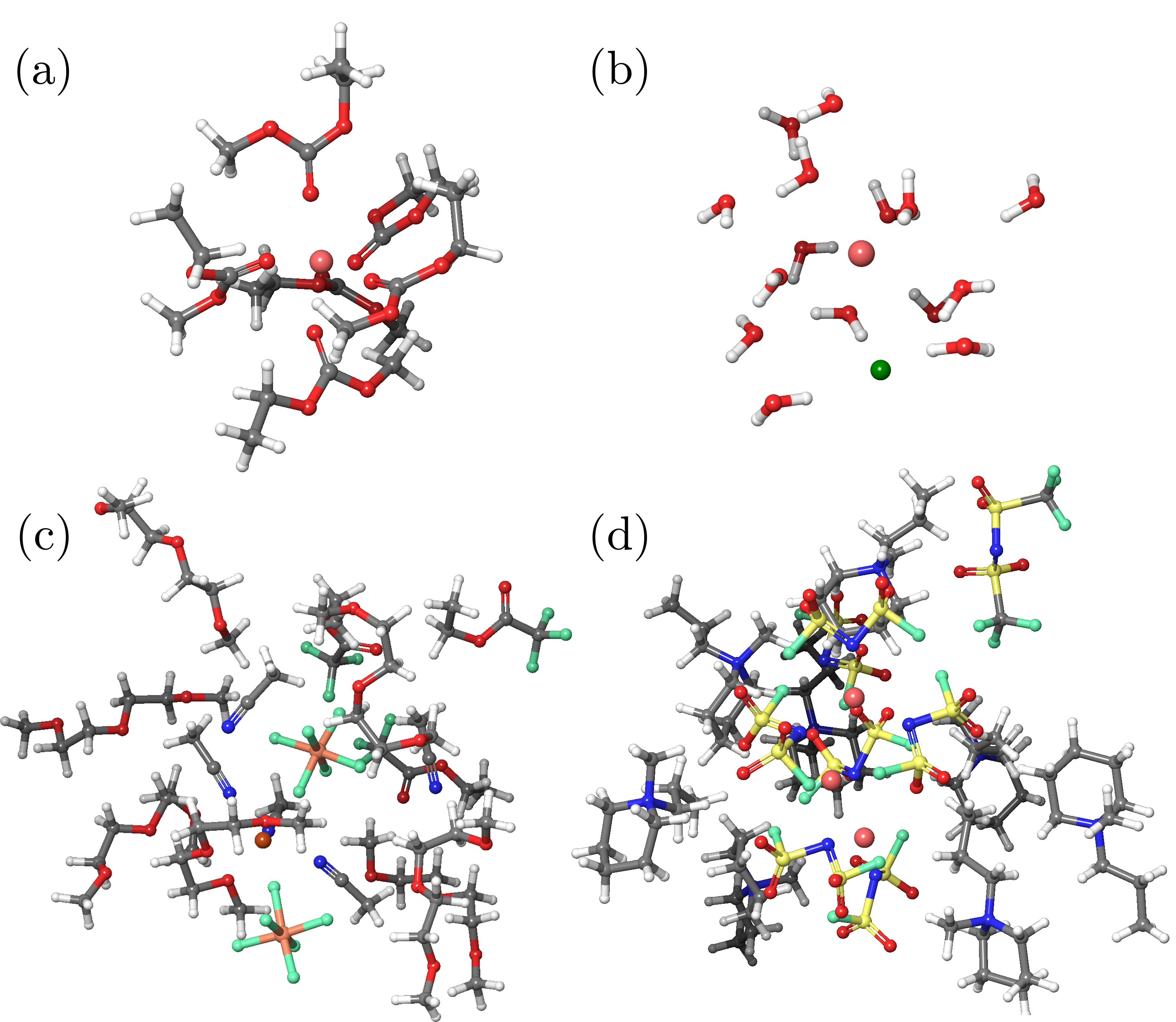}
    \caption{Examples of solvation shells extracted from MD simulations. (a) Li$^+$ ion in ethylene carbonate (EC) ethyl methyl carbonate (EMC), (b) Na-Cl pair in water, (c) $\rm Cu^{2+}$ paired with arsenic hexaflouride in an organic mixture, and (d) Sodium ions in ionic liquid.}
    \label{fig:shellextraction}
\end{figure}

\paragraph{Shell Extraction.} As MD simulations of bulk electrolytes often have considerable spatial and temporal redundancy, it is necessary to downselect diverse structures for inclusion in the final dataset. For each electrolyte system, the original MD trajectory was subsampled such that there are 100 equally spaced frames (every 2.5ns for first generation systems and every 0.2 ns for second generation); all of these frames were used as starting points for extraction. Where applicable, shells centered around each solute molecule were then extracted. To make the extraction procedure simple and scalable, we chose system-agnostic initial shell radii of 3\angs{}, 5\angs{}, and 7\angs{}, representing first, second, and third coordination shells. All molecules which had any atom within this distance from the central solute were extracted as part of the shell. Systems larger than a maximum atom count of 350 atoms were discarded. The extracted shells were then sorted into groups of conformers. From each group, a final set of 20 diverse structures were selected according to an approximate maximization of the minimum Root Mean Squared Deviation (RMSD) between structures (a random structure was chosen as seed and the most dissimilar structure by RMSD to all the structures previously taken were added iteratively to the list of accepted structures). We chose this approximate scheme since the Max-Min Diversity problem - the task of selecting a subset of $m$ elements from a set of $n$ elements such that the sum of the distances between the chosen elements is maximized - is in general NP hard \cite{erkut1990discrete,ghosh1996computational}. To explicitly target solvent-solvent interactions, the extraction procedure was repeated for solvent-centered shells; for this procedure, only shells that contained no solute were retained. These extracted clusters were then randomly, uniformly sampled for each central solute or solvent species, in the case of first generation systems. In the case of second generation systems, all ion-centered shells were taken, along with a sufficient number of solvent systems to more or less equal the ion-centered clusters. No attempt was made to compare clusters by RMSD between electrolyte system runs with the same species (e.g., between multiple systems with the same solvent for solvent-centered shells). As such, there may be lower diversity than exhaustive application of the Max-Min Diversity algorithm (either exact or approximate) over all electrolyte systems would imply. Samples of the results for the shell extraction can be found in \Cref{fig:shellextraction}.

\subsection{Small Molecule Generation}
\label{sec:methods:smallmol}

A library of small, ``electrolyte-like'' molecules were generated by selecting functional groups from a predefined collection and substituting these functional groups quasirandomly onto ``core'' structures. This library primarily serves to sample diverse bonded interactions. Note that the term ``electrolyte-like'' is used because many of the generated molecules are likely not useful for any practical electrolytes and may not even be synthesizable, but the ``core'' structural motifs reflect common electrolyte structures.

A collection of 77 ``core'' structures were selected by surveying the literature, in particular focusing on diverse battery chemistries and ionic liquids. Most cores contain only main-group elements, though there are several containing transition metals (\textit{e.g.}, Nb and Ta). We note that 36 of the 77 cores were taken from the Ionic Liquid --- Electrochemical Stability Window (IL-ESW) dataset developed by Moraes et al. Cores have as few as one site that can be substituted onto and as many as twelve. Seventy-two core structures were classified as in-distribution (ID), and five were designated to be out-of-distribution (OOD).

Similarly, 240 diverse functional groups containing only main-group elements were collated, including approximately 130 taken from the IL-ESW. Of these 240 functional groups, 228 were designated as ID and twelve were designated as OOD.

The general procedure for generating a small molecule library is as follows: for a given core, choose a random number of randomly selected sites ($n$) to substitute, where $1 \leq n \leq n_{\rm max}$ and $n_{\rm max}$ is the total number of substitution sites. For each of the $n$ sites, select a functional group. If the molecule currently has $m$ atoms, only functional groups smaller than $N - m$ atoms, where $N$ is a threshold size (here $N = 90$ atoms), are allowed. Among the set of allowed functional groups, the selection is weighted using $\text{weight} = 1/m_{\rm sub}$, where $m_{\rm sub}$ is the number of atoms of the substituent functional group. If no functional groups are small enough, then a hydrogen atom is added at the site. Likewise, hydrogen atoms are added to all $n_{\rm max} - n$ sites that were not chosen. This procedure is then repeated a set number of times $p$, generating a sub-library for the particular core.

Once a sub-library consisting of $p$ attempt molecules is constructed, it is then filtered. First, duplicate structures are identified by comparing InChI strings and removed. Molecules can also be removed because they contain certain chemical motifs, identified via SMARTS matching using RDKit. Specifically, molecules with uncommon motifs (\textit{e.g.}, P-Br bonds) are removed with a probability $1 - 0.8^x$, where $x$ is the number of uncommon motifs present. Molecules with more than one motif found in energetic materials (\textit{e.g.}, O-O or N-N bonds) are removed, as are all molecules with nitro groups bound to nitrogen (``N-N(=O)O'' in SMARTS). For molecules that pass all filters, 3D structures are generated from SMILES strings using the Software for Chemical Interaction Networks Molassembler package. If Molassembler could not generate a valid 3D structure in 25 attempts, for instance because of steric clashes in molecules with multiple bulky groups, the molecule was discarded.

Library generation was conducted in three stages. In the first stage, only ID cores and ID functional groups were used, and 3,000 attempts were made per core. In the second stage, ID cores were used with OOD functional groups with 250 attempts per core; each molecule generated in this stage includes at least one OOD functional group. Finally, in the third stage, OOD cores were used with both ID and OOD functional groups, and 3,000 attempts were again made per core. Molecules generated in the third stage were not required to include OOD functional groups. Molecules generated from the first stage are considered ID, and molecules from the second and third stages are OOD. In total, roughly 149,000 molecules were generated using this procedure, with 249,000 attempts made. The geometry of each molecule generated using this procedure was optimized for up to five steps. 

\subsection{Molecular Clusters}
\label{sec:methods:clusters}

The structures obtained from classical MD (Section \ref{sec:methods:md}) provide valuable information regarding non-covalent and non-bonded interactions but contain a relatively small number of unique species limited by the availability of force-field parameters. To supplement the MD-generated data, quasi-random molecular clusters including more diverse species were constructed.

The Architector package was used to pose initial structures for complexes. Architector can randomly dock a molecule or ion around another molecule and then leverage the semi-empirical quantum chemistry method GFN2-xTB to relax the random complex, providing a reasonable starting point for DFT optimization.

Cluster generation involved two sets of species: a set of central molecules around which complexes were generated, and a set of solvating species placed around the central molecules. In this work, the set of central molecules included small electrolyte-like molecules described in Section \ref{sec:methods:smallmol} with 50 or fewer atoms, and the set of solvating species, listed in the Appendix, resembles the set of molecules and ions included in MD simulations but includes, for instance, more unique metal ions.

Clusters are generated in three ways: 1) dimers; 2) full solvation-like structures; and 3) random clusters. All clusters were limited to a maximum size of 180 atoms.

A dimer is constructed by placing one random, valid solvating species around the central molecule. If the central molecule is charged, then solvating species of like charge to the central molecule are not considered, to ensure that the clusters do not separate due to electrostatic repulsion.

For full solvation-like structures, a single, neutral solvating species (the solvent) was randomly selected. Then, a target solvation shell size was selected on a Gaussian distribution, with $\mu = m + 60$ and $\sigma = 30$, where $m$ is the number of atoms in the central molecule. If the randomly selected budget was below $m + 20$ atoms, then it was shifted to $m + 20$. Likewise, if the budget was above the overall maximum size of 180 atoms, then it was set to 180. Then, $s$ copies of the solvent were placed around the central molecule, where $s = \lceil(\text{budget} - m)/m_{\rm solvent}\rceil$ and $m_{\rm solvent}$ is the number of atoms in the solvent.

Random clusters began with the determination of a size budget, using the same procedure as the full solvation-like structures. Then, solvating species are added one at a time. In each iteration, the pool of solvating species is reduced to include only those that are small enough to respect the size budget ($m_\text{solvating species} \leq \text{budget} - m$, where $m_\text{solvating species}$ is the number of atoms in the solvating species), and then one species is randomly selected from this reduced set, weighted by size ($\text{weight} = 1/m_\text{solvating species}$).

For each initial molecule, one dimer, one full solvation-like structure, and three random clusters were constructed using only ID solvating species. In addition, for 10\% of initial molecules, either a dimer, a full solvation-like structure, or a random cluster was generated including at least one OOD solvating species; this data was labeled OOD.

\subsection{Water Clusters}

We augmented the water-containing frames from above with 150,000 snapshots of 70 water molecules. 
To capture water structures more accurately, we use the AMOEBA forcefield \cite{Shi2013, Ponder2010, Ren2003}, which includes polarization effects absent from the non-polarizable forcefields, e.g., OPLS, used in the MD simulations of bulk electrolytes and flexible water molecules. 
These snapshots were created by simulating a box of water in OpenMM with Langevin dynamics and a Monte Carlo barostat at 300K and 1 atm for 150 ns, taking frames every 1 ps, following a 100 ps equilibration period. The integration time step was 2 fs and a cubic box 2 nm on a side was used. In each frame, the 70 water molecules closest to the center of the box were saved as a cluster.

\subsection{Noble Gases}

Though noble gases rarely form bonds with other elements, they nonetheless do appear variously in chemical applications, often entrained in a larger system. 
Instead of excluding the noble gases or including only a few examples, we aimed to represent them in environments that are most likely to reveal their intermolecular interactions, despite their relative lack of reactivity.
Hence, a 40\angs{} cubic box was filled with five atoms of each of the non-radioactive noble gases (He, Ne, Ar, Kr, Xe) and 5925 TIP4P-D \cite{piana2015water} water molecules. 
This system was equilibrated over 160 ps and then 500 ns of NPT MD with Desmond using the OPLS4 forcefield, taking frames every 250 ps. All molecules within 4\angs{} of each gas molecule were extracted as inputs for DFT (10,000 for each element), containing a noble gas atom and several water molecules.
\section{Molecular Dynamics with MLIPs}\label{sec:methods:mlmd}
To increase the configurational diversity of our inputs, we incorporated several ML sampling approaches capture reactive interactions that are unattainable by our non-reactive classical MD force fields. For this, we trained three independent EquiformerV2 (EqV2) \cite{equiformer_v2} models using an early snapshot of \omol{} across the different domains: biomolecules, metal complexes, and electrolytes. A pretrained MACE-MP-0 model was also used to provide more diversity.

For metal complexes, we generated input structures following the same pipeline described in \ref{app:mc:arch}. We then ran 1-2 ps of MD with the corresponding metal complex-trained EqV2 model from which 5 samples were randomly selected and used for DFT calculations. The same procedure was also performed with MACE-MP-0. MD was performed at a constant temperature of 300K with a time step of 1 fs. A small sample of data was also generated using a rattled Boltzmann sampling \cite{barroso-luque_open_2024} with $\mu=0A$, $\sigma=0.2A$, $N=1000$, and $T=500K$. Similarly, five random samples were selected for DFT calculation. For biomolecules, we generated protein interfaces as described in Section \ref{app:bio:ppi}, but without running classical MD. The EqV2 model trained on biomolecules was then used to run 2 ps of MD following the same procedure as above. The initial structure was then evaluated with DFT, along with two randomly selected samples. Electrolyte samples were generated in a similar manner; inputs were first generated in the same manner as described in Section \ref{app:elyte:desmond} with a new set of random electrolyte boxes. The corresponding EqV2 electrolytes model was used to run 2 ps of MD. The initial structure, along with four randomly selected samples, were then used as inputs for DFT calculations.

While we do not expect these early models to generate accurate MD trajectories, they provide a means to generate reactive configurations. In a similar vein, chemically unreasonable structures generated by these models along the MD trajectory can be particularly useful as training data to help the model learn what to avoid. The above procedure resulted in 1.6M metal complexes, 1.3M electrolytes, and 1.1M biomolecules structures calculated with DFT coming from ML sampling. 
\section{Main-group Molecules}\label{sec:methods:maingroup}
\subsection{Interpolated Reactivity Datasets}\label{sec:methods:interp_rxn}

To increase the number of reactivity samples within \omol, we utilize several reactivity datasets. Robustly sampling the reactivity of even simple systems can be quite challenging. Feasible reactions often require atoms to be carefully arranged, while remote moieties can strongly influence energetics. Moreover, the vast majority of reactions in the chemical literature describe overall chemical transformations which consist of several steps, rather than the elementary reaction steps that trace the movement of atoms. To sample structures containing reactivity, we use the RGD1 dataset~\cite{zhao_comprehensive_2023} of reactant-transition state-product triples, and the PMechDB~\cite{tavakoli_pmechdb_2024} and RMechDB~\cite{tavakoli_rmechdb_2023} databases of elementary reaction steps of polar and radical reactions, respectively. For RGD1, we carry out an interpolation in internal coordinate space from reactant to transition state to product, creating a series of frames that are likely near the minimum energy path. For RMechDB and PMechDB, we generate 3D structures, assign equivalent atom-numbering with the Schr\"odinger Suite, and then use the AFIR procedure as described in Section \ref{sec:dataset:mc:react}. All of these DFT inputs were run in the UKS formalism. This is described in more detail in sec. \ref{sec:app:react:mechdb}.

\subsection{Noble Gases}

Though noble gases rarely form bonds with other elements, they nonetheless do appear variously in chemical applications, often entrained in a larger system. 
Instead of excluding the noble gases or including only a few examples, we aimed to represent them in environments that are most likely to reveal their intermolecular interactions, despite their relative lack of reactivity.
Hence, a 40\angs{} cubic box was filled with five atoms of each of the non-radioactive noble gases (He, Ne, Ar, Kr, Xe) and 5925 TIP4P-D \cite{piana2015water} water molecules. 
This system was equilibrated over 160 ps and then 500 ns of NPT MD with Desmond using the OPLS4 forcefield, taking frames every 250 ps. All molecules within 4\angs{} of each gas molecule were extracted as inputs for DFT (10,000 for each element), containing a noble gas atom and several water molecules.

Additionally, MD boxes of 50 atoms of each noble gas and 50 each molecules of methane, ammonia, carbon monoxide, carbon dioxide, nitrogen, oxygen, hydrogen, and fluorine, or ozone, acetylene, ethylene, and benzene or 50 Li+, 50 Na+, 50 K+, 25 Mg2+, 100 F-, 50 Br-, 50 I- were generated and similarly equilibrated and then run with the above NPT MD procedure at 50, 100, and 150 atm pressure (this high pressure was to ensure a variety of structures were obtained with non-trivial interactions (i.e. within 6A) between noble gases and molecules/ions.

\subsection{Exotic main group}
Similar to the NPT procedure described for noble gases, small exotic main group species were also sampled at 50 atom. This consisted of a box containing diberyllium, H2NBH2, H3NBH3, allene, cumulene, PF3, BH3, S=PF3, AlF3, AlCl3, SeF2, BeF2, ClF, BrF, AlF, AlCl, AlH, BrCl, ICl, and IBr.

The Crystallographic Open Database was also mined for non-metal containing (most metal complexes in COD were held out as a test set) molecules containing the elements Ge, As, Se, Sb, In, Te, or containing a bond between any two of B, P, or As, or containing a three sulfurs bonded (i.e. S-S-S). The purpose was to extract exotic main-group compounds with low representation in other datasets or unusual bonding. Structure so obtained include borane clusters, P4 molecules, and a variety of heavy-main group substituted organic structures. Additionally, COD structures containing no metals and only the elements C, H, N, O, S, F, Cl, Br, I, and Si (after removing common solvents) were also extracted; among these structures were (doped and otherwise) fullerenes, corrannulenes, and other extended $\pi$-systems (purely organic molecules with crystal structures is biased towards more exotic systems than typical organic chemistry). Due to the challenging nature of these types of crystal structures, both systems with and without disorder were considered (for disordered systems, only the highest occupancy disorder groups were retained).

\subsection{Heavy main-group enumeration}
\subsubsection{Architector}

Certain heavy main-group elements were treated as "metals" in the Architector-based scheme described in sec. \label{{app:mc:arch}}. The oxidation state and coordination numbers are given in Table \ref{tab:hmg_metals}. Three sets of Architector setting were employed. One restricted the total number of atoms to 20 atoms, seeking idiosyncratic small heavy main-group systems (e.g. SbF5). Another removed ligands that themselves contained heavy main-group elements and allowed up to 120 atoms. A final set restricted to 120 atoms had no restrictions on the ligands used.

\begin{table}
\centering
\caption{\textbf{Heavy main-group elements, oxidation states, and coordination numbers sampled for OMol25 Architector main-group enumeration.}}
\label{tab:hmg_metals}
\noindent
  \begin{tabular}{lrl}
    \hline
    metal & oxidation state & coordination numbers \\
    \hline
Si & 4 & 4,5,6 \\
Ge & 2 & 2,3,4 \\
Ge & 4 & 2,3,4,5,6 \\
Se & 2 & 2,3,4 \\
Se & 4 & 2,3,4,5,6 \\
Se & 6 & 2,3,4,5,6 \\
Te & 2 & 2,3,4 \\
Te & 4 & 2,3,4,5,6 \\
Te & 6 & 2,3,4,5,6 \\
Sb & 3 & 2,3,4,5,6 \\
Sb & 5 & 2,3,4,5,6 \\
As & 3 & 2,3,4,5,6 \\
As & 5 & 2,3,4,5,6 \\
    \hline
  \end{tabular}
\end{table}

\subsubsection{ANI-2X substitution}

Random substitutions of molecules from the ANI-2X dataset were made. First, a selection of SMARTS patterns were written to describe common motifs for O, N, and C. Specifically, ROH, ROR', and R=O for oxygen, primary, secondary, and tertiary amines, imines, and terminal nitride (e.g. -C\#N) for nitrogen, and primary, secondary, tertiary (and in the case of $sp^3$, quaternary) $sp^3$ and $sp^2$ carbons, along with R=C-R' and R\#C for carbon. These were randomly replaced with Se/Te, P/As/Sb, or Si/Ge, respectively. Primary sites were givens 1/4 the weight as other sites owing to their relative ubiquity (there are a lot of terminal methyl groups) and lower structural diversity (they all have mostly H as ligand). Between one and four substitutions were made per molecule, weighted 4:4:2:1 (i.e. approximately four times as many systems have one substitution as four substitutions). To compensate for larger atomic radii in the heavy main group elements. Short (max 10 steps) force-field relaxations were applied to only those bonds attached to substituted atoms (that is, most of the molecule remained frozen except for the changing group). To ensure that the area away from equilibrium was also sampled, 15\% of the systems had a random bond containing a substituted atom stretched or compressed by between 10 and 20\%.

\subsection{Charge, spin, and proton decoration}

Protonation/deprotonation of both ANI-2X and the heavy-atom substituted version of ANI-2X were effected with the Epik package. This included more exotic protonations such as of benzene rings. The SPICE2, ANI-2X, main group COD, heavy-atom substituted ANI-2X were also randomly sampled in +1 and -1 charge states (as doublets), the +2 charge states (in both singlet and triplet spin states), and the neutral charge state but in the triplet spin state. These high-energy structures are needed to ensure that models trained on \omol{} behave appropriately for more extreme (but not inconceivable) states.

\section{Reactivity}

Reactive geometries were generated via four distinct approaches. 

\subsection{RGD1}\label{sec:app:react:rgd}

The RGD1 dataset \cite{zhao_comprehensive_2023} comprises 126,857 reactant, product, and transition state triplets for neutral, closed-shell species containing C, H, O, and N and up to ten heavy atoms generated via a combination of xTB and DFT simulations. For each reaction, we performed a geodesic interpolation \cite{10.1063/1.5090303} from reactant to transition state and from transition state to product, with each interpolation containing ten configurations. Removing one copy of the duplicate transition state structure, which is present in both interpolations, yielded 19 snapshots per reaction or a total of 2,410,283 snapshots across all reactions.

\subsection{RMechDB and PMechDB}\label{sec:app:react:mechdb}

RMechDB \cite{tavakoli_rmechdb_2023} contains over 5,300 elementary radical reaction steps, \textit{i.e} each with a single transition state. PMechDB \cite{tavakoli_pmechdb_2024} contains over 100,000 elementary polar reaction steps in a similar format. While these reactions are mainly organic, some do contain metals. 

Each reaction is provided as a SMIRKS string. We use Schrodinger's Fast3D to generate initial 3D geometries of the reactant and product structures from these. For structures without any metals, we then perform a UFF optimization with an fmax of 0.1 eV/\angs{} and at most 50 relaxation steps. If an RMSD greater than 100\angs{} is observed due to UFF optimization, we revert to the unoptimized structure. We then perform a GFN-FF optimization with an fmax of 0.1 eV/\angs{} and at most 50 relaxation steps on the results from the previous step and for those structures with metals.

Total charge and spin are then detected via Architector utilities. In practice, this only modifies charge and spin for species containing metals for which the charge and spin in RMechDB or PMechDB is suspect. For species with multiple metals, we enforce a low-spin (antiferromagnetic) configuration, which is justified by the fact that the reactions in the MechDBs with multiple metals typically have the metals close together. Species with isolated O$_2$ molecules have their multiplicity increased by two to account for the fact that O$_2$ is a triplet in the ground state.

We then apply a custom implementation of the applied force induced reaction (AFIR) procedure \cite{Sameera2016}, built atop Architector and ASE utilities, with MACE-MP-0 \cite{batatia2024foundationmodelatomisticmaterials} as our source of energy and forces to create frames which approximate the reaction pathway. Note that, while the assigned total charge and spin multiplicity have no impact on the energy and force evaluation, they do impact the DFT run on the snapshots that come out of the AFIR procedure.

Our AFIR implementation starts by identifying the broken and formed bonds based on the Architector connectivity graphs of the reactant and product structures. We then apply constraints in ASE that push together pairs of atoms that are supposed to form bonds in the reaction and pull apart atoms in bonds that break with a constant force. We then carry out an optimization of the reactant structure under these constraints using ASE's LBFGSLineSearch, with an fmax cutoff of 0.15 eV/\angs{}, an initial force constant of 0.1 eV/\angs{}, and at most 50 relaxation steps. If LBFGSLineSearch fails to converge, we switch to BFGSLineSearch instead, which remedies the vast majority of optimization convergence issues. Following optimization convergence, we check the minimum distance between any two atoms in any structure along the optimization trajectory. If a minimum distance less than 0.7\angs{} is observed, the AFIR procedure is terminated, as this was found to only occur when a pathological region of the PES with this MLIP was encountered. If no problematic interatomic distances are found, then we check if the intended bonds have broken and formed along the optimization trajectory. We use a bond breaking distance cutoff of 1.5\angs{} and a bond formation distance cutoff of 1.2\angs{}. We note that we do not check for or take steps to prevent the breakage or formation of bonds besides those intended. If the intended bonds have not all broken or formed, then we increment the constraint force constants by 0.2 eV/\angs{} and run another optimization starting from the end of the preceding optimization trajectory. This iterative procedure repeats until either all intended bonds have broken and formed, a problematic minimum distance is found, or force constants exceed 4 eV/\angs{} (\textit{i.e.}, at the 20th AFIR iteration). This creates one optimization trajectory with various force constants from reactant to product.

We then subsample the optimization trajectories to be simulated with DFT. We are most interested in structures in which bonds are actively breaking/forming, as such structures are unlikely to be captured in other areas of \omol, and we seek to avoid selecting multiple structures that are too similar. We begin selecting structures with the highest energy structure from along the optimization trajectory.

For each structure in the trajectory, we calculate the average force magnitude (without the AFIR forces) and the change in energy with respect to the previous and following frame. We also compute the minimum RMSD between this frame and all snapshots already selected. We select this frame if:
\begin{enumerate}
    \item $max(abs(F), deltaE) > 0.6$ and RMSD > 0.03 \angs{}
    \item $max(abs(F), deltaE) > 0.4$ and RMSD > 0.04 \angs{}
    \item $0.6 > max(abs(F), deltaE) > 0.4$ and RMSD > 0.04 \angs{}
    \item $0.4 > max(abs(F), deltaE) > 0.3$ and RMSD > 0.07 \angs{}
    \item $0.3 > max(abs(F), deltaE) > 0.18$ and RMSD > 0.1 \angs{}
    \item $0.18 > max(abs(F), deltaE)$ and RMSD > 0.14 \angs{}
\end{enumerate}

This procedure was empirically developed by testing on many different example reactions and based on the desire to sample approximately ten snapshots per reaction and the observation that the RMSD between adjacent snapshots in the geodesic interpolations described previously almost never went below 0.03 Angstroms. Running the AFIR and sampling procedure on all reactions yielded 67,302 RMechDB and 1,175,609 PMechDB total snapshots for DFT simulation.

\subsection{Metal Reactivity}

Metal complex template reactions were taken from MOR41 \cite{Dohm2018}, ROST61 \cite{Maurer2021}, and MOBH35 \cite{Semidalas2022}. Each reaction was examined, and of the 137 total reactions from these three sources, 14 were removed.

\begin{table}
    \centering
    \caption{Reactions removed from previous metal reactivity datasets.}
    \resizebox{0.93\textwidth}{!}{%
    \begin{tabular}{l|l}
     \textbf{Reactions removed} & \textbf{Reason} \\
     \hline
     \vspace{-10pt}\\
     ROST17, ROST18, ROST28, ROST58, MOBH18, MOBH20 & Duplicative of other reactions when swapping ligands \\
     ROS25, ROST36, ROST43 & Too many reactants/products implying not elementary\\
     ROST42 & Contained half a product species\\
     ROST55, MOBH21 & No bonds breaking or forming\\
     MOR32 & Too many active bonds implying not elementary
    \end{tabular}
    }
\end{table}

Schrodinger tools were then used to perform atom mapping and, when necessary, assemble reactant and product complexes from component species structures. All endpoint complexes were then manually inspected and validated, during which we realized that we could obtain two additional reactions by manually moving the location of the placed I2 species for MOR36 and MOR23. Thus, in total we ended up with 125 template reactions. We then execute a procedure to perform ligand swaps and metal swaps in order to generate hundreds of thousands of unique reactions from our 125 templates.

We visually and programatically inspected all reaction templates, assigning possible metal charges in order to facilitate swapping with like-charged metals while simultaneously identifying easily swapped ligands \cite{coordcomplexsampling2025}. Briefly, metals were removed and ligand charges were assigned using a modified OpenBabel formal charge assignment scheme accounting for the metal-bound atom valence (routines found in Architector). Visual inspection of the ligand charges resulting in manual overriding of 10-15 ligand charges assigned. With ligand charges assigned, metal oxidation states were selected that were both (1) common oxidation states according to the mendeleev package and (2) kept the total charge of the reaction complex less than +5 and greater than -3 to minimize charge deviations from the intended templates. 

To reduce the complexity of swapping ligands, we chose to only perform ligand swaps with ligands curated from the metal complex generation ligand set with matching charge state and denticity. Further, to minimize overlapping atoms and steric clashes with reaction sites, only a single ligand was swapped at a time, only ligands with denticity 2 or 1 were swapped, and no ligands involved in any bond breaking/forming in the template reaction were allowed to be swapped. Additionally, if there are multiple ligands with identical symmetries to a reaction site (\textit{e.g.}, single ligand dissociation from homoleptic octahedral complex, all 4 ligands \textit{cis-} to the dissociating ligand) then all such ligands were treated by a single swappable ligand site.

To perform ligand swaps, we leveraged Architector 3D functionalization capabilities, removing the ligand that is being swapped out, and adding the ligand that being swapped in, and relaxing with UFF while keeping the rest of the atom indices and positions fixed in both the reactant and product complexes. Spin states were determined for the swapped metal/oxidation states using mendeleev and any additional charge/spin present on the ligands was taken from the pre-tabulated values.

We note that 99 of our reaction templates contained one or more swappable ligands, while 34 did not. Given that those without any swappable ligands yielded only 1436 reactions after metal swapping, we decided to reserve those 1436 reactions for use as a test set. After enumerating all reactions with swappable ligands, we obtained $\sim$1.1M possible reaction templates. Some reaction templates contained vastly more possibilities after swapping ligands and metals, for example, reactions where 4 different monodentate ligands and 3 possible metal oxidation states could be swapped resulted in hundreds of thousands of possible reactions, while reactions with only 1 swappable bidentate ligand and 1 possible metal oxidation state resulted in only $\sim$2K possible reactions. To sample as evenly as possible across reactions given a computational budget of $\sim$5M AFIR snapshots we selected 250k reactions as a target, and selectively down-sampled including all swapped reactions with fewer than 3164 possible swaps, and 3164 randomly sampled swaps from reaction templates with more than 3164 possible swaps. After performing the swaps and removing reactions that produced overlapping atoms or distorted structures, we obtained 246,785 valid reactions, not including the hold-out test set of 1436 reactions.

All reactions obtained from ligand and metal swapping were subjected to the AFIR procedure described previously. This path was subsampled to generate DFT inputs of metal complexes in reactive geometries using both the highest and lowest spin configuration. The AFIR procedure can produce different results when run from reactant to product and product to reactant. Thus, when running the AFIR, we randomly swapped reactant and product with a 50\% probability to increase overall structural diversity.  From our 246,785 total ligand and metal swapped reactions, we obtained 5,145,142 snapshots from AFIR to be simulated with DFT which were run in up to two spin states (highest and lowest).

\subsection{Electrolyte Reactivity}
Electrolyte reactions were taken from previous work on reaction networks and mechanistic studies of electrolyte decomposition and solid-electrolyte interphase formation \cite{Xie2021, Barter2023, SpotteSmith2022, SpotteSmith2022a, SpotteSmith2023}. Schrodinger tools were used to construct 3D reactant and product complexes when they were not already available. When they were already available, Schrodinger tools were used to construct alternate versions of endpoint complexes. We removed any reactant-product pair whose RMSD was less than 0.3\angs{}. If we had both an original and Schrodinger endpoint structure, we retained both in our template set if the RMSD between them was greater than 0.4\angs{} and the sum of that RMSD and the RMSD of reactant to product was greater than 1.4. This procedure was empirically developed via testing and manual inspection of many different example reactions. If these cutoffs were not met, then only the Schrodinger endpoint structures were used.

We performed metal swaps for all reaction templates with at least one metal ion. Li was allowed to be swapped with all eight other metals with common +1 oxidation states according to mendeleev (i.e. Na, K, Cs, Cu, Ag, Rb, Tl, and Hg). Mg was allowed to be swapped with other metals with common +2 or +3 oxidation states according to mendeleev (+2: Ca, Zn, Be, Cu, Ni, Pt, Co, Pd, Ag, Mn, Hg, Cd, Yb, Sn, Pb, Eu, Sm, Cr, Fe, V, Ba, Sr, +3: La, Ce, Pr, Nd, Pm, Sm, Eu, Gd, Tb, Dy, Ho, Er, Tm, Yb, Lu, Al, Ga, In, Tl, Bi, Sc, Cr, Fe, Co, Y, Ru, Rh, Ir, Au). The reaction's charge and/or spin was adjusted accordingly. In order to account for the differences in atomic radii between swapped metals, the distance between the metal and each other distinct species in the reaction complex was scaled by the ratio of the new and old van der Waals radii. A total of 146,962 reactions were obtained following metal swapping. 

We attempted to apply our AFIR procedure to these electrolyte reactions, but neither MACE-MP-0 nor even B97-3c had an accurate description of the potential energy surface in the presence of undercoordinated metals/radicals, resulting in poor quality snapshots. The geodesic procedure used in Appendix \ref{sec:app:react:rgd} also proved insufficient. Thus, we instead employed the recently developed Path Optimization with a Continuous Representation Neural Network aka Popcornn method \cite{Khegazy2025}, using the geodesic model potential and loss function but with a continuous path representation during optimization. 17 snapshots were then sampled uniformly from the optimized path, and the middle 15 were selected for simulation with DFT (i.e. endpoints were discarded). Thus, approximately 2.2 million snapshots were generated by running Popcornn on all electrolyte reactions. 

In addition to reactive geometries that ended up in the test set due to composition-based splitting, we also generated two additional sets of reactive snapshots for use in testing which attempt to be explicitly out-of-distribution: Geodesic interpolations of 1782 organic reactions which are neither in Transition-1x nor RGD1, and AFIR snapshots of the 1436 reactions obtained from our metal complex templates which had no swappable ligands. These two sets in total contain approximately 67k snapshots for use as an OOD reactivity evaluation set. 
\section{Community Datasets}

SPICE \cite{Eastman_spice_2023,eastman_nutmeg_2024}, Transition-1x \cite{schreiner_transition1x_2022}, ANI-2x \cite{devereux_extending_2020}, and OrbNet Denali \cite{christensen_orbnet_2021} datasets were recalculated at the OMol level of theory without any modifications. Solvated protein fragments~\cite{https://doi.org/10.5281/zenodo.2605372} structures were randomly split into three groups: 10\% had an electron added, 10\% had an electron removed, the remaining 80\% were unchanged. These systems were then optimized with Sella with up to a maximum of two geometry optimization steps. ANI-1xBB \cite{zhang_ani-1xbb_2025} structures were removed if an isolated atom of elements B, C, N, O, Si, P, S, or Se was found, where an atom was considered to be isolated if the nearest other atom was greater than 1.8 times the sum of the two elements' covalent radii. Structures were also filtered if an isolated \ce{O2} or \ce{S2} species were identified based on a bonding graph generated by Architector, due to the ambiguity about the correct spin state for the system. All remaining structures were simulated with the standard \omol{} procedure for open-shell singlets. A random sample of 10\% of the structures were additionally simulated as triplets. GEOM~\cite{axelrod_geom_2022} structures were taken from $\sim$300k unique molecule families. For each family, $\sim$30\% of the conformers were selected for DFT calculations. For evaluations, unique families of conformers were selected and of the selected structures, 50\% had their atomic positions rattled with displacements sampled from a Gaussian distribution of $\mu=0$ and $\sigma=0.1$\angs. Structures were then optimized to a max per-atom force of 0.05 eV/\angs{} using Sella.
\section{Evaluation Details}

\subsection{Protein-ligand Interactions}\label{sec:app:eval:protein_ligand}

Protein-ligand pockets that were added to the BioLiP2 dataset between May 31, 2024 and Mar 13, 2025 were processed by the procedure described in Appendix \ref{sec:app:bio:prot-lig}. Drug-like pockets were identified by scoring the ligands by the QED method as implemented in RDKit~\cite{landrum_rdkitrdkit_2025}, retaining those with a value greater than 0.5, and requiring the number of rotatable bonds to be between 3 and 12, inclusive. Ligand-pocket interaction energy and interaction forces are defined as the difference between the energy/forces of the ligand-pocket complex and the isolated ligand and isolated pocket, with atom positions of each isolated component unchanged from the ligand-pocket complex and all calculations performed in vacuum. 

\subsection{Ligand Strain}\label{sec:app:eval:lig_strain}

The protein-ligand pockets described above also furnished the bioactive conformations of the systems used to the ligand strain evaluation. The local minimum of the ligand-in-pocket geometry is obtained by taking the union of:
\begin{enumerate}
    \item CREST, starting from the bioactive conformer
    \item RDKit's conformer generation combined with MM94FF optimization
    \item Schr\"odinger's MacroModel
\end{enumerate}

All of these conformers are re-optimized with xTB and de-duplicated if the maximum distance between corresponding atoms was less than 0.25\angs{} and the xTB energies are within 1 kcal/mol. The resulting pool of conformers were DFT-optimized with Sella and an fmax of 0.01 eV/\angs. The 5 lowest energy conformers and a random set of 5 were retained for the evaluation. That is, there are 10 conformers for each bioactive conformation.

\subsection{Conformers}\label{sec:app:eval:conformers}

A subset of GEOM molecule families is held out of train for constructing the evaluation set. For each family, optimizations are performed on all conformers with Sella and a max per-atom force of 0.01 eV/\angs. The 5 lowest energy conformers and a random set of 5 were retained for the evaluation. That is, there are 10 conformers for each molecule family. Linear sum assignment based on RMSD yields a mapping between DFT and MLIP optimized structures which minimizes the sum of the RMSDs across the ensemble, which allows for the scenario of different initial geometries finding the same minimized structure during DFT and MLIP optimization. The average of the RMSDs under the identified mapping is the ensemble RMSD evaluation metric. In order to construct an additional metric which incorporates the increased importance of the low energy structures which dominate the thermal ensemble, we define the Boltzmann weighted ensemble RMSD as the dot product between the vector of DFT Boltzmann weights and the vector of RMSDs under the identified mapping. Since the Boltzmann weights sum to 1, no further averaging is required. While the resulting metric does not account for the difference in conformer energies between DFT and the MLIP, such differences are directly evaluated via the $\Delta E$ and $\Delta E_{reopt}$ evaluation metrics described fully in the main text. 

We note that the purpose of the reoptimization component of the conformer evaluation is to have some metrics which are somewhat less tied to the sensitivity of the optimizer. Such sensitivity may cause optimizations from the same far-from-minima conformation to converge to different local minima, even when the potential energy surfaces the two optimizations are conducted on are extremely similar. By additionally starting MLIP optimizations from DFT-minimized structures, we probe how far MLIP minima are from DFT minima in a manner which should reduce the impact of optimizer sensitivity compared to optimizations which start from far-from-minima conformations. 

\subsection{Protonation Energies}\label{sec:app:eval:protonation}

Structures for the protonation energy evaluation are taken from Johnston et al.~\cite{johnston_epik_2023}, which includes families of organic molecules with variable protonation states. We subject all structures to tightly converged (fmax = 0.01 eV/A) geometry optimizations with Sella for both DFT and the MLIP being evaluated. For each structure, we can calculate the RMSD between the MLIP and ORCA optimized geometries. We further identify all pairs of structures within a family which differ by exactly one proton and calculate the $\Delta E$ between the optimized structures separately with DFT and the MLIP, where the MAE between the DFT and MLIP $\Delta E$ values thus informs upon the MLIP's ability to calculate protonation energy. 

\subsection{Unoptimized IE/EA and Spin Gap}\label{sec:app:eval:ip_ea}

while vertical IE and EA are typically calculated via energy differences between an initially relaxed structure and that same structure with an increased or decreased charge, we are equally interested in the model's ability to predict energy and forces at different charges and energy differences between charge states of the same geometry irrespective of whether or not a geometry is at a PES minima. For our ``unoptimized'' IE/EA evaluation task, we randomly select 4000 electrolyte structures from our test set with up to two open-shell metals, 1000 metal complexes constructed with Architector from our test set, and 4000 metal complexes from COD. Note that all of these systems are clusters of species, involving both covalent and non-covalent/coordination interactions, where the addition or removal of an electron is more complex than for a single small, isolated, fully connected molecule. For systems that are initially singlets, both charge-increased and charge-decreased snapshots are calculated as doublets. For systems that are initially doublets or a higher multiplicity, charge-increased and charge-decreased snapshots are calculated both with original multiplicity + 1 and original multiplicity - 1 and $\Delta E$ values are calculated between the original charge/spin and all four modified charge/spins. Our evaluation metrics are thus energy and force MAEs for charge-modified structures and both $\Delta E$ MAEs and $\Delta F$ MAEs between principal and charge modified structures.

As with IE/EA, we again opt for vertical spin gaps between static structures without any optimization. We randomly select 1000 electrolyte structures from our test set, 3000 metal complexes constructed with Architector from our test set, and 5000 metal complexes from COD, each constrained to have exactly one open-shell metal and at least two viable spin states. The full ladder of viable spin states are simulated for each complex, and $\Delta E$ and $\Delta F$ values are calculated between the highest possible spin and each other spin value.

\subsection{Distance Scaling: Short-range and Long-range Interactions}\label{sec:app:eval:scaling}

For our distance scaling evaluation task, we randomly select 2500 electrolyte structures from our test set without any open-shell metals, 1000 electrolyte structures from our test set with one open-shell metal, 1000 metal complexes constructed with Architector from our test set, 3000 metal complexes from COD, and 2000 PDB fragment structures from our test set. All systems are clusters of at least two non-covalently interacting species. For PDB fragments and both types of electrolyte samples, we scale the magnitude of the vectors between the center of mass of the full structure and the center of mass of each disconnected component, which modifies the distances between all disconnected components while keeping the intramolecular distances/interactions fixed. We randomly sample two scale factors in the compressive regime between 0.7 and 0.95, and ten scale factors in the short-range expansive regime 1.05 to 1.8. For PDB fragments and closed-shell electrolytes, we additionally sample ten scale factors in the long-range expansive regime 1.8 to 3.0. We do not do any long-range scaling for open-shell electrolyte because the resulting structures would contain a fully isolated open-shell metal, for which we have observed ORCA can struggle to find the lowest energy SCF solution. Given this, for COD and Architector metal complexes we simply scale the distance between the central metal and one coordinating ligand. More specifically, we employ Architector tools and uniformly sample ten points with the metal-ligand distance increased 0.4 Angstroms each up to 4 Angstroms. In the future, we hope to expand this range to include one snapshot in the compressive regime and multiple beyond four Angstroms.

In order to rigorously define which snapshots are in the ``short-range'' regime versus the ``long-range'' regime, we construct the neighbor graph for each structure with a 6\angs cutoff, and snapshots with multiple disconnected graph components are designated ``long-range''. Given an entire scan at variable scaled distances, we then identify the lowest energy short-range structure and calculate $\Delta E$ and $\Delta F$ values for each other structure relative to the that structure. If only long-range snapshots exist for a given scan, then we instead reference against the lowest energy long-range snapshot. We consistently reference against a short-range snapshot when one is available to ensure that we are not mixing long-range errors into short-range metrics.

\section{Baseline Models and Results}\label{sec:app:models}
eSEN~\cite{fu_2025_esen} and GemNet-OC~\cite{gasteiger2022gemnetocdevelopinggraphneural} were used as the primary baseline models for this work. eSEN represents the current state-of-the-art equivariant model on many community materials and molecular tasks including Matbench-Discovery~\cite{riebesell2023matbench} and SPICE-MACE-OFF~\cite{kovacs2023mace}. We train three variants of this model: small (eSEN-sm), medium (eSEN-md), and large (eSEN-lg). GemNet-OC was also trained, representing an accurate invariant model, that was previously state-of-the-art on the OC20~\cite{oc20} benchmarks. MACE~\cite{batatia2022_mace} was also trained, but only for the charge-neutral split as significant modifications would have been required to train across varying charges and spins.

\subsection{Training}\label{sec:app:training}
All models were trained on Nvidia H100 80GB GPU cards, with an AdamW optimizer, a learning rate of 8e-4, and a cosine learning rate scheduler. eSEN-sm was trained with FP32 precision. eSEN-md and eSEN-lg used a multi-stage scheme, where BF16 is used for the entirety of training followed by $\leq$ 1 epoch of finetuning at FP32 at a learning rate of 1e-4. We found this to provide more stable training for larger model sizes. GemNet-OC was also trained with BF16. Batch sizes generally varied to balance model sizes and reasonable training times. All models were trained for 12 epochs on the full dataset and 80 epochs on the 4M. On the neutral split, eSEN and MACE were trained for similar number of optimization steps, but had very different epoch equivalents. MACE incorporated a multi-stage training scheme, beginning with an energy coefficient of 10 but then finetuning with a coefficient of 50 near the end of training.

To account for variable total charge and spin in \omol~, we incorporate the same architectural change to all models to fairly compare the original baseline models. We first randomly initialize two embeddings based on the total charge and spin of the system, with output dimensions corresponding to the hidden channels of the respective model. The embeddings are concatenated and fed through a single linear layer. This embedding is added to the node embeddings and then subsequently at every layer of message passing. Although we recognize that this may not be an ideal approach for all architectures, we found it to give very positive results and provided a naive baseline for future works.

All energies were referenced using a "heat of formation" (HOF) reference:
\begin{align*}
    E_{ref} &= E_{DFT} - \sum_{i}^N \big[E_{i,DFT} - \Delta H_{f,i}\big] 
\end{align*}

Where $E_{DFT}$ is the total system energy coming from ORCA, $E_i$ is the energy of an isolated atom, $N$ is the number of atoms in the system, and $\Delta H_{f,i}$ is the heat of formation of atom species $i$ \cite{mendeleev}. Energies are then linearly referenced, following the same procedure in the OC22 dataset~\cite{oc22}.

\subsection{Total Energy Test Results}\label{sec:app:total_res}
Total energy results on the test splits and the corresponding breakdown results are provided in Tables \ref{tab:test_ood_total} and \ref{tab:test_breakdown_total}.

\subsection{Validation Results}\label{sec:app:val_res}
Results on the out-of-distribution composition validation set are provided in Table \ref{tab:val_breakdown_total}.

\subsection{Additional Results}\label{sec:app:addl_res}
Alongside the main test results, we provide additional out-of-distribution results in Table \ref{tab:test_ood_extra}. These include the TorsionNet500~\cite{rai_torsionnet_2022} benchmark dataset, curated out-of-distribution cations, solvents, and ions+solvents datasets. For each split, we evaluated the performance in predicting the energy and forces of the structure.

\begin{table}[h!]
\caption{Out-of-distribution composition \textbf{validation} results. Alongside the total metrics, results are also broken across biomolecules, electrolytes, metal complexes, and neutral organics. Total energy and force mean absolute error metrics are reported across the two different training splits - All and 4M.}
\begin{threeparttable}
\centering
\resizebox{\textwidth}{!}{%
\begin{tabular}{ll SS SS SS SS SS}
\multicolumn{12}{c}{Val-Comp} \\ \midrule
&
&
\multicolumn{2}{c}{Biomolecules} &
\multicolumn{2}{c}{Electrolytes} &
\multicolumn{2}{c}{Metal Complexes} &
\multicolumn{2}{c}{Neutral Organics} &
\multicolumn{2}{c}{Total} \\
\cmidrule(l){3-12} 
Dataset & Model & \mcc{Energy}   & \mcc{Forces}    
& \mcc{Energy}   & \mcc{Forces}     
& \mcc{Energy}   & \mcc{Forces}  
& \mcc{Energy}   & \mcc{Forces}  
& \mcc{Energy}   & \mcc{Forces}  \\
\midrule

\multirow{7}{*}{OMol-0} 
 & eSEN-sm-d. & 2.24 & 0.145 & 2.03 & 0.217 & 3.35 & 0.763 & 0.79 & 0.319 & 2.06 & 0.229 \\
 & eSEN-sm-cons. & 2.00 & 0.106 & 1.69 & 0.186 & 3.06 & 0.665 & 0.53 & 0.256 & 1.77 & 0.190 \\
 & eSEN-md-d. & 1.17 & 0.060 & 1.12 & 0.101 & 2.25 & 0.461 & 0.38 & 0.130 & 1.15 & 0.110 \\
 & GemNet-OC-r6 & 0.93 & 0.103 & 0.98 & 0.154 & 2.42 & 0.589 & 0.55 & 0.245 & 1.04 & 0.165 \\
 & GemNet-OC & 0.54 & 0.089 & 0.71 & 0.138 & 2.38 & 0.579 & 0.49 & 0.239 & 0.78 & 0.150 \\
 & MACE-OMol-L-0 & 6.12 & 0.153 & 4.62 & 0.255 & 3.75 & 0.731 & 1.31 & 0.348 & 4.56 & 0.252 \\
 & UMA-S-1.1 (OMol) & 1.55 & 0.130 & 1.39 & 0.219 & 2.76 & 0.707 & 0.56 & 0.311 & 1.45 & 0.221 \\
 & UMA-M-1.1 (OMol) & 1.36 & 0.075 & 1.05 & 0.134 & 2.16 & 0.507 & 0.34 & 0.146 & 1.14 & 0.138 \\
\midrule
\multirow{5}{*}{4M}  
 & eSEN-sm-d. & 2.94 & 0.187 & 3.11 & 0.291 & 4.45 & 0.933 & 1.38 & 0.465 & 2.99 & 0.300 \\
 & eSEN-sm-cons. & 2.90 & 0.142 & 2.71 & 0.257 & 3.61 & 0.815 & 0.97 & 0.390 & 2.65 & 0.256 \\
 & eSEN-md-d. & 1.62 & 0.078 & 1.81 & 0.150 & 3.29 & 0.630 & 0.77 & 0.214 & 1.78 & 0.156 \\
 & GemNet-OC-r6 & 1.47 & 0.135 & 1.89 & 0.216 & 3.53 & 0.775 & 1.11 & 0.382 & 1.84 & 0.227 \\
 & GemNet-OC & 0.91 & 0.120 & 1.30 & 0.194 & 3.42 & 0.755 & 0.95 & 0.360 & 1.33 & 0.207 \\
\bottomrule
\end{tabular}%
}
\begin{tablenotes}
  \item \kindatiny Energy (kcal/mol), Forces (kcal/mol/\angs)
\end{tablenotes}
\end{threeparttable}
\label{tab:val_breakdown_total}
\end{table}
\begin{table}[h!]
\caption{Structure to energy and force results across \textbf{additional test} splits. \textbf{Energy per atom} and force mean absolute error metrics are reported across the two different training splits - All and 4M.}
\begin{threeparttable}
\centering
\resizebox{\textwidth}{!}{%
\begin{tabular}{ll SS SS SS SS}
\toprule
&
&
\multicolumn{2}{c}{TorsionNet500} &
\multicolumn{2}{c}{OOD Cations} &
\multicolumn{2}{c}{OOD Solvents} &
\multicolumn{2}{c}{OOD Ions+Solvents} \\
\cmidrule(l){3-10} 

Dataset & Model & \mcc{Energy}   & \mcc{Forces}    
& \mcc{Energy}   & \mcc{Forces}     
& \mcc{Energy}   & \mcc{Forces}  
& \mcc{Energy}   & \mcc{Forces}  \\
\midrule

\multirow{7}{*}{OMol-0} 
 & eSEN-sm-d. & 0.07 & 0.062 & 1.73 & 0.147 & 1.11 & 0.166 & 0.93 & 0.085 \\
 & eSEN-sm-cons. & 0.04 & 0.043 & 0.77 & 0.111 & 0.61 & 0.135 & 0.42 & 0.058 \\
 & eSEN-md-d. & 0.03 & 0.022 & 0.80 & 0.062 & 0.58 & 0.082 & 0.44 & 0.036 \\
 & GemNet-OC-r6 & 0.06 & 0.057 & 0.79 & 0.100 & 0.68 & 0.118 & 0.59 & 0.065 \\
 & GemNet-OC & 0.06 & 0.060 & 0.56 & 0.093 & 0.56 & 0.111 & 0.44 & 0.061 \\
 & UMA-S-1.1 (OMol) & 0.04 & 0.054 & 0.96 & 0.146 & 0.66 & 0.154 & 0.53 & 0.080 \\
 & UMA-M-1.1 (OMol) & 0.03 & 0.021 & 0.50 & 0.080 & 0.41 & 0.098 & 0.25 & 0.042 \\
\midrule
\multirow{5}{*}{4M}  
 & eSEN-sm-d. & 0.12 & 0.094 & 2.67 & 0.207 & 1.51 & 0.212 & 1.40 & 0.122 \\
 & eSEN-sm-cons. & 0.06 & 0.069 & 1.40 & 0.165 & 0.87 & 0.179 & 0.76 & 0.086 \\
 & eSEN-md-d. & 0.05 & 0.031 & 1.35 & 0.089 & 0.90 & 0.123 & 0.73 & 0.048 \\
 & GemNet-OC-r6 & 0.10 & 0.083 & 1.33 & 0.141 & 1.03 & 0.164 & 0.99 & 0.091 \\
 & GemNet-OC & 0.09 & 0.081 & 1.02 & 0.127 & 0.82 & 0.151 & 0.65 & 0.083 \\
\bottomrule
\end{tabular}%
}
\begin{tablenotes}
  \item \kindatiny Energy (kcal/mol/atom), Forces (kcal/mol/\angs)
\end{tablenotes}
\end{threeparttable}
\label{tab:test_ood_extra}
\end{table}
\clearpage
\subsection{Wiggle150}

The Wiggle150 evaluation was recently introduced as a high-quality benchmark on strained conformers which uses CCSD(T)/CBS as the reference level of theory~\cite{brew_wiggle150_2025}. It consists of 50 conformers of three molecules. We evaluated the baseline models trained on \omol{} and compare to results taken from the Wiggle150 paper. As the \omol~dataset is computed with the $\omega$B97M-V functional and with a smaller basis than was used in Wiggle150 (def2-QZVP), it cannot be more accurate than this level of theory.

The errors of the baseline models against the reference theory in \omol{} are sufficiently small that all such models have errors less than or equal to 1 kcal/mol. This level of accuracy against coupled-cluster theory was achieved only by $\omega$B97M-V and two double-hybrid functionals in the original Wiggle150 paper.

\begin{table}[h!]
\caption{Comparison of baseline models trained on \omol{} and a sample of methods from the Wiggle150 benchmark~\cite{brew_wiggle150_2025}.$\dagger$ GPU-accelerated timings are also provided for baseline models.}
\begin{threeparttable}
\centering
\resizebox{0.8\textwidth}{!}{%
\begin{tabular}{l|c|c|c}
\toprule
Method & RMSE (kcal/mol) & MAE (kcal/mol) & time (s) \\ \midrule
DSD-PBEP86 & 0.90 & 0.72 & 2580 \\
DLPNO-MP2 & 1.13 & 0.88 & 15,300 \\
$\omega$B97M-V & 1.18 & 0.87 & 1990 \\
$\omega$B97X-V & 2.72 & 2.41 & 1930 \\
$\omega$B97X-D3 & 3.00 & 2.49 & 1870 \\
B3LYP-D3BJ & 1.84 & 1.41 & 1370 \\
M06-2x & 2.42 & 1.97 & 1460 \\
PBE-D3BJ & 5.68 & 4.91 & 244 \\
GFN2-xTB & 15.2 & 14.6 & 3.21 \\
AIMNet2 & 3.13 & 2.39 & 0.0170 \\
ANI-2X & 5.40 & 3.05 & 0.008 \\
Orb-V2-D3 & 11.1 & 9.25 & 0.216 \\
MACE-MP0 & 28.5 & 26.6 & 0.272 \\
\textbf{eSEN-sm-d.} &\textbf{1.31} & \textbf{1.00}& 0.0612 (0.024$\dagger$)\\
\textbf{eSEN-sm-cons.} &\textbf{1.25} & \textbf{0.91} & 0.142 (0.063$\dagger$)\\
\textbf{eSEN-md-d.} &\textbf{1.28} & \textbf{0.92} & 0.218 (0.039$\dagger$) \\
\textbf{GemNet-OC} & \textbf{1.25} & \textbf{0.90} & 0.192 (0.059$\dagger$) \\ \bottomrule
\end{tabular}%
}
\begin{tablenotes}
  \item $\dagger$ GPU-accelerated timings are also provided for baseline models.
\end{tablenotes}
\end{threeparttable}
\label{tab:wiggle150}
\end{table}

\clearpage

\subsection{Hyperparameters}\label{sec:app:params}
eSEN model and training parameters for the small, medium, and large models are provided in Table \ref{tab:params_esen}. GemNet-OC model and training parameters were taken from \href{https://github.com/facebookresearch/fairchem/blob/main/configs/oc22/s2ef/gemnet-oc/gemnet_oc.yml}{the original OC20 configuration} and also provided in Table \ref{tab:params_goc}. Two configurations of GemNet-OC were trained, the baseline with a cutoff radius of 12 and a cutoff radius of 6, to be consistent with eSEN-sm+md. GemNet-OC required doubling the batch size to train the All set, as compared to its 4M variant. MACE parameters are provided in Table \ref{tab:mace_params}.

\begin{table}[h!]
\caption{Model hyperparameters for the three eSEN model configurations trained in this work.}
\centering
\resizebox{0.7\textwidth}{!}{%
\begin{tabular}{l|c|c|c}
\toprule
Hyperparameters & eSEN-sm & eSEN-md & eSEN-lg \\ \midrule
\# sphere channels & 128 & 128 & 256 \\
lmax & 2 & 4 & 6 \\
mmax & 2 & 2 & 2 \\
max neighbors & 30 & 30 & 30 \\
cutoff radius & 6 & 6 & 12 \\
\# edge channels & 128 & 128 & 256 \\
distance function & gaussian & gaussian & gaussian \\
\# distance basis & 64 & 128 & 256 \\
\# layers & 4 & 10 & 16 \\
\# hidden channels & 128 & 128 & 256 \\
normalization type & rms\_norm\_sh & rms\_norm\_sh & rms\_norm\_sh \\
activation type & gate & gate & gate \\
ff\_type & spectral & spectral & spectral \\
\multicolumn{1}{l|}{} & \multicolumn{1}{l|}{} & \multicolumn{1}{l|}{} & \multicolumn{1}{l}{} \\
\# gpus & 32 & 64 & 128 \\
batch size ( \# atoms) & 89600 & 89600 & 44800 \\
energy coeff. & 10 & 10 & 10 \\
force coeff. & 5 & 5 & 5 \\ \bottomrule
\end{tabular}
}
\label{tab:params_esen}
\end{table}
\begin{table}[h!]
\caption{Model hyperparameters for training MACE.}
\centering
\resizebox{0.4\textwidth}{!}{%
\begin{tabular}{@{}lc@{}}
\multicolumn{2}{c}{MACE} \\ \midrule
Hyperparameters &  \\ \midrule
max\_ell & 3 \\
correlation & 3 \\
max\_L & 1 \\
num\_channels & 512 \\
num\_interactions & 3 \\
num\_radial\_basis & 8 \\
r\_max & 7 \\
irreps & 16x0e \\
batch size & 256 \\
energy coefficient & 10 (50 finetuning) \\
force coefficient & 10 \\ \bottomrule
\end{tabular}%
}
\label{tab:mace_params}
\end{table}

\begin{table}[ht!]
\caption{Model hyperparameters for training GemNet-OC.}
\centering
\resizebox{0.4\textwidth}{!}{%
\begin{tabular}{l|c}
\multicolumn{2}{c}{GemNet-OC} \\ \midrule
Hyperparameters & \\ \midrule
num\_spherical & 7 \\
num\_radial & 128 \\
num\_blocks & 4 \\
emb\_size\_atom & 256 \\
emb\_size\_edge & 512 \\
emb\_size\_trip\_in & 64 \\
emb\_size\_trip\_out & 64 \\
emb\_size\_quad\_in & 32 \\
emb\_size\_quad\_out & 32 \\
emb\_size\_aint\_in & 64 \\
emb\_size\_aint\_out & 64 \\
emb\_size\_rbf & 16 \\
emb\_size\_cbf & 16 \\
emb\_size\_sbf & 32 \\
num\_before\_skip & 2 \\
num\_after\_skip & 2 \\
num\_concat & 1 \\
num\_atom & 3 \\
num\_output\_afteratom & 3 \\
cutoff & 6,12 \\
cutoff\_qint & 6,12 \\
cutoff\_aeaint & 6,12 \\
cutoff\_aint & 6,12 \\
max\_neighbors & 30 \\
max\_neighbors\_qint & 8 \\
max\_neighbors\_aeaint & 20 \\
max\_neighbors\_aint & 1000 \\
RBF & gaussian \\
envelop & polynomial \\
CBF & spherical\_harmonics \\
SBF & legendre\_outer \\
output\_init & HeOrthogonal \\
activation & silu \\
quad\_interaction & True \\
atom\_edge\_interaction & True \\
edge\_atom\_interaction & True \\
atom\_interaction & True \\
num\_atom\_emb\_layers & 2 \\
num\_global\_out\_layers & 2 \\
 &  \\
\# gpus & 64 \\
batch size ( \# systems) & 512,1024 \\
energy coefficient & 1 \\
force coefficient & 10 \\ \bottomrule
\end{tabular}%
}
\label{tab:params_goc}
\end{table}
\clearpage

\end{document}